\documentclass[pra,superscriptaddress,reprint,amsmath,amssymb,nofootinbib]{revtex4-1}

\usepackage{graphicx}
\usepackage{dcolumn}
\usepackage{subfiles} 
\usepackage{bm}
\usepackage[usenames,dvipsnames]{xcolor}
\usepackage[pdftex,plainpages=false,colorlinks=true]{hyperref}
\usepackage[toc,page,title,titletoc,header]{appendix} 
\usepackage{multirow}
\usepackage{tabularx}
\usepackage[bottom]{footmisc}
\usepackage{caption}
\usepackage{subcaption}
\usepackage[bottom]{footmisc}
\usepackage{soul}

\usepackage{properties-HETEJ1900}
\usepackage{properties-IGRJ00291}
\usepackage{properties-SAXJ1808}
\usepackage{properties-XTEJ1814}
\usepackage{properties-XTEJ0929}

\usepackage{amssymb}% http://ctan.org/pkg/amssymb
\usepackage{pifont}% http://ctan.org/pkg/pifont

\newcommand{\cm}{\ding{51}}%
\newcommand{\xm}{\ding{54}}%

\newcommand{\Tdrift}{T_{\mathrm{drift}}}
\newcommand{\Tobs}{T_{\mathrm{obs}}}
\newcommand{\Tasc}{T_{\mathrm{asc}}}
\newcommand{\asini}{a_{\mathrm{0}}}
\newcommand{\fgw}{f_{\mathrm{gw}}}
\newcommand{\fgwdot}{\dot{f}_{\mathrm{gw}}}
\newcommand{\Porb}{P}
\newcommand{\Norb}{N_{\mathrm{orb}}}
\newcommand{\fact}{f_\mathrm{act}}
\newcommand{\fqu}{f_\mathrm{qu}}
\newcommand{\fdotact}{\dot{f}_{\mathrm{act}}}
\newcommand{\fdotqu}{\dot{f}_{\mathrm{qu}}}
\newcommand{\hsd}{h_{\mathrm{0,sd}}}
\newcommand{\fdot}{\dot{f}}
\newcommand{\fstar}{f_{\star}}
\newcommand{\fstardot}{\dot{f}_{\star}}

\newcommand{\TSFT}{T_{\mathrm{SFT}}}
\newcommand{\fdotacc}{\dot{f}_{\mathrm{acc}}}

\newcommand{\fline}{f_{\mathrm{line}}}

\newcommand{\Qstar}{Q^{\star}}

\newcommand{\SthPSG}{S_\mathrm{th}^{\mathrm{G}}}

\newcommand{\SthPSOT}{S_\mathrm{th}^{\mathrm{OT}}}

\newcommand{\So}{S_{\mathrm{\cup}}}
\newcommand{\Sa}{S_{\mathrm{a}}}
\newcommand{\Sb}{S_{\mathrm{b}}}
\newcommand{\Sth}{S_{\mathrm{th}}}
\newcommand{\fo}{f_{\mathrm{\cup}}}

\newcommand{\fb}{f_{\mathrm{b}}}

\newcommand{\Fstat}{\mathcal{F}}
\newcommand{\Jstat}{\mathcal{J}}

\newcommand{\Hz}{\mathrm{Hz}}
\newcommand{\Hzps}{\mathrm{Hz}~\mathrm{s}^{-1}}
\newcommand{\days}{\mathrm{d}}
\newcommand{\kpc}{\mathrm{kpc}}

\newcommand{\m}{\mathrm{m}}
\newcommand{\s}{\mathrm{s}}
\newcommand{\lts}{\mathop{\mbox{lt-s}}}

\newcommand{\NTasc}{N_{\Tasc}}
\newcommand{\Nasini}{N_{\asini}}
\newcommand{\NPorb}{N_{\Porb}}
\newcommand{\NTot}{N_{\mathrm{tot}}}

\newcommand{\TascOTwo}{T_{\mathrm{asc,O2}}}

\newcommand{\alphaN}{\alpha_{\NTot}}

\newcommand{\IaD}{~\cite{GallowayEtAl:2005,Galloway:2006}}
\newcommand{\SaD}{~\cite{SAXJ1808GallowayCumming:2006}}
\newcommand{\XbD}{~\cite{XTEJ0929Distance:2017}}
\newcommand{\XaD}{~\cite{XTEJ1814KraussEtAl:2005,XTEJ1814:StrohmayerEtAl:2003}}

\newcommand{\htorque}{h_{\mathrm{torque}}}
\newcommand{\Flux}{F_{\mathrm{X}}}
\newcommand{\FluxUnit}{\mathrm{erg}\,\mathrm{cm}^{-2}\,\mathrm{s}^{-1}}

\begin{document}

\preprint{}

\title{Search for gravitational waves from five low mass X-ray binaries in the second Advanced LIGO observing run with an improved hidden Markov model}

\author{Hannah Middleton}
\email{hannah.middleton@unimelb.edu.au}
\affiliation{%
 School of Physics, University of Melbourne, Parkville, Vic, 3010, Australia\\
}%
\affiliation{
 OzGrav-Melbourne, Australian Research Council Centre of Excellence for Gravitational Wave Discovery, Parkville, Victoria, 3010, Australia
}

\author{Patrick Clearwater}
\affiliation{%
 School of Physics, University of Melbourne, Parkville, Vic, 3010, Australia\\
}%
\affiliation{
 OzGrav-Melbourne, Australian Research Council Centre of Excellence for Gravitational Wave Discovery, Parkville, Victoria, 3010, Australia
}
\affiliation{
 Data61, Commonwealth Scientific and Industrial Research Organisation, Corner Vimiera \&  Pembroke Roads, Marsfield NSW 2122, Australia
}

\author{Andrew Melatos}
\affiliation{%
 School of Physics, University of Melbourne, Parkville, Vic, 3010, Australia\\
}%
\affiliation{
 OzGrav-Melbourne, Australian Research Council Centre of Excellence for Gravitational Wave Discovery, Parkville, Victoria, 3010, Australia
}

\author{Liam Dunn}
\affiliation{%
 School of Physics, University of Melbourne, Parkville, Vic, 3010, Australia\\
}%
\affiliation{
 OzGrav-Melbourne, Australian Research Council Centre of Excellence for Gravitational Wave Discovery, Parkville, Victoria, 3010, Australia
}

\date{\today}

\begin{abstract}
Low mass X-ray binaries are prime targets for continuous gravitational wave searches by ground-based interferometers.
Results are presented from a search for five low-mass X-ray binaries whose spin frequencies and orbital elements are measured accurately from X-ray pulsations: \HETEaName, \IGRaName, \SAXaName, \XTEbName, and \XTEaName.
Data are analysed from Observing Run 2 of the Advanced Laser Interferometer Gravitational-wave Observatory (LIGO).
The search algorithm uses a hidden Markov model to track spin wandering, the $\Jstat$-statistic maximum likelihood matched filter to track orbital phase, and a suite of five vetoes to reject artefacts from non-Gaussian noise. 
The search yields a number of low-significance, above threshold candidates consistent with the selected false-alarm probability. 
The candidates will be followed up in subsequent observing runs. 
\begin{description}

\item[DOI]
\end{description}
\end{abstract}

\maketitle

\section{Introduction}
\label{sec:introduction}

Ground-based gravitational-wave observatories such as Advanced LIGO~\cite{AdvancedLIGO:2015} and Advanced Virgo~\cite{AdvancedVirgo:2015} are searching for persistent, periodic gravitational wave (GW) signals.
One of the key targets for these continuous-wave (CW) searches are rotating neutron stars.  
At the present time, GWs have been observed from transient events comprising binary black hole ~\cite{GW150914, GW151226, GW170104, GW170814, GWTC-1:2018} and binary neutron star~\cite{GW170817,GW170817multi} coalescences, but no CW signal has been detected.
However the GW spectrum has only just been opened. 
Future CW detections will shed light on a number of fundamental physics questions, including the properties of bulk nuclear matter~\cite{Riles:2013,AnderssonEtAl:2011}.

Low mass X-ray binaries (LMXBs) are of particular interest for CW searches. 
LMXBs are composed of a compact object (such as a neutron star or stellar mass black hole) with a low mass stellar companion (typically $\lesssim 1M_{\odot}$)~\cite{xraybinaries:1997,LewinVanDerKlis:2006,CasaresEtAl:2017}. 
Within these binaries, matter is transferred from the low mass star to the neutron star, spinning it up. 
Electromagnetic (EM) observations show that these accreting neutron stars rotate below the centrifugal break-up frequency. 
Torque balance between spin up via accretion and spin down via GW emission is one way to explain the frequency distribution observed~\cite{BildstenTB:1998}.

Scorpius X-1 is a prime LMXB target for CW searches by Advanced LIGO and Advanced Virgo.
Scorpius X-1 shines brightly in X-rays, indicating a high accretion rate and potentially strong CW emission. 
Extensive searches have been made for Scorpius X-1 with multiple search pipelines and data sets from Initial LIGO and Initial Virgo ~\cite{SearchFStatS2:2007, SearchCStatS5:2015, SearchTwoSpecS6VSR2VSR3:2014, SearchRadiometerS5:2011, MeadorsEtAlS6LMXBSearch:2017}\footnote{A search for LMXB XTE J1751-305 was also made with Initial LIGO~\cite{MeadorsEtAlS6LMXBSearch:2017}.}.
Scorpius X-1 was also targeted in the first Advanced LIGO Observing Run (O1) from September 2015 to January 2016~\cite{ScoX1ViterbiO1:2017, SearchRadiometerO1:2017, SearchCrossCorrO1:2017}, and the second Advanced LIGO Observing Run (O2)~\cite{ScoX1ViterbiO2,RadiometerO1O2:2019}. 
O2 started in November 2016 with the LIGO instruments and was joined by Virgo for the final month before the run ended in August 2017. 
To date, no CW detection has been made. 
However upper limits have been placed on the GW strain from Scorpius X-1, the best being the cross-correlation search which gives a $95$\% confidence upper limit on the strain of $h_0^{95\%} \lesssim 2.3 \times 10^{-25}$ at $175\,\Hz$~\cite{SearchCrossCorrO1:2017,MeadorsEtAl:2018}.

Searches for CWs from Scorpius X-1 and other LMXBs face a number of challenges. 
First, the rotation frequency of the compact object wanders significantly during the observation.
A hidden Markov model (HMM) has the ability to track the wandering efficiently and accurately.
Following Refs.~\cite{ScoX1ViterbiO2,SuvorovaEtAl:2017,SuvorovaEtAl:2016}, this is the approach we apply in this paper.
A second challenge is that the rotation frequency is sometimes unknown from EM observations.
For Scorpius X-1, no EM pulsations have been observed from the system~\cite{WangEtAlScoX1Ephem:2018}.
Wide band searches need to be carried out, e.g. the recent O2 Scorpius X-1 search spanned  $60$--$650\,\Hz$~\cite{ScoX1ViterbiO2}.

Many LMXBs do have EM observations of pulsations, whose frequencies are measured with an accuracy of $\sim 10^{-8}\,\Hz$. 
It is these targets we focus on for this search. 
EM observations of pulsations greatly reduce the computational cost of the search, making these targets appealing for CW searches despite their lower X-ray brightness in comparison to Scorpius X-1. 

In this work we search data from O2, focusing only on data from the two LIGO observatories (due to the shorter duration of the Virgo data in O2). 
We present a search for five LMXB targets with well known rotation frequencies. 
The search method is identical to that used in Ref.~\cite{ScoX1ViterbiO2}. 
We briefly review it in Sec.~\ref{sec:SearchOverview}. 
The target list is described in Sec.~\ref{sec:targets} and the searched parameter ranges in Sec.~\ref{sec:targetParameterRanges}. 
We describe the application of this search to LMXB targets in Sec.~\ref{sec:o2Search} and Sec.~\ref{sec:vetos}. 
The results are presented in Sec.~\ref{sec:results}, followed by a brief discussion of the expected strain from LMXBs in Sec.~\ref{sec:expectedStrain}. 
The conclusions are summarized in Sec.~\ref{sec:conclusions}.

\section{Search Algorithm}
\label{sec:SearchOverview}

The LMXB search follows the same procedure as the O2 search for Scorpius X-1~\cite{ScoX1ViterbiO2}. 
Here we briefly review the search method, which is described in full in Refs.~\cite{ScoX1ViterbiO2,ScoX1ViterbiO1:2017,SuvorovaEtAl:2016,SuvorovaEtAl:2017}.

\subsection{HMM}
\label{sec:HMM}
In a Markov process, the probability of occupying the current state depends only on the previous state. 
In a hidden Markov process, the state is unobservable.
The LMXB targets of this search have observed spin frequencies. 
However, we note that a drift may exist between the rotation of the crust (where EM pulsations originate) and the core (which may or may not dominate the GW-emitting mass or current quadruple)~\cite{CrabSearch:2008,SuvorovaEtAl:2016}.
The CW frequency is therefore hidden.

Following the notation in Ref.~\cite{ScoX1ViterbiO2}, we label the hidden state variable as $q(t)$. 
This state transitions between some set of allowed values $\{q_1,\dots,q_{N_Q}\}$ at times $\{t_0,\dots,t_{N_T}\}$.
The probability of jumping from $q_i$ at time $t_n$ to $q_j$ at $t_{n+1}$ is given by a transition matrix $A_{q_jq_i}$ which depends only on $q(t_n)$. 
Measurements are made of some observable with allowed values $\{o_i,\dots,o_{N_o}\}$, and an emission matrix $L_{o_jq_i}$ relates the likelihood that an observation of $o_j$ relates to a hidden state $q_i$.
In a CW search, the observable is the interferometer data or some data product generated from it, e.g. a Fourier transform, or detection statistic.

The total duration of the observation is $\Tobs$. 
When searching for LMXBs, the observation is divided into $N_T$ equal parts, each of length $\Tdrift = \Tobs / N_T$.
Identically to Ref.~\cite{ScoX1ViterbiO2}, we take $\Tdrift=10$ days (other HMM searches use a shorter $\Tdrift$ depending on the type of target~\cite{MillhouseStrangMelatos:2020,PostMergerRemnantSearch:2019,SunEtAlSNR:2018}). 
For each segment, $L_{o_jq_i}$ is calculated from some frequency domain estimator, such as the $\Fstat$-statistic or $\Jstat$-statistic as discussed in Sec.~\ref{sec:jstat}.

Given an estimator, the probability that an observation $O=\{ o(t_0),\dots,o(N_T) \}$ is associated with a particular hidden path $Q=\{q(t_0),\dots,q(N_T) \}$ is
\begin{align}
P(Q|O) =& L_{o(t_{N_T}) q(t_{N_T})} A_{q(t_{N_T}) q(t_{N_T-1})} \dots L_{o(t_1) q(t_1)} \nonumber \\
& \times A_{q(t_1)q(t_0)} \Pi_{q(t_0)},
\end{align}
where $\Pi_{q(t_0)}$, the prior (i.e. the probability that the system starts in $q_i$ at $t=t_0$), is taken to be uniform.
Our objective is to find the optimal hidden path $\Qstar$ maximising $P(Q|O)$. 
The Viterbi algorithm~\cite{Viterbi:1967} achieves this in a computationally-efficient way by avoiding an exhaustive search of all paths. 
It is applied in Refs.~\cite{ScoX1ViterbiO2,SuvorovaEtAl:2016,SuvorovaEtAl:2017,ScoX1ViterbiO1:2017} to CW searches.

The Viterbi detection score~\cite{ScoX1ViterbiO1:2017} for a given path is defined as the number of standard deviations the path's log likelihood exceeds the mean log likelihood of all paths in a user-selected frequency sub-band containing the path's end state. 
The quantity $\delta_{q_i}(t_{N_T})$ is defined as the likelihood of the most likely path ending in state $q_i$ at step $N_T$.
The mean and standard deviation of $\ln \delta_{q_i}(t_{N_T})$, marginalized over a sub-band, are given by
\begin{eqnarray}
\mu_{\ln \delta (t_{N_T})} &=& \frac{1}{N_Q} \sum^{N_Q}_{i=1} \ln \delta_{q_i} (t_{N_T}), \\
\sigma^2_{\ln \delta(t_{N_T})} &=& \frac{1}{N_Q} \sum^{N_Q}_{i=1} \left[ \ln \delta_{q_i} (t_{N_T}) - \mu_{\ln \delta}(t_{N_T})\right]^2.
\end{eqnarray}
The Viterbi score for the path with the highest likelihood at step $N_T$, i.e. $\delta_{q^{\star}}$ for $q^{\star} = \mathrm{arg~max}_i \delta_{q_i}(t_{N_T})$, is then
\begin{equation}
S = \frac{\ln \delta_{q^{\star}}  - \mu_{\ln \delta}(t_{N_T}) }{ \sigma_{\ln \delta}(t_{N_T}) }.
\end{equation}
As in Refs.~\cite{ScoX1ViterbiO1:2017,ScoX1ViterbiO2}, we use the Viterbi score as our detection statistic throughout this paper.

\subsection{$\mathcal{J}$-statistic}
\label{sec:jstat}
A frequency domain estimator is used to convert the detector data into the probability that a signal is present at a frequency $f$. 
CW searches are carried out over many months, so the estimator must account for the motion of the Earth around the Solar System barycenter. 
The $\Fstat$-statistic~\cite{JKS:1998} is an example of an estimator used as a matched filter in CW searches for isolated neutron stars. 

In a binary the signal is Doppler modulated by the binary motion.
The orbital component of the phase varies with time $t$ as
\begin{equation}
\Phi_\mathrm{s}(t) = -2 \pi \fstar \asini \sin \Omega \left( t - t_a \right),
\label{eqn:jstatdoppler}
\end{equation}
where $\fstar$ is the rotation frequency of the star, $\asini$ is the projected semi-major axis, $\Omega = 2\pi/\Porb$ is the orbital angular velocity with orbital period $\Porb$, and $t_a$ is some reference time usually chosen to be the time of passage of the ascending node $\Tasc$. 
The $\Jstat$-statistic introduced in Ref.~\cite{SuvorovaEtAl:2017} extends the $\Fstat$-statistic matched filter to include binary orbital modulation. 
The orbital motion spreads the $\Fstat$-statistic power into orbital sidebands spaced by $1/\Porb$ and centred on $\fstar$. 
The $\Jstat$-statistic weights and sums these sidebands given a set of three binary parameters: $\Porb$, $\asini$ and $\Tasc$. 
The sum is performed coherently with respect to orbital phase. 
We make the assumption of circular orbits.
As in Ref.~\cite{ScoX1ViterbiO2}, we use the $\Jstat$-statistic as our estimator for this search.

\section{Targets}
\label{sec:targets}

The targets of this search are LMXBs.
In LMXBs, $\fstar$ is measured from X-ray observations of pulsations or burst oscillations~\cite{GWLMXBsWattsEtAl:2008}.
Accreting millisecond pulsars (AMSPs) are a subclass of LMXBs that can exhibit intermittent X-ray pulsations. 
These sources make interesting targets for CW searches because, in many cases, $\fstar$ is measured to better than $10^{-8}\,\Hz$. 
(Again, we emphasise that the signal frequency is not necessarily equal to $\fstar$; see Sec.~\ref{sec:HMM}).
Generally, AMSPs are transient; they have `active' (outburst) and `quiescent' phases. 
We denote the spin frequency of the star in its active and quiescent phases as $\fact$ and $\fqu$ respectively and use $\fstar$ as a general label for the spin frequency in either phase, whenever there is no need to distinguish between $\fact$ and $\fqu$.
As discussed in Sec.~\ref{sec:expectedStrain}, the frequency derivative $\fstardot$ has implications for the anticipated signal strength.

The traditional picture is that the active phase is associated with accretion onto the neutron star. 
Pulsations are often observed during outburst, whereupon $\fact$ and the $\fdotact$ can be measured directly. 
Active phases can last from weeks to years. 
Some sources pulsate persistently throughout the active phase, whilst others pulsate intermittently~\cite{WattsEtAl:2009}. 
In the active phase, $\fdotact$ is set by the accretion torque~\cite{GhoshLambPethick:1977}.

During quiescence, pulsations are not observed. 
However, $\fdotqu$ can be inferred from the difference in $\fact$ measured during the neighbouring active epochs.
In quiescence, $\fdotqu$ is usually set by magnetic dipole braking, although there may still be accretion taking place during these periods~\cite{MelatosMastrano:2016}.

CW emission is possible during both active and quiescent phases. 
Torque balance is possible when the gravitational radiation reaction torque cancels the accretion torque. 
It is also possible for the star to spin up or down under the action of positive or negative hydromagnetic accretion torques~\cite{GhoshLambPethick:1977,BildstenEtAl:1997}. 
Radiation reaction may contribute to a negative $\fdotact$ or $\fdotqu$. 
Equally, if a positive $\fdotact$ or $\fdotqu$ is observed, say due to accretion, it may outweigh and therefore mask a negative torque due to gravitational radiation reaction.

Deformations in the neutron star which are not aligned with the rotation axis produce CW emission. 
Neutron star mountains can be formed by magnetic stresses misaligned with the spin axis~\cite{BonazzolaGourgoulhonMagMountains:1996, MelatosPayneMagMountains:2005, HaskellEtAlMagMountains:2008, MelatosMastranoMagMountains:2012} and thermo-elastic deformations of a relativistic star~\cite{JohnsonMcDanielOwensElasticDeformation:2013}.
Another emission mechanism involves r-modes (Rossby waves due to the Coriolis effect). 
The r-mode oscillations are predicted to be excited by radiation-reaction instabilities and can persist into quiescence~\cite{rmodesOwenEtAl:1998,rmodesAnderson:1998}.

An overview of the known accreting neutron stars, as well as their prospects as GW targets, can be found in Ref.~\cite{GWLMXBsWattsEtAl:2008}. 
Below we briefly introduce the five sources we analyse in this paper.
The LMXB targets in this search have distances between $3.4$--$8.0\,\kpc$; see also sections \ref{sec:targetSumHETEa}--\ref{sec:targetSumXTEa}.
All targets are further away than Scorpius X-1 ($2.8 \pm 0.3\,\kpc$~\cite{WangEtAlScoX1Ephem:2018}). 
However, they are less than three times further away, so the wave strain should be comparable to Scorpius X-1, if the quadrupole moment and $\fstar$ are comparable too.
The binary properties are collated in Table~\ref{tab:targetDetails}.

\begin{table*}
\caption{\label{tab:targetDetails} 
Target list: position (right ascension and declination), orbital period ($\Porb$), projected semi-major axis ($\asini$ in light-seconds), time of ascension ($\Tasc$), and frequency of observed pulsations ($\fstar$). 
Quoted errors (in parentheses) are the $1\sigma$ uncertainties apart from \XTEaName, where they indicate the $90$\% confidence interval from Ref.~\cite{XTEJ1814PapittoEtAl:2007}.
}
\begin{tabularx}{\textwidth}{llllllll}
\hline
\hline
\\
Target   & RA & Dec  & $\Porb$($\s$)  & $\asini$ ($\lts$)  & $\Tasc$ (GPS time)  & $\fstar$ ($\Hz$)  & Refs. \\
\\
\hline
\\
\HETEaName & \HETEaRA & \HETEaDec & $\HETEaPorbE$ & $\HETEaAsiniE$ & $\HETEaObsTascE$ & $\HETEaNSFreqE$     & \cite{HETEJ1900KaaretEtAl:2006}\\
\IGRaName  & \IGRaRA  & \IGRaDec  & $\IGRaPorbE$  & $\IGRaAsiniE$  & $\IGRaObsTascE$  & $\IGRaNSFreqE$      & \cite{IGRJ00291SannaEtAl:2017,IGRJ00291Discovery:2004}\\
\SAXaName  & \SAXaRA  & \SAXaDec  & $\SAXaPorbE$  & $\SAXaAsiniE$  & $\SAXaObsTascE$  & $\SAXaNSFreqXMME$   & \cite{SAXJ1808HartmanEtAl:2008,SAXJ1808SannaEtAl:2017}\\
\XTEbName  & \XTEbRA  & \XTEbDec  & $\XTEbPorbE$  & $\XTEbAsiniE$  & $\XTEbObsTascE$  & $\XTEbNSFreqE$      & \cite{XTEJ0929Discovery:2002} \\
\XTEaName  & \XTEaRA  & \XTEaDec  & $\XTEaPorbE$  & $\XTEaAsiniE$  & $\XTEaObsTascE$  & $\XTEaNSFreqE$      & \cite{XTEJ1814PapittoEtAl:2007} \\
\\
\hline
\hline
\end{tabularx}
\end{table*}

\subsection{\HETEaName} 
\label{sec:targetSumHETEa}
\HETEaName~was first observed in outburst in 2005 by HETE-II (High Energy Transient Explorer-2)~\cite{HETEJ1900VanderspekEtAl:2005}. 
It has distance estimates of $\sim 4.3\,\kpc$~\cite{SuzukiEtAlHETEDist:2007} and $4.7\pm 0.6\,\kpc$~\cite{GallowayEtAlHETEDist:2008}.
Early observations by RXTE (Rossi X-ray Timing Explorer) revealed X-ray pulsations which were detected continuously for $22\,\days$ but became intermittent and then undetectable~\cite{HETEJ1900KaaretEtAl:2006,HETEJ1900GallowayEtAl:2008}. 
A spin-orbit model~\cite{ManchesterTaylorPulsars:1977} was used to compute a fit for the orbital parameters and pulsation frequency, yielding $\HETEaNSFreqE\,\Hz$ for the latter quantity.
On MJD $53559$ (8 July 2005) a brightening in the source flux was observed as well as a shift in frequency to $377.291596(16)\,\Hz$, after which pulsations became suppressed~\cite{HETEJ1900KaaretEtAl:2006}. 
The source remained in outburst without observed pulsations for $\sim 10\,\mathrm{years}$ until its return to quiescence in 2015~\cite{HETEJ1900DegenaarEtAl:2017}. 
In this paper we use the timing solution in Ref.~\cite{HETEJ1900KaaretEtAl:2006} computed from the period before the spin jump, when pulsations were observed.
There is no frequency derivative measured.

\subsection{\IGRaName}
\label{sec:targetSumIGRa}
\IGRaName~is the fastest known AMSP at $\IGRaNSFreqE\,\Hz$.
Distance estimates yield a lower limit of $4\,\kpc$ and an upper limit of $6\,\kpc$ from Refs.~\cite{GallowayEtAl:2005} and~\cite{Galloway:2006} respectively.
It was discovered in a $14\,\days$ outburst in 2004~\cite{IGRJ00291Discovery:2004,IGRJ00291TorresEtAl:2008}. 
Searches in the RXTE All Sky Monitor data indicate marginal evidence for two prior outbursts during 1998 and 2001~\cite{IGRJ00291PriorBurstsATel:2004,IGRJ00291DeFalcoEtAl:2017}.
A double outburst was observed in 2008, lasting $9\,\days$ in August and $15\,\days$ in September~\cite{IGRJ00921DoubleOutburst:2011,IGRJ00291PapittoEtAl:2011}.
The most recent outburst in 2015 lasted $25\,\days$.
A timing solution for the spin and orbital parameters was computed from the 2015 outburst in Ref.~\cite{IGRJ00291SannaEtAl:2017,IGRJ00291DeFalcoEtAl:2017}.
Several estimates of $\fstardot$ exist from active and quiescent periods (see Table~\ref{tab:expectedStrain}).
In this paper we use the timing solution from Ref.~\cite{IGRJ00291SannaEtAl:2017}.

\subsection{\SAXaName}
\label{sec:targetSumSAXa}
\SAXaName~is a regular outburster discovered in 1996 by the \emph{BeppoSAX} satellite~\cite{BeppoSAX:1997} with an estimated distance in the range $3.4$--$3.6\,\kpc$~\cite{SAXJ1808GallowayCumming:2006}.
Eight outbursts have been observed, the most recent of which was in 2019~\cite{BultEtAll:2019}.
As the 2019 outburst occured after O2, we use the most recent outburst prior to O2 which began in April 2015~\cite{SAXJ1808SannaEtAl:2017}. 
In Ref.~\cite{SAXJ1808SannaEtAl:2017} the spin and orbit parameters are computed from observations of the 2015 outburst by XMM-Newton and NuSTAR (Nuclear Spectroscopic Telescope Array). 
XMM-Newton and NuSTAR yield frequencies of $\SAXaNSFreqXMME\,\Hz$ and $\SAXaNSFreqNuSTARE\,\Hz$ respectively. 
Several observations of $\fstardot$ have been made in both active and quiescent phases (see Table~\ref{tab:expectedStrain}).
In this paper, we use the timing solution from Ref.~\cite{SAXJ1808SannaEtAl:2017}.

\subsection{\XTEbName}
\label{sec:targetSumXTEb}
\XTEbName~was discovered in outburst during two months in April--June 2002 by RXTE~\cite{XTEJ0929Discovery:2002}, the only outburst observed to date~\cite{XTEJ0929Distance:2017}. 
It has an estimated distance $>7.4\,\kpc$~\cite{XTEJ0929Distance:2017}.
The spin and orbital parameters were computed from RXTE timing data.
There is also an estimate of $\fstardot$ during the active phase~\cite{XTEJ0929Discovery:2002}.
The pulsation frequency is $\XTEbNSFreqE\,\Hz$ with $\fdotact = -9.2(4)\times 10^{-14}\,\Hzps$ (spin down). 
In this paper we use the timing solution in Ref.~\cite{XTEJ0929Discovery:2002}.

\subsection{\XTEaName}
\label{sec:targetSumXTEa}
\XTEaName~was discovered in outburst in 2003~\cite{XTEJ1814Discovery:2003} by RXTE. 
The outburst lasted $53\,\days$ and is the only one observed. 
Distance estimates range from $3.8\,\kpc$~\cite{XTEJ1814KraussEtAl:2005} to $\sim 8\,\kpc$~\cite{XTEJ1814:StrohmayerEtAl:2003}.
The spin and orbital parameters were computed via timing analysis.
Pulsations at $\XTEaNSFreqE\,\Hz$ were observed with $\fdotact = -6.7(7)\times 10^{-14}\,\Hzps$. 
In this paper we use the timing solution from Ref.~\cite{XTEJ1814PapittoEtAl:2007}.

\section{Spin, orbital, and astrometric parameters}
\label{sec:targetParameterRanges}

A targeted CW search requires the sky position [right ascension (RA) and declination (Dec)] of the source, needed for the $\Jstat$-statistic to account for the motion of the Earth with respect to the target.
To apply the $\Jstat$-statistic, three binary orbital parameters are also necessary: the orbital period $\Porb$, the projected semi-major axis $\asini$, and the orbital phase $\phi_a$. 
The phase of the orbit from X-ray observations is often quoted as the time of the ascending node $\Tasc$ where the Doppler-shifted frequency of the neutron star is lowest (the phase is also sometimes quoted as the time of inferior conjunction of the companion star, $T_{90} = \Tasc + \Porb/4$~\cite{GWLMXBsWattsEtAl:2008,XTEJ0929Discovery:2002}).
In this search we use $\Tasc$ for all targets.
EM observations of pulsations constrain the neutron star spin frequency $\fstar$.

The electromagnetically determined search parameters are summarized in Table~\ref{tab:targetDetails}.
Observations of X-ray pulsations during active phases are able to directly constrain $\fstar$ to high precision. 
The uncertainties in $\fstar$ are typically small, as Table~\ref{tab:targetDetails} shows. 
However, signal frequency is not necessarily identical to $\fstar$ (see also Sec.~\ref{sec:HMM} and Sec.~\ref{sec:searchDescriptionFreqBinning}). 
Timing solutions inferred from the Doppler-shifted pulsations allow the orbital parameters to be constrained (see $\Porb$, $\asini$, and $\Tasc$ in Table~\ref{tab:targetDetails}).

There are several mechanisms which can lead to CW emission from a rotating neutron star as described in Sec.~\ref{sec:targets}. 
Thermal or magnetic `mountains' emit at $2\fstar$ and possibly $\fstar$~\cite{BildstenTB:1998}.
R-mode oscillations emit at $\sim 4\fstar/3$~\cite{rmodesOwenEtAl:1998,rmodesAnderson:1998,Lee:2010}. 
Pinned superfluids emit at $\fstar$~\cite{MelatosDouglassSimula:2015} and $2 \fstar$~\cite{Jones:2010}. 
We also search for signals  at $\fstar/2$, where harmonics may exist.
In summary, we search bands containing $\{1/2, 1, 4/3, 2\}\,\fstar$ for each target as discussed in Sec.~\ref{sec:searchDescriptionFreqBinning}.

Identically to Ref.~\cite{ScoX1ViterbiO2}, we choose a sub-band size of $\sim 0.61\,\Hz$ (see Sec.~\ref{sec:o2Search} for details).
Previous CW searches have used sub-bands in the range $0.01$--$1\,\Hz$ depending on the target, algorithm, and search type (e.g. all sky, targeted)~\cite{BetzwieserEtAlCrabSearch:2009,knownPulsarO2:2019,SearchCrossCorrO1:2017,EinsteinATHomeAllSky:2017,EinsteinATHomeTargetet:2019}.

Recent developmental work on r-mode searches recommends scanning a relatively wide frequency range around the $4\fstar/3$ value~\cite{CarideEtAlrmodeSearch:2019}. 
For the targets considered here, we calculate the recommended search band using Eq. (17) in Ref.~\cite{CarideEtAlrmodeSearch:2019}.
\XTEaName~has the narrowest band ($253$--$291\,\Hz$), and \IGRaName~has the widest band ($669$--$940\,\Hz$).
The $\sim 0.61\,\Hz$ sub-bands searched in this paper are deliberately chosen to be narrower than these ranges. 
An exhaustive, broadband, r-mode search across hectohertz frequencies lies outside the scope of this paper, whose central objective is to conduct fast, narrowband searches at a selection of sensible harmonics of $\fstar$, taking advantage of well-measured EM observables in LMXBs for the first time. 
We postpone a broadband r-mode search to future work (see also Ref.~\cite{FesikPapa:2020} for a recent r-mode search). 

\begin{table*}
\caption{\label{tab:TascRange}
Propagated time of ascension just before the start of O2, along with the error (in parentheses) and the search interval. 
The error in the second column is the $1\sigma$ uncertainty of the propagated $\Tasc$ except for \XTEaName, where it is the $90$\% interval. 
The search intervals in the third column are the $\pm 3\sigma$ range apart from \XTEbName. 
For \XTEbName, the search interval is equal to the orbital period, covering $\Tasc \pm \Porb/2$.
 }
\begin{tabular}{p{3.7cm}p{4cm}p{4cm}}
\hline
\hline
\\
Target            & $\TascOTwo$ (GPS time) & Search range (GPS time)\\
\\
\hline
\\
\HETEaName&$\HETEaTascGPSPropE$&$\HETEaTascLow$--$\HETEaTascHigh$ \\
\IGRaName &$\IGRaTascGPSPropE$ &$\IGRaTascLow$--$\IGRaTascHigh$   \\
\SAXaName &$\SAXaTascGPSPropE$ &$\SAXaTascLow$--$\SAXaTascHigh$   \\
\XTEbName &$\XTEbTascGPSPropE$ &$\XTEbTascLow$--$\XTEbTascHigh$   \\
\XTEaName &$\XTEaTascGPSPropE$ &$\XTEaTascLow$--$\XTEaTascHigh$   \\
\\
\hline
\hline
\end{tabular}
\end{table*}

The search ranges for the orbital parameters are based on the uncertainty in the EM measurement. 
In many cases, the orbital parameters are known to high accuracy, reducing the computational cost. 
The error in $\Tasc$ is typically $<1\,\s$. 
However, the uncertainty in both $\Tasc$ and $\Porb$ means that the extrapolation becomes more unreliable the further it extends. 
If $\Tasc$ is measured several years before O2, we can use $\Porb$ to calculate a time when the binary returns to the same position in its orbit close to the O2 start time (at $T_{\mathrm{O2,start}}=1\,164\,562\,334$).
To propagate the combined error, we compute the number of orbits $\Norb$ between the observed $\Tasc$ and the time of ascension just before the start of O2 from 
\begin{equation}
\TascOTwo = \Tasc + \Norb \Porb,
\end{equation}
and the error for the propagated $\Tasc$ is 
\begin{equation}
\sigma_{T_{\mathrm{asc,O2}}}=\left[\sigma_{\Tasc}^2+(\Norb \sigma_{\Porb})^2\right]^{1/2},
\end{equation}
where $\sigma_{\Porb}$ and $\sigma_{\Tasc}$ are the errors on $\Porb$ and $\Tasc$ respectively. 
For all targets we choose to use $3\sigma$ uncertainties, except for \XTEbName~where we search a $\Tasc$ range equal to its orbital period ($\Tasc\pm\Porb/2$).
This search range achieves good coverage of the parameter space whilst keeping the computational cost manageable. 
The $\TascOTwo$ values used in the search are given in Table~\ref{tab:TascRange}.

\section{Searching O2 data}
\label{sec:o2Search}

Most CW searches of LIGO data, including this one, begin with short Fourier Transforms (SFTs) of time segments of the data.
Each SFT has duration $\TSFT = 1800\,\mathrm{s}$ (see Appendix~\ref{app:sftLengths}).
For each target, the first step is to compute the $\Fstat$-statistic `atoms', defined in Refs.~\cite{PrixAtoms:2011,SuvorovaEtAl:2017}. 
The data are split into $N_{T} = 23$ segments of duration $\Tdrift = 10\,\days$. 
The atoms are computed using the fixed values of RA and Dec in Table~\ref{tab:targetDetails}, which are typically known for LMXBs.

\begin{table*}
\caption{\label{tab:thresholds}
Sub-band frequencies, number of templates and associated Gaussian and off-target Viterbi score thresholds. 
The second column shows the sub-band frequencies for $\{\fstar/2, \fstar, 4\fstar/3, 2\fstar\}$ where the value displayed is the start of the sub-band, which is $\sim 0.61\,\Hz$ wide.
For each sub-band, the third and fourth columns show the number of $\Porb$ and $\Tasc$ templates searched respectively ($\NPorb$ and $\NTasc$) (note: $\Nasini = 1$). 
The fifth column shows the total number of templates searched ($\NTot$).
The final columns show the Viterbi score thresholds. 
The sixth column shows the Viterbi score threshold from identical searches on $100$ synthetic Gaussian noise realisations $\SthPSG$. 
The seventh column shows the Viterbi score threshold from identical searches on $100$ randomly selected sky positions in real data $\SthPSOT$. 
The thresholds are for a $0.30$ false alarm probability per sub-band. 
}
\begin{center}
\begin{tabular}{p{3.7cm}p{2.9cm}p{1.1cm}p{1.1cm}p{1.7cm}p{2.2cm}p{2.2cm}}

\hline
\hline
\\
Target & Sub-band start     & \multicolumn{3}{l}{Number of templates} & \multicolumn{2}{l}{Threshold Viterbi score} \\
       & frequency ($\Hz$)  & $\NPorb$ & $\NTasc$ & $\NTot$           & $\SthPSG$ & $\SthPSOT$   \\
\\
\hline
\\
\HETEaName & $\HETEaFBa$ & $\HETEaNPa$ & $\HETEaNTa$ & $\HETEaNTota$ & \HETEaSthGpsa & \HETEaSthOTpsa \\
           & $\HETEaFBb$ & $\HETEaNPb$ & $\HETEaNTb$ & $\HETEaNTotb$ & \HETEaSthGpsb & \HETEaSthOTpsb \\
           & $\HETEaFBc$ & $\HETEaNPc$ & $\HETEaNTc$ & $\HETEaNTotc$ & \HETEaSthGpsc & \HETEaSthOTpsc \\
           & $\HETEaFBd$ & $\HETEaNPd$ & $\HETEaNTd$ & $\HETEaNTotd$ & \HETEaSthGpsd & \HETEaSthOTpsd \\
\\
\IGRaName  & $\IGRaFBa$ & $\IGRaNPa$ & $\IGRaNTa$ & $\IGRaNTota$ & \IGRaSthGpsa & \IGRaSthOTpsa \\
           & $\IGRaFBb$ & $\IGRaNPb$ & $\IGRaNTb$ & $\IGRaNTotb$ & \IGRaSthGpsb & \IGRaSthOTpsb \\
           & $\IGRaFBc$ & $\IGRaNPc$ & $\IGRaNTc$ & $\IGRaNTotc$ & \IGRaSthGpsc & \IGRaSthOTpsc \\
           & $\IGRaFBd$ & $\IGRaNPd$ & $\IGRaNTd$ & $\IGRaNTotd$ & \IGRaSthGpsd & \IGRaSthOTpsd \\
\\
\SAXaName  & $\SAXaFBa$ & $\SAXaNPa$ & $\SAXaNTa$ & $\SAXaNTota$ & \SAXaSthGpsa & \SAXaSthOTpsa \\
           & $\SAXaFBb$ & $\SAXaNPb$ & $\SAXaNTb$ & $\SAXaNTotb$ & \SAXaSthGpsb & \SAXaSthOTpsb \\
           & $\SAXaFBc$ & $\SAXaNPc$ & $\SAXaNTc$ & $\SAXaNTotc$ & \SAXaSthGpsc & \SAXaSthOTpsc \\
           & $\SAXaFBd$ & $\SAXaNPd$ & $\SAXaNTd$ & $\SAXaNTotd$ & \SAXaSthGpsd & \SAXaSthOTpsd \\
\\
\XTEbName  & $\XTEbFBa$ & $\XTEbNPa$ & $\XTEbNTa$ & $\XTEbNTota$ & \XTEbSthGpsa & \XTEbSthOTpsa \\
           & $\XTEbFBb$ & $\XTEbNPb$ & $\XTEbNTb$ & $\XTEbNTotb$ & \XTEbSthGpsb & \XTEbSthOTpsb \\
           & $\XTEbFBc$ & $\XTEbNPc$ & $\XTEbNTc$ & $\XTEbNTotc$ & \XTEbSthGpsc & \XTEbSthOTpsc  \\
           & $\XTEbFBd$ & $\XTEbNPd$ & $\XTEbNTd$ & $\XTEbNTotd$ & \XTEbSthGpsd & \XTEbSthOTpsd  \\
\\
\XTEaName  & $\XTEaFBa$ & $\XTEaNPa$ & $\XTEaNTa$ & $\XTEaNTota$ & \XTEaSthGpsa & \XTEaSthOTpsa\\
           & $\XTEaFBb$ & $\XTEaNPb$ & $\XTEaNTb$ & $\XTEaNTotb$ & \XTEaSthGpsb & \XTEaSthOTpsb \\
           & $\XTEaFBc$ & $\XTEaNPc$ & $\XTEaNTc$ & $\XTEaNTotc$ & \XTEaSthGpsc & \XTEaSthOTpsc \\
           & $\XTEaFBd$ & $\XTEaNPd$ & $\XTEaNTd$ & $\XTEaNTotd$ & \XTEaSthGpsd & \XTEaSthOTpsd \\
\\
\hline
\hline
\end{tabular}
\end{center}
\end{table*}

\subsection{Number of orbital templates}
The next step is to define the search grid for each target in $\Porb$, $\asini$, and $\Tasc$ to compute the $\Jstat$-statistic.
It is assumed that $\Porb$, $\asini$, and $\Tasc$ remain within the same bin throughout the search. 
When performing a gridded search, it is unlikely that the true signal parameters fall exactly on a grid point or template; there is some mismatch between the signal parameter and the template parameter. 
The grid is marked out so as to keep the mismatch to an acceptable level whilst keeping the number of templates low enough to be computationally feasible. 
We follow the same procedure as in Ref.~\cite{ScoX1ViterbiO2}.
Using Eq. (71) of Ref.~\cite{LeaciPrix:2015}, the number of $\Porb$, $\asini$, and $\Tasc$ templates are
\begin{eqnarray}
\NPorb  &=& \frac{\pi\sqrt{2}}{2} \mu_{\mathrm{max}}^{-1/2} f \asini \frac{\gamma \Tdrift}{\sqrt{12}} \frac{2\pi}{\Porb^2} \Delta \Porb, \label{eqn:NPorb} \\
\Nasini &=& \frac{\pi\sqrt{2}}{2} \mu_{\mathrm{max}}^{-1/2} f \Delta\asini, \label{eqn:Nasini}\\
\NTasc  &=& \frac{\pi\sqrt{2}}{2} \mu_{\mathrm{max}}^{-1/2} f \asini \frac{2\pi}{P} \Delta \Tasc, \label{eqn:NTasc} 
\end{eqnarray}
where $\mu_{\mathrm{max}}$ is the maximum allowed mismatch, which we choose to be $10\%$ ($\mu_{\mathrm{max}}=0.1$) and $\gamma$ is defined in general in Eq. (67) of Ref.~\cite{LeaciPrix:2015}.
The factor $\gamma$ is a refinement factor introduced because the data are processed in $23$ separate segments; in the special case of the O2 data considered here where the segments are contiguous in time, we have $\gamma = N_T = 23$~\cite{LeaciPrix:2015}.
The values $\Delta\Porb$, $\Delta\asini$, and $\Delta\Tasc$ are the $3\sigma$ error bars on the EM measurements of $\Porb$, $\asini$, and $\Tasc$ respectively.
We make a conservative estimate of $\NPorb$, $\Nasini$, and $\NTasc$ by setting $f$ equal to the largest frequency value in each sub-band. 
The grid is uniformly spaced in each sub-band. 

For the five targets in this paper, we find $N_{\asini} < 1$ formally.
Hence we search over $\Porb$ and $\Tasc$ with $\asini$ held fixed. 
In contrast, for the O2 search for Scorpius X-1, the frequency dependent number of templates ranges from $\Nasini = 768$ and $\NTasc = 78$ for $60\,\Hz$ to $\Nasini = 8227$ and $\NTasc = 824$ for $650\,\Hz$~\cite{ScoX1ViterbiO2}. 
For seven sub-bands, we find $\NPorb <1$. 
The third and fourth columns of Table~\ref{tab:thresholds} show $\NPorb$ and $\NTasc$ respectively for each target and sub-band. 
Where Eq.~\ref{eqn:NPorb} predicts an even number for $\NPorb$, we round up by one to ensure that the central value from EM observations is searched (e.g. where $\NPorb=2$, we search $\NPorb=3$). 
The fifth column shows the total number of templates $\NTot = \NPorb \Nasini \NTasc = \NPorb \NTasc $ for each target and sub-band.

\subsection{Frequency binning}
\label{sec:searchDescriptionFreqBinning}
In Ref.~\cite{ScoX1ViterbiO2}, the search band is divided into equal sub-bands of width $\Delta f_{\mathrm{band}} = 2^{20} \Delta f_{\mathrm{drift}}=0.6068148\,\Hz$. 
The choice of a power of two for $\Delta f_{\mathrm{band}} / \Delta f_{\mathrm{drift}}$ speeds up the computation of the Fourier transform~\cite{DunnEtAl:2019}. 
We adopt the same binning strategy here\footnote{
The choice of sub-band varies between CW searches. 
An initial search for the Crab pulsar searched a range of $10^{-2}\,\Hz$ around $2\fstar$~\cite{BetzwieserEtAlCrabSearch:2009}, a search of O2 data for known pulsars used sub-bands in the range $0.06$--$0.81$$\,\Hz$ depending on the target~\cite{knownPulsarO2:2019}, the cross-correlation O1 search for Scorpius X-1 used $0.05\,\Hz$ sub-bands~\cite{SearchCrossCorrO1:2017}, and the Einstein@Home search used $0.05\,\Hz$ and $1\,\Hz$ sub-bands for recent all-sky and targeted searches respectively~\cite{EinsteinATHomeAllSky:2017,EinsteinATHomeTargetet:2019}.
}. 
Every ten days (i.e. $\Tdrift$), the frequency of the signal can increase or decrease by $\Delta f_{\mathrm{drift}} = 5.787037 \times 10^{-7}\,\Hz$, or remain the same. 
For each target we search the $\sim 0.61\,\Hz$ sub-bands which contain $\fstar/2$, $\fstar$, $4\fstar/3$ and $2\fstar$ (see Sec.~\ref{sec:targetParameterRanges}).
One advantage of the Viterbi algorithm is its speed, which allows us to search $\sim 0.61\,\Hz$ cheaply. 
The sub-band boundaries are identical to those used in Ref.~\cite{ScoX1ViterbiO2}; therefore the EM frequency may not be at the centre of the sub-band.  
There is no guarantee that the GW-emitting quadrupole and the X-ray emitting crust are exactly locked together; theoretical estimates of the balance between the nuclear pinning and Magnus forces in a neutron star predict a crust-core lag, for example~\cite{LinkEpsteinCrustCoreLag:1991,MelatosCrustCoreLag:2012}.
The starting frequencies of the sub-bands for each target are shown in the second column of Table~\ref{tab:thresholds}.

\subsection{Thresholds}
\label{sec:thresholds}

As described in Sec.~\ref{sec:jstat}, the $\Jstat$-statistic is applied to the SFTs to account for the Doppler modulation of the binary signal (Eq.~\ref{eqn:jstatdoppler}) using the template defined by the orbital parameters $\Porb$, $\asini$, and $\Tasc$. 
The Viterbi algorithm is then applied to find the best path through the $2^{20}$ frequency bins over $N_T=23$ segments for each template.

The result of the search is a Viterbi score $S$, as described in Sec.~\ref{sec:HMM}, corresponding to the most likely path for each orbital template ($\Porb$, $\asini$, $\Tasc$) and sub-band.
A path is a detection candidate if its Viterbi score exceeds a threshold $\Sth$ corresponding to a desired false alarm threshold. 
As the distribution of $S$ in noise-only data is unknown analytically, Monte-Carlo simulations are used to establish $\Sth$.

For our purposes, each sub-band is searched $\NTot$ times as listed in Table~\ref{tab:thresholds}. 
The more templates are searched, the more likely that a single template results in an above-threshold score due solely to statistical fluctuation (i.e. a false alarm). 
The probability of experiencing a false alarm during a search is called the false alarm probability (FAP).
The FAP can be defined as the probability of experiencing a false alarm when searching a single template ($\alpha$), or the probability of a false alarm when searching all $\NTot$ templates that constitute a search of a whole sub-band ($\alphaN$). 
For example, if we set $\alpha = 10^{-3}$ then the FAP of a sub-band amounts to $\approx 10^{-3} \NTot$ for $\NTot < 100$. 
The two probabilities $\alpha$ and $\alphaN$ are related by
\begin{equation}
\alphaN = 1 - \left( 1 - \alpha \right)^{\NTot}.  \label{eqn:FAP}
\end{equation}
We can therefore set $\alphaN$ and compute $\alpha$.

Previous, comparable, CW searches for Scorpius X-1 have set the FAP per sub-band between $0.01$ and $0.10$, which yields an expected $\sim 10$ candidates across the full band spanning $\sim 0.5\,{\rm kHz}$ and containing $\sim 10^2$ sub-bands~\cite{SearchCStatS5:2015,ScoX1ViterbiO1:2017,ScoX1ViterbiO2}. 
In this paper, where we search only four sub-bands per independent target, it is arguably too conservative to set the FAP in the above range, especially when one expects that any astrophysical signal lies near the detection limit of the experiment based on spin-down and torque-balance arguments (see Refs.~\cite{Riles:2013,BildstenTB:1998,PapaloizouPringleTB:1978,WagonerTB:1984} and Sec.~\ref{sec:expectedStrain}). 
We therefore set $\alphaN = 0.30$ in the analysis which follows. 
Looking forward to the results in Sec.~\ref{sec:results}, it turns out that $\alphaN=0.30$ yields ten above-threshold candidates in four of the targets, consistent with expectations, which are then screened by the vetoes in Sec.~\ref{sec:vetos}. 
Ten candidates is a modest number which keeps the veto workload manageable, while enhancing the chance of a detection. 
Naturally the reader is invited to set the FAP lower if desired. 
For example, if one sets the FAP to $0.10$ per sub-band, the search yields zero above-threshold candidates across all five targets (see Sec.\ref{sec:results}), again consistent with expectations.

It now remains to determine $\Sth$, which we do in two ways: in synthetic Gaussian noise, and in real data to represent the detector noise more faithfully. 
We generate Gaussian noise realisations for each target and sub-band to calculate the Gaussian noise threshold $\SthPSG$ directly. 
The realisations are generated using the tool \texttt{lalapps\_Makefakedata\_v4} in the LIGO Scientific Collaboration Algorithm Library (LAL)~\cite{lalsuite}\footnote{In the O2 Scorpius X-1 search~\cite{ScoX1ViterbiO2}, realisations were computed in seven sub-bands covering $60$--$650\,\Hz$, and $\Sth$ was extrapolated for computational efficiency.}.
We use $100$ realisations for each target and sub-band and the search is performed for each realisation. 
The $\alphaN = 0.30$ threshold from Gaussian noise realisations is shown in the sixth column of Table~\ref{tab:thresholds} for each target and sub-band. 
Typically $\SthPSG$ increases with $\NTot$, after allowing for statistical variations between sub-bands.

In reality, the LIGO noise is not Gaussian; it contains persistent harmonic features (lines).
Some bands are particularly corrupted.
In order to correct for this, we also perform the search at each $\Porb$, $\asini$, and $\Tasc$ template and sub-band for $100$ random off-target sky positions (varying RA and Dec) using the real O2 data. 
The off-target thresholds per sub-band $\SthPSOT$ are higher than the Gaussian thresholds, if the sub-band is noisy. 
The results for $\alphaN = 0.30$ are shown in the seventh column of Table~\ref{tab:thresholds}.
The scores in columns six and seven are rounded down to one decimal place to avoid rejecting marginal templates due to rounding errors. 
We see that there is little difference between $\SthPSG$ and $\SthPSOT$ except in the $\HETEaFBc\,\Hz$ and $\IGRaFBa\,\Hz$ sub-bands which contain known instrumental artefacts.

\section{Vetoes}
\label{sec:vetos}

Templates may produce Viterbi scores above the thresholds defined in Sec.~\ref{sec:thresholds}. 
We examine whether there are reasonable grounds to systematically veto these candidates as non-astrophysical sources. 
In Sec.~\ref{sec:vetoCriteria} we lay out the veto criteria following the method and notation of Ref.~\cite{ScoX1ViterbiO2}. 
Four vetoes are copied from Refs.~\cite{ScoX1ViterbiO1:2017,ScoX1ViterbiO2}. 
The off-target veto is new for this paper. 
In Sec.~\ref{sec:vetoScenarios} we explain how to classify the results of vetos 2 and 3.

\subsection{Veto descriptions}
\label{sec:vetoCriteria}

\subsubsection*{1. Known lines}
\label{sec:vetoLines}
The detector output contains many harmonic features (instrumental lines), which have been identified as noise as part of the detector characterisation process~\cite{CovasEtAl:2018}. 
The physical sources of these noise lines are varied.
Sometimes the effect can be mitigated, but not always.

A candidate in proximity to a known noise line~\cite{GWOSC} at frequency $\fline$ is vetoed, if the optimal HMM path $\fo(t)$ satisfies $|\fo(t) - \fline| < 2 \pi \asini \fo(t) / \Porb$ for any $t$ in the range $0 \leq t \leq \Tobs$.

\subsubsection*{2. Single interferometer}
The second veto is applied by searching the data from each interferometer separately. 
If a signal is astrophysical in origin, and if it is relatively strong, it should be found in the analysis of both detectors individually.
If an astrophysical signal is relatively weak, it may be found in neither detector individually, even though it is found in the combined data.
The interpretation of the Viterbi scores for single interferometer vetoes are described in Sec.~\ref{sec:vetoScenarios}.

\subsubsection*{3. $\Tobs/2$}
The third veto is applied by splitting the data in two segments and searching the two intervals separately. 
Again, if the signal is astrophysical and strong, it should be present in both halves of the data. 
If the signal is weak it may be below the threshold in both halves.

The O2 data are split so that the first segment covers $140$ days and the second $90$ days. 
This division is copied from Ref.~\cite{ScoX1ViterbiO2} and is chosen so that the effective observing time (accounting for the duty cycle of the interferometers) is approximately equal in the two segments.

As with the single interferometer veto, the interpretation of Viterbi scores from the $\Tobs/2$ veto is described in Sec.~\ref{sec:vetoScenarios}.

\subsubsection*{4. Off target search}
The fourth veto, which is new (cf. Refs.~\cite{ScoX1ViterbiO1:2017} and~\cite{ScoX1ViterbiO2}), is applied by searching in an off-target sky position. 
If the off-target search returns a score above threshold, the origin of the signal is likely to be instrumental. 
In this paper off-target means the position of the target plus $10\,\m$ and plus $10'$ for RA and Dec respectively.

\subsubsection*{5. $\Tdrift$}
The last veto is applied by analysing the frequency wandering of the Viterbi path~\cite{ScoX1ViterbiO1:2017}. 
A signal whose wandering timescale exceeds $\Tdrift$ should return a higher $S$, when $\Tdrift$ is increased to the observed wandering time-scale. 
This veto cannot be applied if the wandering timescale is already close to $\Tdrift$, which is the case in Ref.~\cite{ScoX1ViterbiO2} and also for this search (see Sec.~\ref{sec:results} and Appendix~\ref{app:completeSearchResults}).

\subsection{Veto scenarios}
\label{sec:vetoScenarios}
The interpretation of the Viterbi scores under the single interferometer and $\Tobs/2$ vetoes divides into four scenarios as in Ref.~\cite{ScoX1ViterbiO2}.
We label the original score by $\So$, the threshold score (see Sec.~\ref{sec:thresholds}) by $\Sth$, and the two veto runs by $\Sa$ and $\Sb$, i.e. the scores from two individual detectors or from the two halves of the data.

\emph{Category A.} One veto search returns a sub-threshold score whilst the other is higher than the original search:
\begin{equation}
\left( \Sa < \Sth \right) \land \left( \Sb > \So \right), 
\end{equation}
where $\land$ denotes Boolean AND. 
If the frequency $\fb$ associated with $\Sb$ is close to that of the original candidate $\fo$,
\begin{equation}
|\fo - \fb | < 2 \pi \asini \fo / \Porb, 
\end{equation}
then we conclude that the signal is likely to be a noise artefact in one detector or one half of the data. 
Category A candidates are vetoed.

\emph{Category B.} The situation is identical to Category A, except that the paths are not close, with $|\fo - \fb| > 2 \pi \asini \fo / \Porb$. 
The two veto searches may not have found the same signal in both detectors or halves of the data. 
This could be due to the search instead finding a noise artefact in one detector or half the data. 
Perhaps the signal is too weak to be detected in only one interferometer or half the data. 
Category B candidates cannot be vetoed.

\emph{Category C.} The candidate exceeds the threshold in both veto searches:
\begin{equation}
\left( \Sa > \Sth \right) \land \left( \Sb > \Sth \right). 
\end{equation}
This could represent a strong astrophysical signal.
Equally it could represent a noise source which is common to both detectors or present in the full observing run. 
Category C candidates cannot be vetoed.

\emph{Category D.} The candidate falls below the threshold in both veto searches:
\begin{equation}
\left( \Sa < \Sth \right) \land \left( \Sb < \So \right). 
\end{equation}
The origin of the combined detection is unclear. 
One possibility is that it is a weak astrophysical signal which requires the full data set to be detectable. 
Category D candidates cannot be vetoed.

\section{O2 Search results}
\label{sec:results}

In this section we present search results for the five LMXB targets listed in Table~\ref{tab:targetDetails}.
Ten templates have scores exceeding the $\alphaN = 0.30$ thresholds set in Sec.~\ref{sec:thresholds}. 
Of these: four have scores above both thresholds (i.e. $\So>\SthPSG$ and $\So>\SthPSOT$), three have $\So>\SthPSG$ only, and three have $\So>\SthPSOT$ only. 
We apply the veto procedure outlined in Sec.~\ref{sec:vetos} to the ten candidates. 
There are two further templates which have scores within $0.4\%$ of $\SthPSOT$ (see Figs.~\ref{fig:IGRaSearch} and~\ref{fig:SAXaSearch} in Appendix~\ref{app:completeSearchResults}).
No other templates are within $1.8\%$ of $\SthPSOT$.
For completeness we add these nearly above-threshold templates as candidates to be considered by the veto procedure, although note they do no meet our formal $\alphaN = 0.30$ FAP target. 
There are similarly close templates to $\SthPSG$ in two other sub-bands ($503.0\,{\rm Hz}$ and $299.0\,{\rm Hz}$ for \HETEaName~and \IGRaName~respectively). 
We do not include these in the veto procedure due to the presence of broad instrumental lines in these sub-bands (see Fig.~\ref{fig:HETEaSearch} and Fig.~\ref{fig:IGRaSearch} in Appendix~\ref{app:completeSearchResults}). 
We follow up 12 templates in total (ten above-threshold and two nearly above-threshold).

\begin{figure}
\begin{center}
\includegraphics[width=0.49\textwidth]{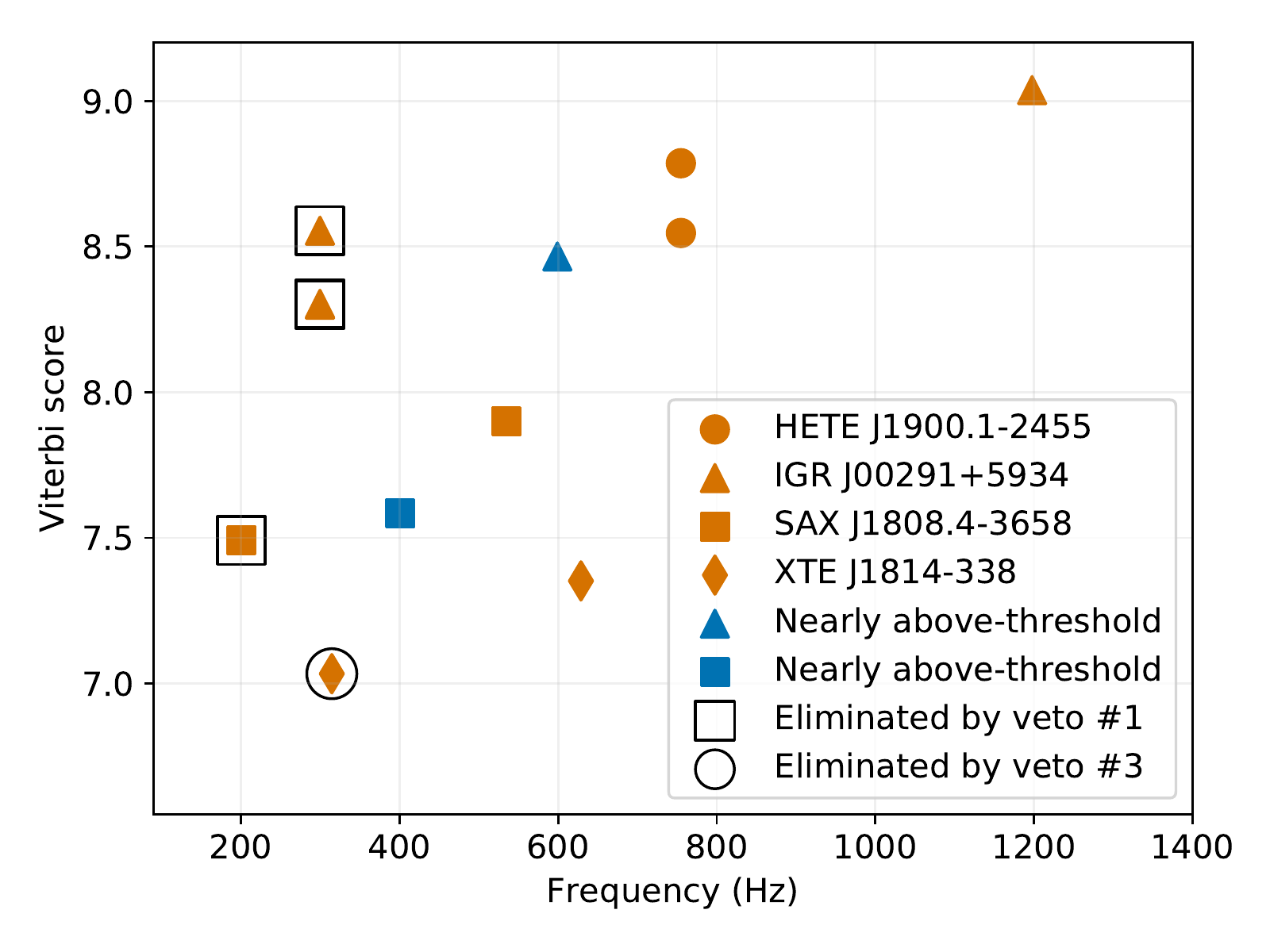}
\end{center}
\caption{\label{fig:candidateSummary}
Summarized search results for all LMXB targets.
The horizontal axis shows the frequency at the end of the best path for the template and the vertical axis shows the associated Viterbi score. 
Each orange marker indicates a template which has resulted in a path with a Viterbi score higher than either of the Gaussian or off-target thresholds or both. 
Results for \HETEaName, \IGRaName, \SAXaName, and \XTEaName~are shown by the circle, triangle, square, and diamond markers respectively. 
There are no above-threshold candidates for \XTEbName. 
The additional blue markers for \IGRaName~and \SAXaName~(at $\sim 400\,\Hz$ and $\sim 600\,\Hz$ respectively) indicate templates with $\So \approx \SthPSOT$ which are included for safety in the face of rounding errors.
Candidates outlined with a black square or circle indicate templates vetoed by veto 1 (known lines) and veto 3 ($\Tobs/2$) respectively. 
We note that there are two almost identical \IGRaName~markers at $\So\approx 8.3$, $f\approx 300\,\Hz$ (indistiguishable in the figure). 
Both are eliminated by veto 1.  
}
\end{figure}

The Viterbi scores and frequencies of the candidates are summarized in Fig.~\ref{fig:candidateSummary}. 
Each marker shows the terminating frequency of a candidate's Viterbi path (i.e. $q^{\star}$ as defined in Sec.~\ref{sec:HMM}) and the associated Viterbi score. 
Candidates for each target are shown by different marker shapes.
The ten above-threshold candidates and two nearly above-threshold templates are shown in orange and blue respectively.
Candidates which are removed through the veto procedure are indicated by the black-square and black-circle outlines for elimination by veto 1 and veto 3 respectively.

The FAP per sub-band in this paper is deliberately set higher than in previous, comparable, CW searches for Scorpius X-1, because the total number of sub-bands is lower, as noted in Sec.~\ref{sec:thresholds}. 
If we set the FAP per sub-band to $0.20$ instead of $0.30$, all candidates except three fall below the Gaussian and off-target thresholds and do not graduate to the veto stage. 
One of these candidates is the highest scoring template in the \IGRaName~$\fstar/2$ sub-band ($\So\approx 8.5$), which is eliminated by veto 1. 
The other two candidates are the two highest scoring templates in the \HETEaName~$2\fstar$ sub-band ($\So\approx 8.5$ and $\So\approx 8.8$) which survive the veto procedure. 
If we set the FAP per sub-band to $0.10$ instead of $0.30$, zero candidates exceed the Gaussian and off-target thresholds. 
The reader is invited to experiment with various choices of FAP when reproducing the results.

The search uses a combination of central processing unit (CPU) and graphical processing unit (GPU) computation.
We work with a GPU implementation of the $\Jstat$-statistic identical to the one in Ref.~\cite{ScoX1ViterbiO2}.
We use the computing facilities of the OzSTAR supercomputer~\cite{OzSTAR}.
OzSTAR has compute nodes with Xeon Gold 6140 CPUs running at $2.3\,{\rm GHz}$~\cite{CPU:Intel} and NVIDIA P100 12GB PCIe GPU cards benchmarked at $\approx 9.3$ TeraFLOPS (floating point operations per second) single-precision performance~\cite{GPU:NVIDIA}.
On OzSTAR the search takes $\approx 5$ CPU-hours and $\approx 23$ GPU-hours (every GPU-hour also requires one CPU-hour). 
The false alarm thresholds take $\approx 1000$ CPU-hours and $\approx 4600$ GPU-hours to perform $100$ searches on Gaussian and real data off-target realisations.

In sections~\ref{sec:HETEaResults}--\ref{sec:XTEaResults} we summarize the search results for each of the five targets. 
The results are laid out in full in Fig.~\ref{fig:HETEaSearch} for one target only, namely \HETEaName, to guide the reader without cluttering the main body of the paper.
The complete search results, including veto outcomes and optimal Viterbi paths are collated for reference and reproducibility in Appendix~\ref{app:completeSearchResults}. 
The O2 search returns five veto survivors. 
A search of LIGO O1 data narrowly targeted at the templates in the sub-bands containing the five veto survivors shows no support for an astrophysical signal.

\subsection{\HETEaName}
\label{sec:HETEaResults}

\begin{figure*}
\begin{center}
\includegraphics[width=.49\textwidth]{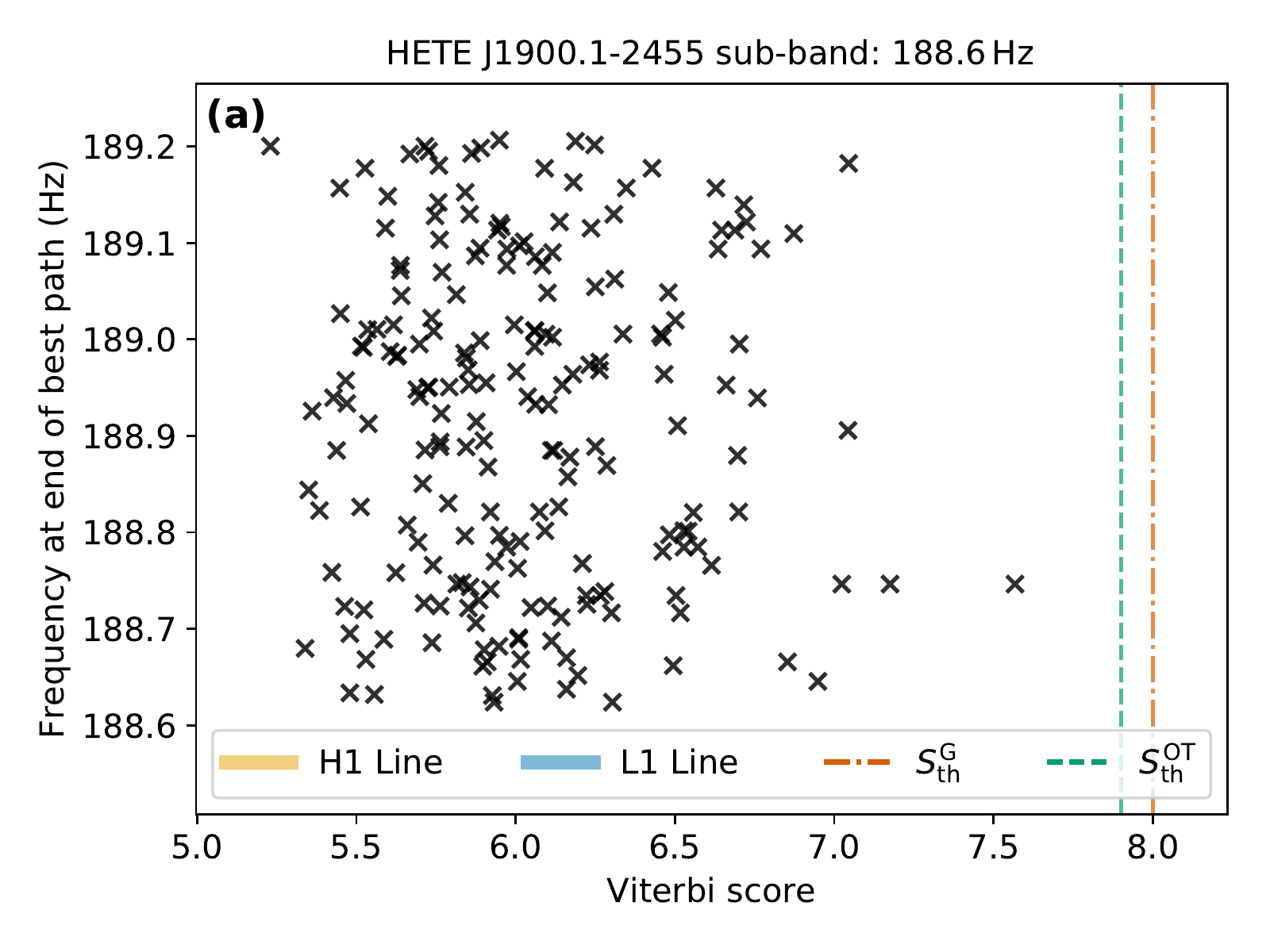}
\includegraphics[width=.49\textwidth]{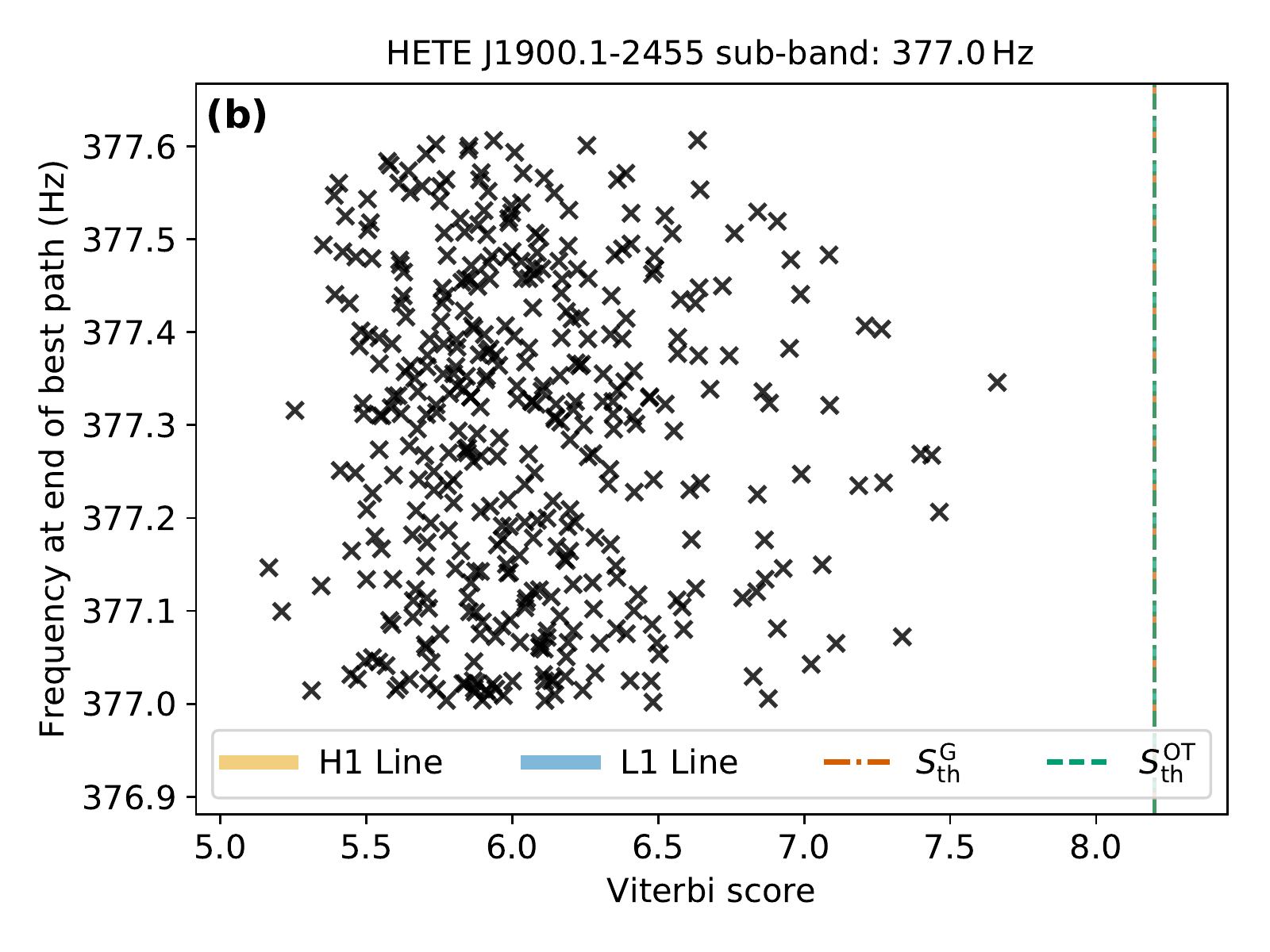}\\
\includegraphics[width=.49\textwidth]{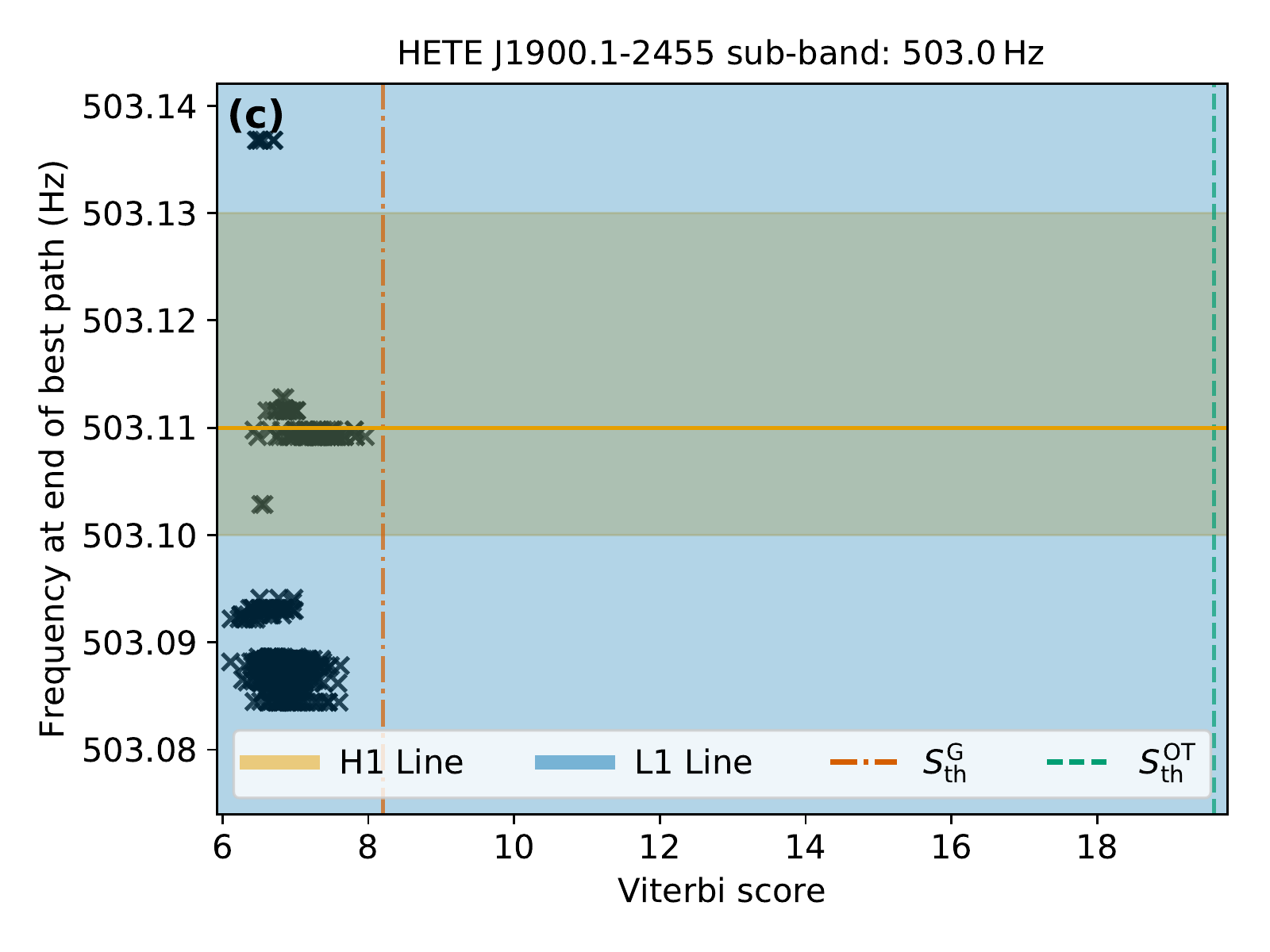}
\includegraphics[width=.49\textwidth]{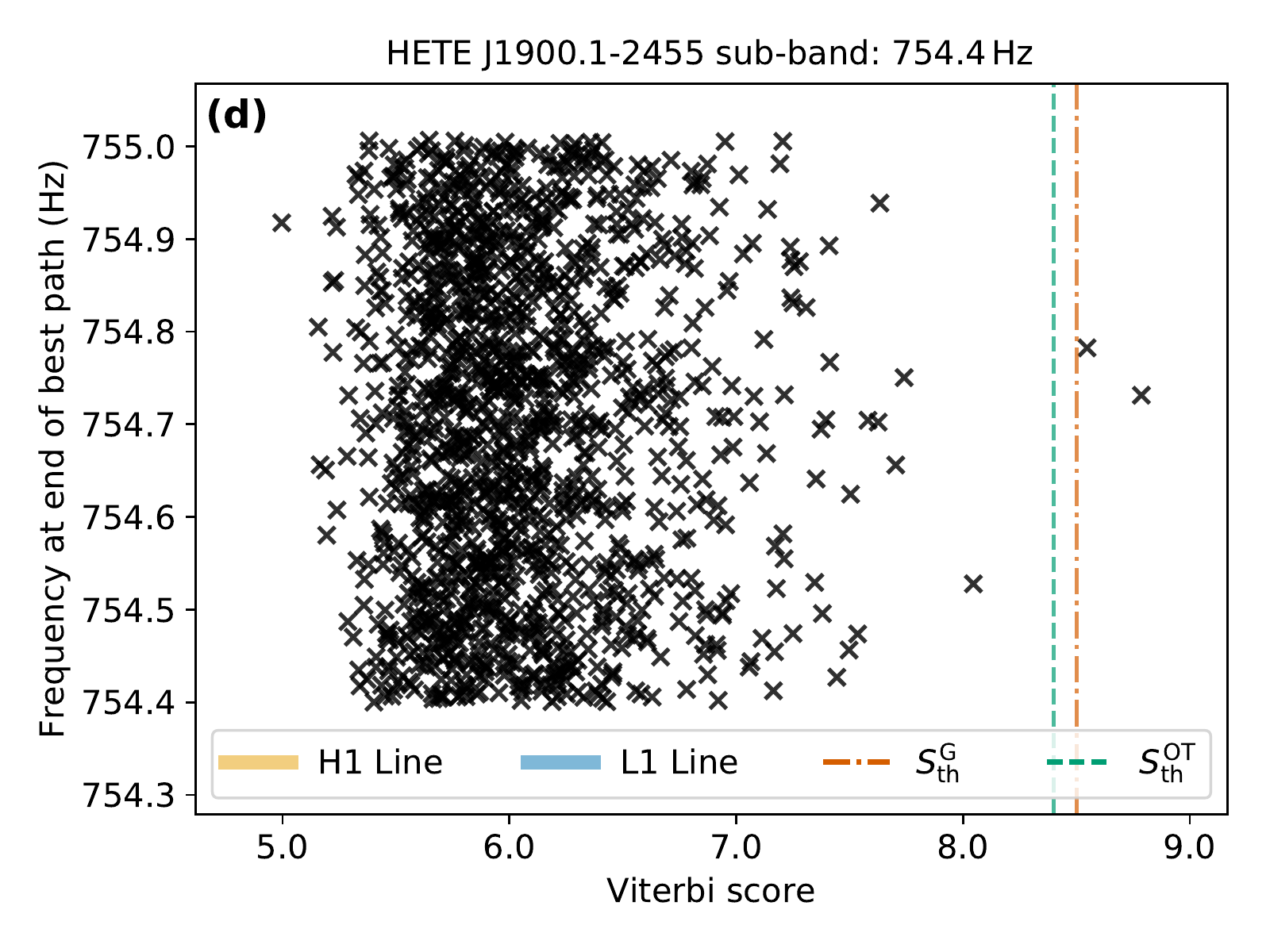}
\end{center}
\caption{\label{fig:HETEaSearch}
Search results for \HETEaName~at the four sub-bands corresponding to $\fstar/2$, $\fstar$, $4\fstar/3$, and $2\fstar$ in panels (a), (b), (c), and (d) respectively. 
Each point marks the terminating frequency of the best path for a template (vertical axis) and the associated Viterbi score (horizontal axis). 
The vertical lines indicate $\Sth$ for a false alarm probability of $0.30$ per sub-band determined from $100$ Gaussian noise realisations $\SthPSG$ (red-dot-dash) and $100$ off-target searches $\SthPSOT$ (green-dash). 
The orange and blue horizontal stripes indicate known instrumental lines in the Hanford and Livingston observatories respectively; the solid-line indicates where the instrumental line peaks and the shading indicates its width. 
In the case of the sub-band starting at $503.0~\Hz$, broad instrumental lines cover the whole frequency range of the plot.
Where there is a loud instrumental line in a sub-band, the search is likely to select paths very close to the instrumental line. 
This means that the paths for a contaminated sub-band can lie entirely within the instrumental line. 
The instrumental lines in sub-band $503.0~\Hz$ are due to violin mode resonances in the detector mirror suspensions~\cite{GWOSC}. 
There are two candidates above both $\SthPSG$ and $\SthPSOT$ in the $2\fstar$ sub-band (bottom-right). 
Similar plots for the other four targets can be found in Appendix~\ref{app:completeSearchResults}.
}
\end{figure*}

The search for \HETEaName~returns two candidates as shown in Fig.~\ref{fig:HETEaSearch}. 
Each marker in Fig.~\ref{fig:HETEaSearch} shows the terminating frequency of the best path for a template and associated Viterbi score. 
The vertical red-dot-dash and green-dash lines show the $\SthPSG$ and $\SthPSOT$ thresholds respectively (for $\alphaN = 0.30$). 
A marker with a higher Viterbi score than the threshold lines indicates an above-threshold candidate. 
The four panels in Fig.~\ref{fig:HETEaSearch} show the search results in each sub-band: $\fstar/2$ (top-left), $\fstar$ (top-right), $4\fstar/3$ (bottom-left), and $2\fstar$ (bottom-right). 
There are two candidates in the $2\fstar$ sub-band and zero candidates in the other three sub-bands. 
The $4\fstar/3$ sub-band is noisy. 
The horizontal line and shaded bands in the bottom-left panel show where there are instrumental lines. 
The solid line indicates the peak of the instrumental line and the transparent shading shows its extent (see Refs.~\cite{GWOSC,CovasEtAl:2018}).
Instrumental lines in the Hanford and Livingston data are shown in orange and blue respectively. 
The instrumental lines in the $4\fstar/3$ sub-band are due to violin modes of the Hanford and Livingston detectors. 
The Hanford line peaks at $503.11\,\Hz$ with a range $503.10$--$503.13\,\Hz$ and the Livingston line peaks at $503.1825\,\Hz$ with range $502.9500$--$503.3000\,\Hz$ (covering the entire plotted region)\cite{GWOSC}.

The candidates in the $2\fstar$ sub-band are above both $\SthPSG$ and $\SthPSOT$. 
Both candidates survive vetoes 1, 2, 3, and 4. 
Veto 5 is not applicable as the frequency wandering timescales of the candidate paths is $\approx \Tdrift$ (see Fig.~\ref{fig:HETEaPaths} in Appendix~\ref{app:completeSearchResults}). 
The frequency paths, Viterbi scores, and veto outcomes for these candidates are further detailed in Appendix~\ref{app:completeSearchResults}. 

\subsection{\IGRaName}
\label{sec:IGRaResults}
The search for \IGRaName~returns four candidates. 
Three are above $\SthPSG$ and one is above $\SthPSOT$ (zero of these are above both thresholds). 
We include a fifth nearly above-threshold template in the veto procedure due to its proximity to $\SthPSOT$ in the face of rounding errors. 
The search results, frequency paths, and veto outcomes are shown in Appendix~\ref{app:completeSearchResults} (Fig.~\ref{fig:IGRaSearch}, Fig.~\ref{fig:IGRaPaths}, and Table~\ref{tab:vetoOutcomes} respectively).

Three of the candidate are in the $\fstar/2$ sub-band. 
They have scores which are above $\SthPSG$ and below $\SthPSOT$. 
All three candidates are eliminated by veto 1 due to a broad instrumental line in the Hanford data (see Fig.~\ref{fig:IGRaSearch} in Appendix~\ref{app:completeSearchResults}).
The other above-threshold candidate is in the $2\fstar$ sub-band with $\So > \SthPSOT$ and $\So < \SthPSG$. 
It survives vetoes 1--4. 
Veto 5 is not applicable. 
The nearly above-threshold template is in the $\fstar$ sub-band. 
It survives vetoes 1--4 and veto 5 is not applicable.

\subsection{\SAXaName}
\label{sec:SAXaResults}
The search for \SAXaName~finds two candidates. 
A third template almost equals the threshold, and we include it in the veto procedure. 
See Appendix~\ref{app:completeSearchResults} for full search results, candidate frequency paths, and veto outcomes. 

One candidate is in the $\fstar/2$ sub-band with $\So<\SthPSG$ and $\So>\SthPSOT$. 
It is eliminated by veto 1 due to proximity to an instrumental line in the Livingston data (see Fig.~\ref{fig:SAXaSearch} in Appendix~\ref{app:completeSearchResults}). 
There is one candidate in the $4\fstar/3$ sub-band. 
It is close to, but just above both thresholds ($\So\gtrsim\SthPSG$ and $\So\gtrsim\SthPSOT$). 
The candidate survives vetoes 1--4 and veto 5 is not applicable. 
The nearly above-threshold template is in the $\fstar$ sub-band with $\So\approx\SthPSOT$.
It survives vetoes 1--4 and veto 5 is not applicable.

\subsection{\XTEbName}
\label{sec:XTEbResults}
The search for \XTEbName~returns zero candidates with $\So>\SthPSG$ or $\So>\SthPSOT$. 
Full search results for \XTEbName~are shown in Fig.~\ref{fig:XTEbSearch} in Appendix~\ref{app:completeSearchResults}.

\subsection{\XTEaName}
\label{sec:XTEaResults}
The \XTEaName~search returns two candidates. 
One candidate is in the $\fstar$ sub-band with $\So<\SthPSG$ and $\So>\SthPSOT$. 
It survives vetoes 1 and 2. 
It is eliminated by veto 3 as the signal is found to be stronger in the later part of O2. 
The other candidate, in the $2\fstar$ sub-band, scores above both thresholds. 
It survives vetoes 1--4. 
Veto 5 is not applicable. 
See Appendix~\ref{app:completeSearchResults} for full search results.

\section{Expected GW strain from LMXBs}
\label{sec:expectedStrain}

\begin{table*}[t]
\caption{\label{tab:expectedStrain}
Spin-down limits on the maximum GW strain inferred from EM observations for \IGRaName, \SAXaName, \XTEbName~and \XTEaName~(\HETEaName~is excluded as there is no frequency derivative measurement). 
The first column summarizes the target name and estimated distance $D$. 
The second column indicates whether the $\fstardot$ observation is from an active or quiescent phase.
The third and fourth columns show the observed frequency derivative $\fstardot$ and the associated spin-down limit $\hsd$ respectively using the minimum $D$ from the first column. 
The final two columns reference the data used for the estimate. 
The $\hsd$ limits marked with * indicate $\fdotact>0$ where the assumption $\fgwdot \approx -\fdotact$ is made. 
}
\begin{tabular}{lrrlll}
\hline
\hline
\\
Target Details              & Active or  & ~Freq. derivative~                & ~Spin-down~                   & ~Notes                          & ~Ref.  \\ 
                            & quiescent  & ~$\fdot$  $(\Hzps)$~              & ~limit $\hsd$~                &  \\
\\
\hline
\\
\IGRaName                      & Quiescent & $-4.1 (1.4)\times 10^{-15}$& ~$ 5.2 \times 10^{-28}$  & Before 2008 outburst            & \cite{IGRJ00291PapittoEtAl:2011} \\ 
$4<D/\kpc<6$ \IaD              & Active    & $+3(5)\times 10^{-12}$     & ~$ 1.4 \times 10^{-26}$* & 2015 outburst                   & \cite{IGRJ00291SannaEtAl:2017} \\ 
                               & Active    & $+5.1 (4)\times 10^{-15}$  & ~$ 5.8 \times 10^{-28}$* & 2008 outburst                   & \cite{IGRJ00291PapittoEtAl:2011} \\ 
\\
\hline
\\
\SAXaName                      & Quiescent & $-5.5(1.2)\times 10^{-16}$ & ~$ 2.7 \times 10^{-28}$  & Five outbursts up to 2008    & \cite{SAXJ1808HartmanEtAl:2009}\\
$D\approx 3.4$--$3.6~\kpc$\SaD & Quiescent & $-1.65(20)\times 10^{-15}$ & ~$ 4.7 \times 10^{-28}$  & Six outbursts up to 2011    & \cite{SAXJ1808PatrunoEtAl:2012}\\
                               & Active    & $+2.6(3)\times 10^{-11}$   & ~$ 5.9 \times 10^{-26}$* & 2015 outburst (XMM-Newton data) & \cite{SAXJ1808SannaEtAl:2017}  \\
                               & Active    & $+1.1(3)\times 10^{-10}$   & ~$ 1.2 \times 10^{-25}$* & 2015 outburst (NuSTAR data)     & \cite{SAXJ1808SannaEtAl:2017}  \\
                               & Quiescent & $-1.5(2)\times 10^{-15}$   & ~$ 4.5 \times 10^{-28}$  & Long-term spin-down             & \cite{SAXJ1808SannaEtAl:2017}  \\
\\
\hline
\\
\XTEbName                      & Active    & $-9.2(4) \times 10^{-14}$  & ~$ 2.4\times 10^{-27}$   & 2002 outburst                   & \cite{XTEJ0929Discovery:2002} \\
$D > 7.4\kpc$\XbD           & \\
\\
\hline
\\ 
\XTEaName                      & Active    & $-6.7 (7)\times 10^{-14}$  & ~$ 3.0 \times 10^{-27}$  & 2003 outburst                   & \cite{XTEJ1814PapittoEtAl:2007} \\
$D \approx 3.8$--$8~\kpc$\XaD  & \\
\\
\hline
\hline
\end{tabular}
\end{table*}

In view of the results in Sec.~\ref{sec:results}, it is useful to ask how strong the signal from a particular source is expected to be, given the EM information available. 
From Eq. (52) in Ref.~\cite{Riles:2013}, the indirect spin-down limit on the maximum GW strain $\hsd$ is
\begin{align}
\hsd =& ~2.5\times 10^{-25} \left( \frac{1\,{\rm kpc}}{D} \right) \nonumber \\
& \times \left[ \left( \frac{1\,{\rm kHz}}{\fgw} \right) \left( \frac{- \fgwdot}{10^{-10}\,\Hzps} \right) \left( \frac{I_{zz}}{I_0}\right) \right]^{1/2},
\label{eqn:hspindownlim}
\end{align}
where $\fgw$ and $\fgwdot$ are the GW frequency and frequency derivative respectively, $D$ is the distance to the source, $I_{zz}$ is the $zz$ component of the moment-of-inertia tensor, and $I_0$ is the moment of inertia of the un-deformed star. 
For our purposes, we make the assumptions $I_{zz}/I_0 \approx 1$, $\fgw \propto \fstar$, and $\fgwdot \propto \fstardot$. 

As described in Sec.~\ref{sec:targets}, EM observations constrain $\fstar$ and $\fstardot$ during both active and quiescent phases.
Many of our targets have a range of estimates from observations of different phases. 
During active periods, the neutron star typically spins up, although this is not always the case; the hydromagnetic accretion torque can be negative~\cite{GhoshLambPethick:1977}. 
In quiescence, the neutron star typically spins down.

In Table~\ref{tab:expectedStrain}, we collate estimates of $\fstardot$ for each of the targets. 
\IGRaName~and \SAXaName~have several values of $\fstardot$ estimated from observations of different active and quiescent phases. 
These two targets both follow the typical picture of active spin up ($\fdotact > 0$) and quiescent spin down ($\fdotqu < 0$).
\XTEbName~and \XTEaName~have only a single observed active phase and therefore each have only one $\fstardot$ estimate. 
Both exhibit active spin down, unlike \IGRaName~and \SAXaName~which show active spin up.
\HETEaName~ is excluded from this calculation, as there is no $\fstardot$ measurement from the short time it exhibited pulsations.

Many of the targets have a range of $D$ estimates, which we summarize in the first column of Table~\ref{tab:expectedStrain}. 
To calculate the maximum $\hsd$, we use the minimum $D$ for each target.

For $\fstardot < 0$, we compute $\hsd$ assuming $\fgwdot \approx \fstardot$. 
The $\hsd$ estimates are collected in the fourth column of Table~\ref{tab:expectedStrain}. 
We find that the maximum $\hsd$ for $\fstardot < 0$ comes from the active phases of \XTEbName~and \XTEaName~with $\hsd \lesssim 2.4\times 10^{-27}$ and $\lesssim 3.0\times 10^{-27}$ respectively. 
For comparison, the Scorpius X-1 O2 search set an upper limit on the detectable wave strain of $3.47 \times 10^{-25}$ at $194.6\,\Hz$ with $95\%$ confidence~\cite{ScoX1ViterbiO2}.

Equation~\ref{eqn:hspindownlim} requires $\fstardot < 0$.
For $\fstardot > 0$, we can make a different order of magnitude estimate for $\hsd$.
When the star is observed to spin up, the positive net torque (dominated by accretion) may mask a negative gravitational radiation reaction torque of a similar order of magnitude. 
In principle, $\fdotact>0$ allows for an arbitrarily large frequency derivative due to accretion, i.e. $\fdotacc > 0$ can be as large as one wishes, as long as we also have $|\fgwdot| = \fdotacc - \fdotact$
One arguably plausible scenario, without excessive fine tuning, is $\fdotact \sim \fdotacc \sim |\fgwdot|$.
On the other hand, for $\fdotact = \fdotacc - |\fgwdot| < 0$, one must have $|\fgwdot| \geq \fdotact$, and setting $|\fgwdot| = \fdotact$ yields a conservative bound. 
For the observations with $\fstardot \geq 0$, we therefore estimate $\hsd$ assuming $\fgwdot = -\fstardot$. 
We find $\hsd$ in the range $10^{-28}$ to $10^{-25}$ for the active phases of \IGRaName~and \SAXaName. 
None of the targets were active during O2.

\begin{table}
\caption{\label{tab:torqueBalance}
Torque-balance limit $\htorque$ on the GW strain based on EM observations. 
The second column is the long-term X-ray flux from table 1 of Ref.~\cite{GWLMXBsWattsEtAl:2008}; see also references therein. 
The third column is the maximum flux from the LMXB catalogue in Ref.~\cite{LiuEtAlLMXBCatalogue:2007}. 
The final column shows the estimated range of $\htorque$ for the listed fluxes calculated using $\fgw=2\fstar$ in Eq.~\ref{eqn:htorque}. 
}
\begin{center}
\begin{tabular}{p{3.2cm}p{1.9cm}p{1.8cm}p{1.3cm}}
\hline
\hline 
\\
Target & \multicolumn{2}{l}{Flux ($\times 10^{-8}\FluxUnit$)} & $\htorque$ \\
       & Long-term & Maximum                                  & ($\times 10^{-27}$) \\
\\
\hline 
\\
\HETEaName & $0.18$  & $0.238$ & $1.9$--$2.2$ \\ 
\IGRaName  & $0.018$ & $0.281$ & $0.5$--$1.9$ \\ 
\SAXaName  & $0.086$ & $0.211$ & $1.3$--$2.0$ \\ 
\XTEbName  & $0.027$ & $0.069$ & $1.0$--$1.7$ \\ 
\XTEaName  & $0.013$ & $0.025$ & $0.6$--$0.8$ \\ 
\\
\hline 
\hline 
\end{tabular}
\end{center}
\end{table}

We can also calculate the maximum signal strength based on the observed X-ray flux assuming that the accretion and gravitational radiation reaction torques balance each other.
The torque-balance limit is~\cite{Riles:2013, BildstenTB:1998, PapaloizouPringleTB:1978, WagonerTB:1984}, 
\begin{align}
\label{eqn:htorque}
\htorque = & ~5 \times 10^{-27} \nonumber \\
           & \times \sqrt{ \left(\frac{600\,\Hz}{\fgw}\right) \left( \frac{\Flux}{10^{-8} \FluxUnit} \right) },
\end{align}
where $\Flux$ is the X-ray flux. 
In the second and third column of Table~\ref{tab:torqueBalance} we list the long-term flux from Table 1 of Ref.~\cite{GWLMXBsWattsEtAl:2008} and the maximum flux from the LMXB catalogue in Ref.~\cite{LiuEtAlLMXBCatalogue:2007} respectively. 
The third column shows the range of $\htorque$ given these values. 
For some of the targets, the $\htorque$ limits are lower than the $\hsd$ limits in Table~\ref{tab:expectedStrain} during active phases. 
However the numbers are comparable to $\hsd$ in quiescence.

\section{Conclusions}
\label{sec:conclusions}

In this paper we present results of a search for continuous gravitational waves from five LMXBs in the LIGO O2 dataset. 
The search uses a hidden Markov model to track spin wandering and the $\Jstat$-statistic matched filter to track orbital phase. 
The LMXBs have electromagnetically measured pulsation frequencies, thereby restricting the parameter space relative to searches for other objects like Scorpius X-1~\cite{ScoX1ViterbiO2,ScoX1ViterbiO1:2017,SearchFStatS2:2007,SearchCrossCorrO1:2017}.
A Gaussian threshold is set using searches on $100$ realisations of Gaussian noise. 
An off-target threshold is set by searching the O2 dataset in $100$ random off-target sky positions. 

We find no candidates above a threshold corresponding to a $0.10$ FAP per sub-band. 
We find ten candidates above a threshold corresponding to a $0.30$ FAP per sub-band. 
After applying vetoes we are left with five candidates (two for \HETEaName~and one each for \IGRaName, \SAXaName, and \XTEaName). 
The survivors are marginally above the $\alphaN = 0.30$ threshold, exceeding it by less than $\approx 0.4$ in Viterbi score. 
The number of survivors is statistically consistent with the number of false alarms expected from a FAP of $0.30$ per sub-band for $20$ sub-bands (i.e. $0.30 \times 20 = 6$). 

It is premature to speculate about the nature of the surviving candidates. 
We recommend that they be followed up in future observations, including LIGO-Virgo Observing Run 3.

\section*{Acknowledgements}
\label{sec:acknowledgements}

The authors are grateful to 
Sofia Suvorova, William Moran, and Robin Evans for their past developmental work on HMMs for continuous wave searches and also to them and Margaret Millhouse, Patrick Meyers, and Julian Carlin for helpful discussions including advice on the off-target threshold and veto procedure; 
Shanika Galaudage and Duncan Galloway for advice on selecting LMXB targets and locating the most accurate EM measurements of the targets' parameters in the literature; 
and Ling Sun for helpful comments on the manuscript. 
We also thank the Continuous Wave Working Group of the LIGO Scientific Collaboration and Virgo Collaboration for their useful discussion. 
This research is supported by the Australian Research Council Centre of Excellence for Gravitational Wave Discovery (OzGrav) (project number CE170100004).
This work used computational resources of the OzSTAR national facility at Swinburne University of Technology and also at the California Institute of Technology. 
OzSTAR is funded by Swinburne University of Technology and the National Collaborative Research Infrastructure Strategy (NCRIS). 
This research has made use of data, software and/or web tools obtained from the Gravitational Wave Open Science Center (https://www.gw-openscience.org), a service of LIGO Laboratory, the LIGO Scientific Collaboration and the Virgo Collaboration. 
LIGO is funded by the U.S. National Science Foundation. 
Virgo is funded by the French Centre National de Recherche Scientifique (CNRS), the Italian Istituto Nazionale della Fisica Nucleare (INFN) and the Dutch Nikhef, with contributions by Polish and Hungarian institutes.
This work has been assigned LIGO document number P1900273.

\bibliographystyle{myunsrt}
\bibliography{LMXBsBib}

\newpage

\appendix

\section{SFT Lengths}
\label{app:sftLengths}
The SFTs used in this search are of duration $\TSFT=1800\,\s$, identical to those used in searches for Scorpius X-1~\cite{ScoX1ViterbiO2}. 
The targets considered in this search have typically much shorter $\Porb$ than Scorpius X-1 ($\Porb=68\,023.86048 \pm 0.0432\s$~\cite{WangEtAlScoX1Ephem:2018}).
A shorter $\Porb$ requires commensurately shorter SFTs, if the $\Fstat$-statistic is used for the HMM emission probability. 
This is not the case for the $\Jstat$-statistic used in this paper, because it weights and sums the Doppler-shifted orbital sidebands based on the orbital parameters of the target in a manner that is coherent with respect to orbital phase.
In this appendix we run a short test to verify that the $\Jstat$-statistic is calculated accurately with $1800\,\s$ SFTs.

Following Eq. (C2) in Ref.~\cite{LeaciPrix:2015}, the maximum length of an SFT for computing the $\Fstat$-statistic is
\begin{equation}
\label{eqnapp:TSFTLim}
\TSFT^2(f) \leq \frac{6 \sqrt{5 \mu_{\mathrm{SFT}}}}{\pi \asini f \Omega^2},
\end{equation}
where $\mu_{\mathrm{SFT}}$ is the SFT mismatch (which we choose to be 0.01).
The LMXB target with the shortest $\Porb$ considered here is \XTEbName~[$\Porb=\XTEbPorbE\,\s$].
For \XTEbName, Eq.~\ref{eqnapp:TSFTLim} gives $\TSFT \leq 177\,\s$ for the highest frequency searched for this target. 

\begin{figure}[h]
\includegraphics[width=.5\textwidth]{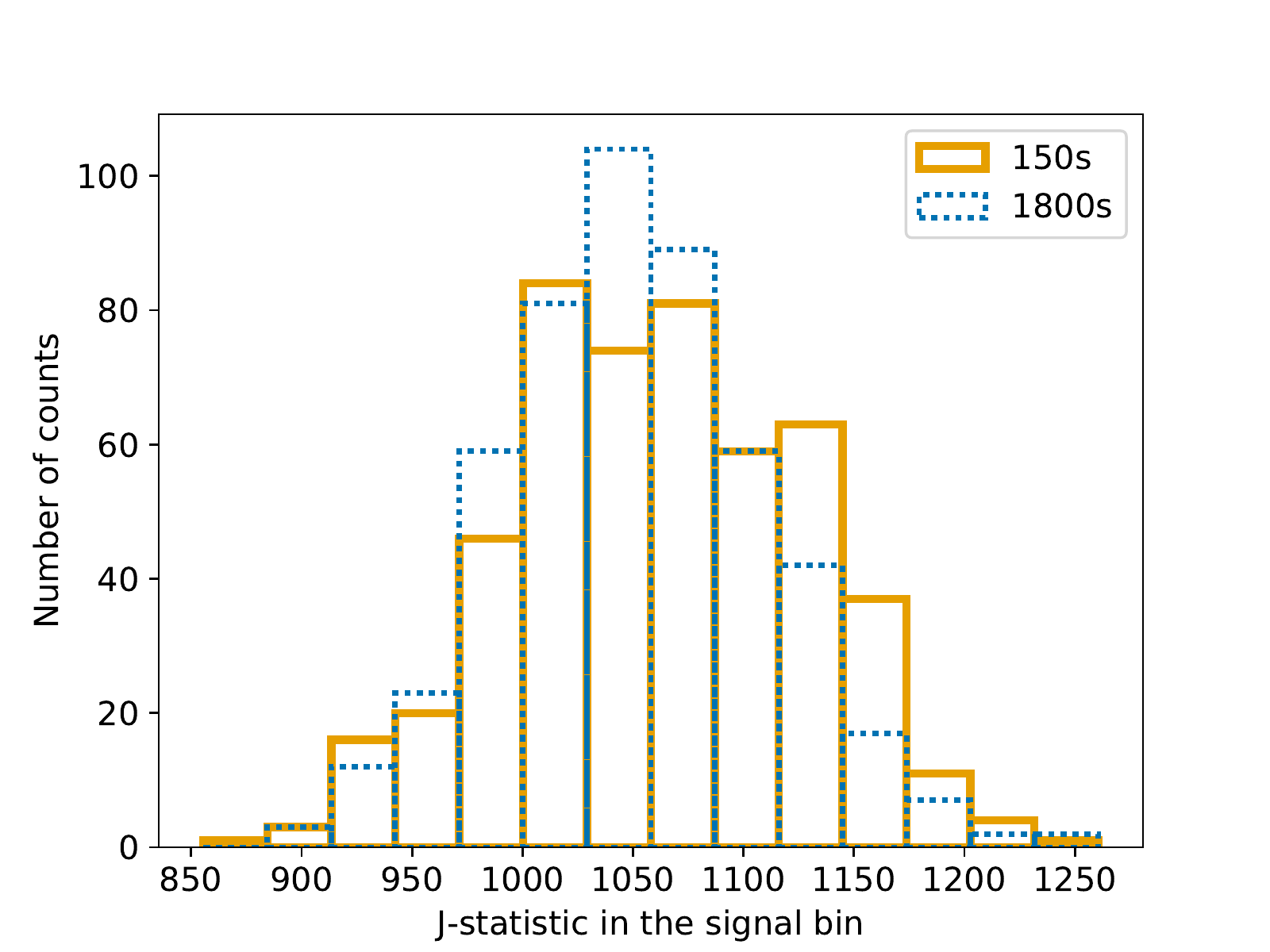}
\caption{\label{fig:SFTLengthResult}
$\Jstat$-statistic versus SFT length for a signal injected with the orbital parameters of \XTEbName, $\fstar = 370.210508594\,\Hz$, and $h_0 = 1\times 10^{-24}$.
The histogram bins refer to the value of the $\Jstat$-statistic in the frequency bin of the injection for $500$ noise realisations. 
The orange-solid and blue-dotted bars correspond to $\TSFT = 150\,\s$ and $1800\,\s$ respectively. 
}
\end{figure}

Using \texttt{lalapps\_Makefakedata\_v4} we generate simulated SFTs containing Gaussian noise and an injected signal with the same orbital parameters as \XTEbName~(see Table~\ref{tab:targetDetails}). 
The signal is injected with a frequency of $2\fstar = 370.210508594\,\Hz$ and a gravitational wave strain of $h_0=1\times 10^{-24}$.
For the purposes of this test we use a $10\,\days$ dataset with $\Tobs=\Tdrift$ and hence $N_T=1$.
Using identical injection parameters, SFTs are generated with $\TSFT = 1800\,\s$ and $150\,\s$ using $500$ Gaussian noise realisations. 
We choose $150\,\s$ for the shorter SFTs to give a whole number of SFTs for a $10\,$day dataset. 
The $\Jstat$-statistic in the frequency bin containing the signal is computed and the $\Jstat$-statistic distributions from the two SFT lengths are compared.

The result is shown in Fig.~\ref{fig:SFTLengthResult}.
The $\Jstat$-statistic distributions for the $1800\,\s$ and $150\,\s$ SFTs are shown by the blue-dot and orange-solid bars respectively. 
We see that the distributions are similar, with a Kullback-Leibler divergence of $0.0015$ using the binning in the figure. 
That is, the $\Jstat$-statistic does not depend on $\TSFT$ in the regime relevant to this paper.

\begin{figure}
\begin{center}
\includegraphics[width=0.49\textwidth]{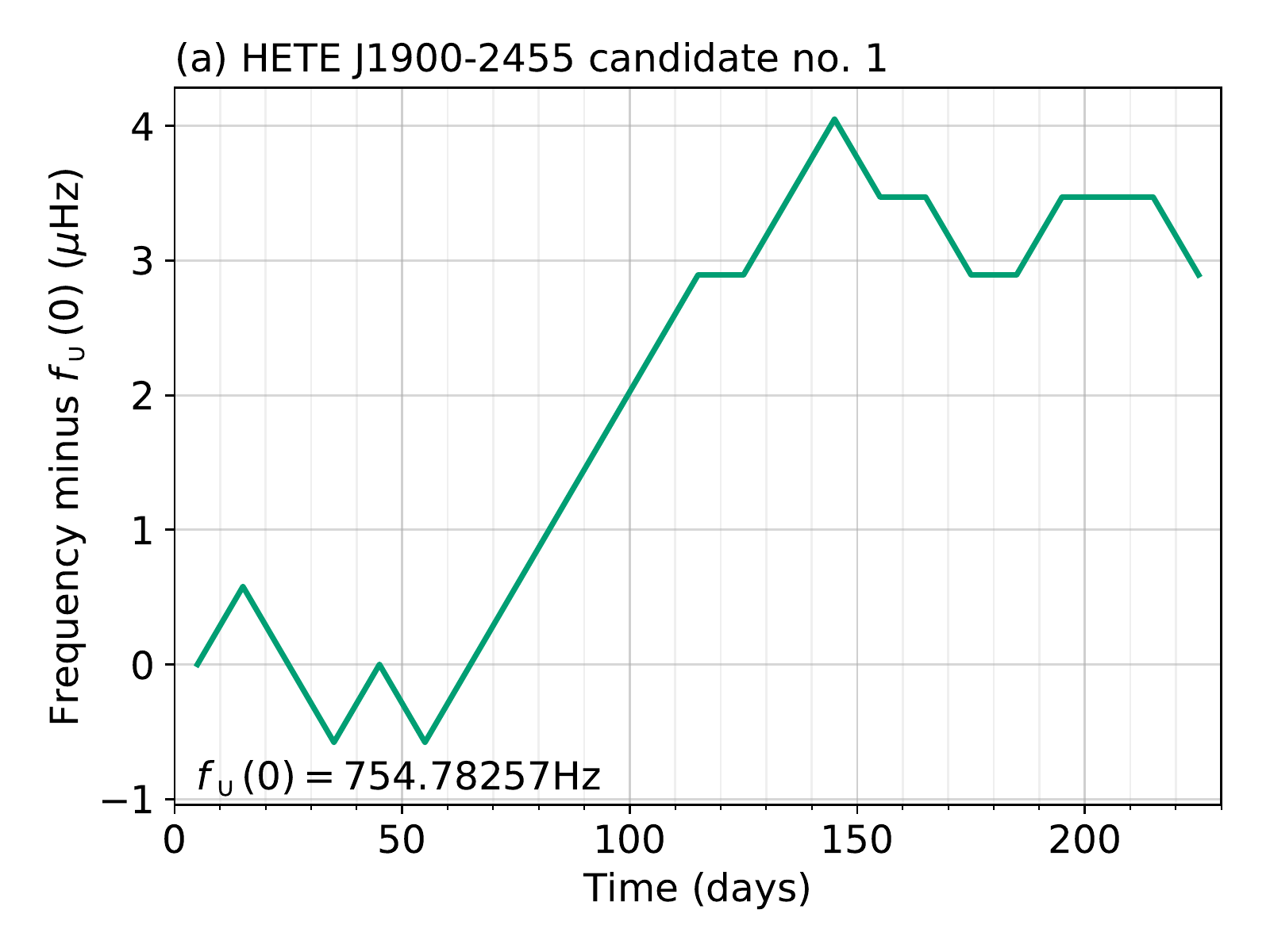}\\
\includegraphics[width=0.49\textwidth]{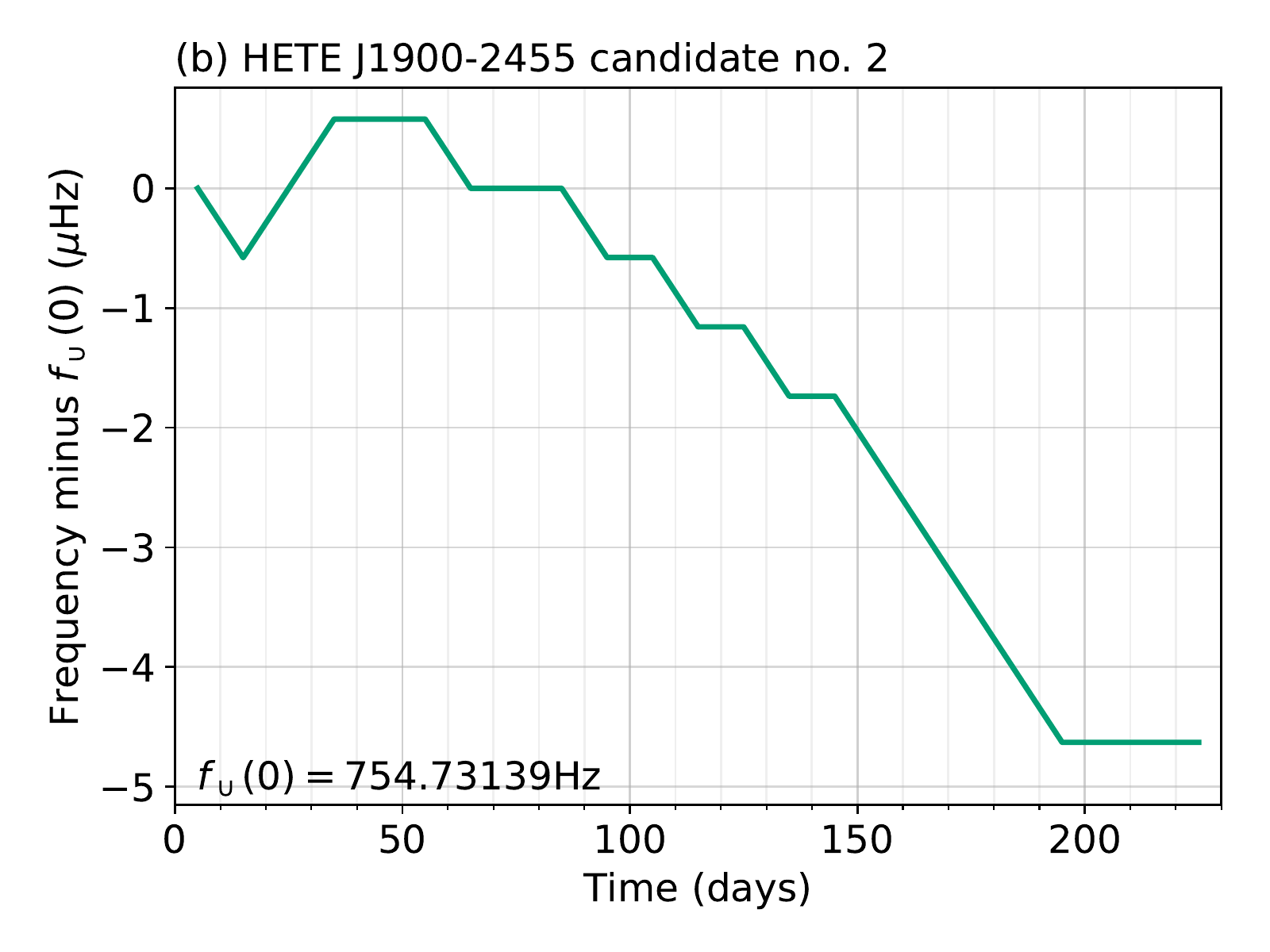}
\end{center}
\caption{\label{fig:HETEaPaths}
Frequency path for the two \HETEaName~candidates. 
The horizontal axis shows the time in days since the begining of O2. 
The vertical axis shows the frequency of the path $\fo$ with the starting frequency of the path subtracted ($\fo(0)$, displayed in the bottom-left corner in the figure). 
}
\end{figure}

\begin{table*}
\caption{\label{tab:vetoOutcomes}
Candidates returned by the search for each of the five LMXBs.
The table shows the orbital period ($\Porb$), projected semi-major axis ($\asini$), and time of ascension ($\Tasc$) for the search template.
The Viterbi score ($\So$) and terminating frequency [$\fo(t=N_T)$] of the template Viterbi path are shown in the fifth and sixth column respectively.
The seventh and eighth columns indicate whether the template's Viterbi score exceeds the Gaussian ($\SthPSG$) and off-target $\SthPSOT$ thresholds respectively (\cm~for yes, \xm~for no). 
The right-most four columns summarize the veto outcomes (\cm~for survives and \xm~for vetoed). 
The two templates marked with * are nearly above-threshold templates which are included in the veto analysis for safety against rounding errors because they have $\So\approx\SthPSOT$.
The veto outcomes for nearly above-threshold templates are indicated in parentheses.  
Without including the nearly above-threshold templates, there are five survivors. 
}
\begin{tabular}{lllllrcccccc}
\hline
\hline\\
Candidate~~~ & $\Porb$ & $\asini$ & $\Tasc$    & $\So$ &  \multicolumn{1}{l}{$\fo(t=N_T)$}  & \multicolumn{2}{l}{Above threshold?} & \multicolumn{4}{c}{~~Survives veto?}\\
number   &  (s)  & (lt-s)   & (GPS time) &     & \multicolumn{1}{l}{($\Hz$)} & $\SthPSG$    & $\SthPSOT$     & ~~\#1 & \#2 & \#3 & \#4 \\\\
\hline\\
\multicolumn{9}{l}{\HETEaName}\\
$1$ & $4995.258$  & $0.01841$ & $1164560065.80899$ & $8.54656$ & $754.78257333$ & \cm&\cm & \cm&\cm&\cm&\cm\\
$2$ & $4995.273$  & $0.01841$ & $1164559919.78652$ & $8.78604$ & $754.73138932$ & \cm&\cm & \cm&\cm&\cm&\cm\\

%                                                                                       & D & D &
\\
\multicolumn{9}{l}{\IGRaName}\\
$1$ & $8844.05$  & $0.0649905$ & $1164557590.28572$ & $8.30371$ & $299.54681334$ & \cm&\xm & \xm&-&-&-\\
$2$ & $8844.095$ & $0.0649905$ & $1164557538.85715$ & $8.55464$ & $299.54681276$ & \cm&\xm & \xm&-&-&-\\
$3$ & $8844.14$  & $0.0649905$ & $1164557127.42857$ & $8.30045$ & $299.55189494$ & \cm&\xm & \xm&-&-&-\\
$4$*& $8844.125$ & $0.0649905$ & $1164557585.14285$ & $8.46619$ & $598.95731631$ & \xm&\xm & (\cm)&(\cm)&(\cm)&(\cm)\\
%                                                                                        & D & D &
$5$ & $8844.112$ & $0.0649905$ & $1164557556.85714$ & $9.03732$ & $1197.76730242$ & \xm&\cm & \cm&\cm&\cm&\cm\\
%                                                                                        & D & D &
\\
\multicolumn{9}{l}{\SAXaName}\\
$1$ & $7249.155$ & $0.06281$ & $1164560394.28571$ & $7.49118$ & $200.35371679$ & \xm&\cm & \xm&-&-&-\\
$2$*& $7249.155$ & $0.06281$ & $1164560531.00000$ & $7.58480$ & $400.60481930$ & \xm&\xm & (\cm)&(\cm)&(\cm)&(\cm)\\
%                                                                                        & D & D &
$3$ & $7249.137$ & $0.06281$ & $1164560405.33333$ & $7.90013$ & $534.45904849$ & \cm&\cm & \cm&\cm&\cm&\cm\\
%                                                                                        & D & D &
\\
\multicolumn{9}{l}{\XTEaName}\\
$1$ & $15388.7229$ & $0.390633$ & $1164547409.69231$ & $7.03362$ & $314.46162299$  & \xm&\cm & \cm&\cm&\xm&-\\
%                                                                                        & D & A &
$2$ & $15388.7229$ & $0.390633$ & $1164547397.23077$ & $7.35219$ & $628.69675855$ & \cm&\cm & \cm&\cm&\cm&\cm\\
%                                                                                        & D & D &
\\
\hline
\hline
\end{tabular}
\end{table*}

\section{Complete search results}
\label{app:completeSearchResults}
This appendix summarises the search results and veto outcomes for the five LMXB targets. 
Table~\ref{tab:vetoOutcomes} lists the ten candidates described in Sec.~\ref{sec:results}.
Columns two, three, and four show the orbital parameters ($\Porb$, $\asini$, and $\Tasc$) of each candidate template. 
The fifth and sixth columns show the Viterbi score and terminating frequency path respectively for the candidate template.
The seventh and eighth columns indicate whether the candidate exceeds $\SthPSG$ and $\SthPSOT$ respectively. 
The final four columns indicate whether the candidate survives vetoes 1--4.
The two additional templates marked with * indicate the nearly above-threshold templates with $\So\approx\SthPSOT$ which we include in the veto procedure to avoid being misled by rounding errors (see also Sec.~\ref{sec:results}).

The search results for \HETEaName~are shown in Fig.~\ref{fig:HETEaSearch} of the main paper. 
Figure~\ref{fig:HETEaPaths} shows the frequency path for the twp~\HETEaName~candidates. 
The horizontal axis shows the time in days and the vertical axis shows the frequency of the path $\fo$ with $\fo(t=0)$ subtracted. 

\begin{figure*}
\begin{center}
\includegraphics[width=.49\textwidth]{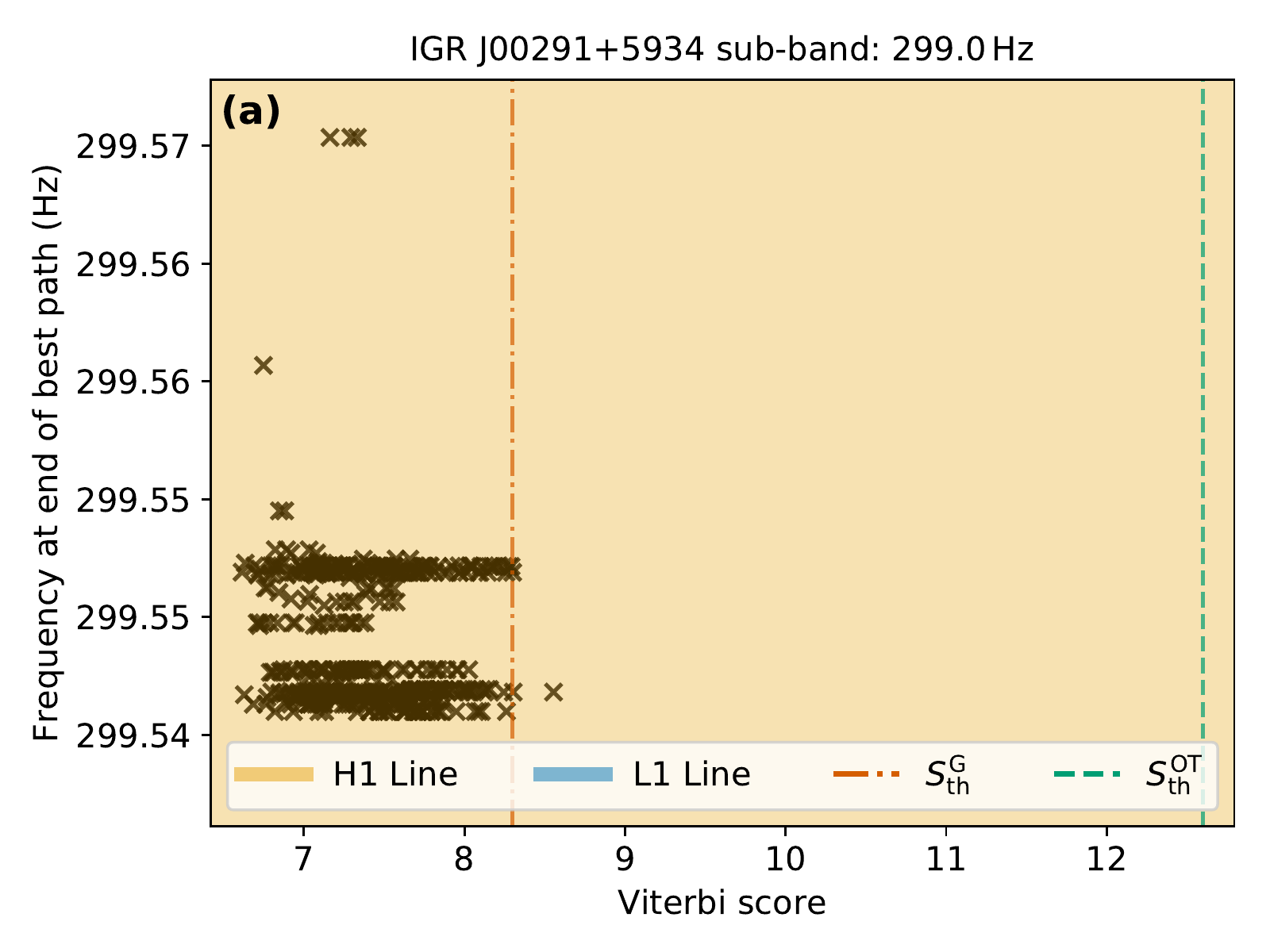}
\includegraphics[width=.49\textwidth]{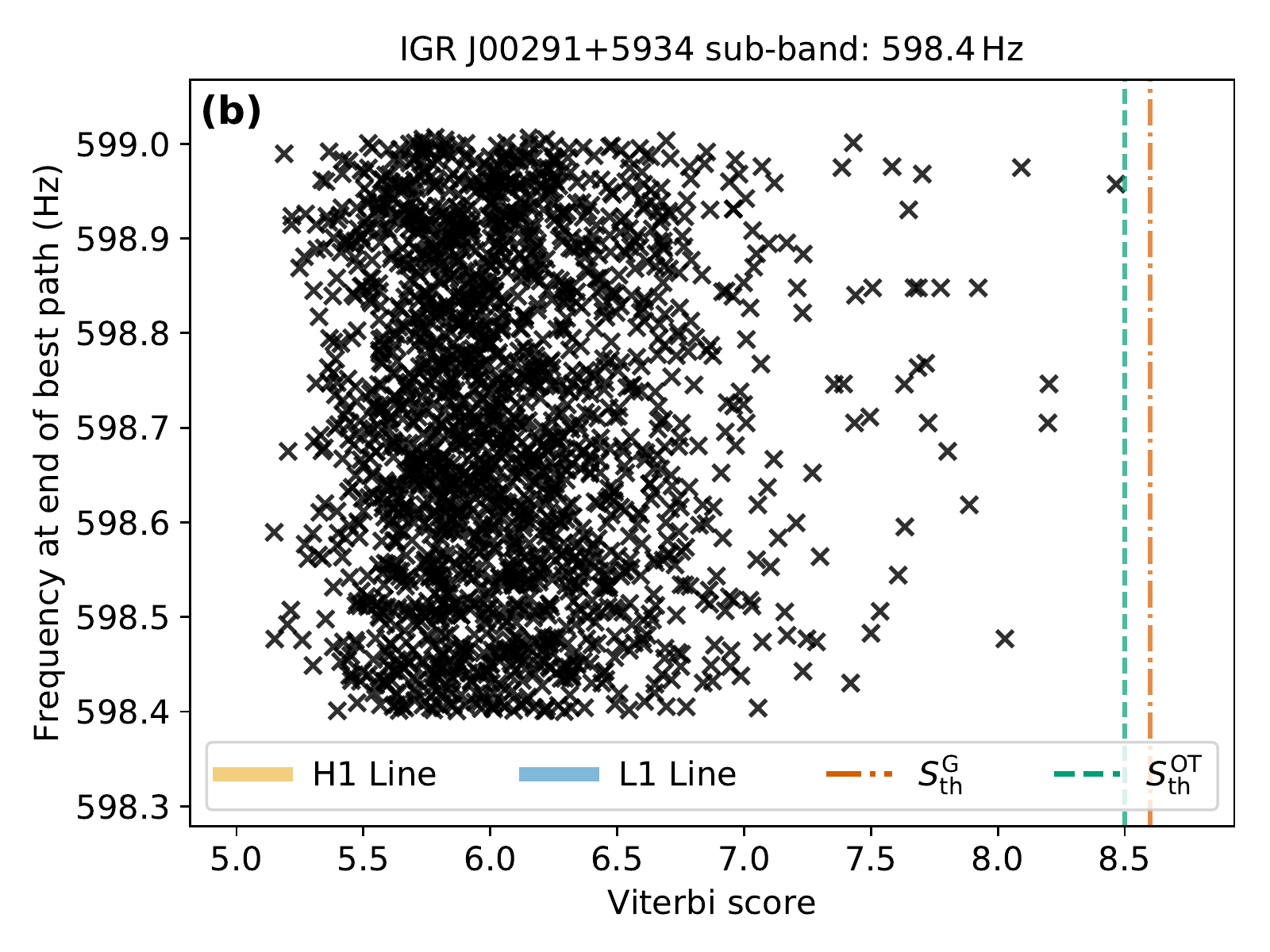}\\
\includegraphics[width=.49\textwidth]{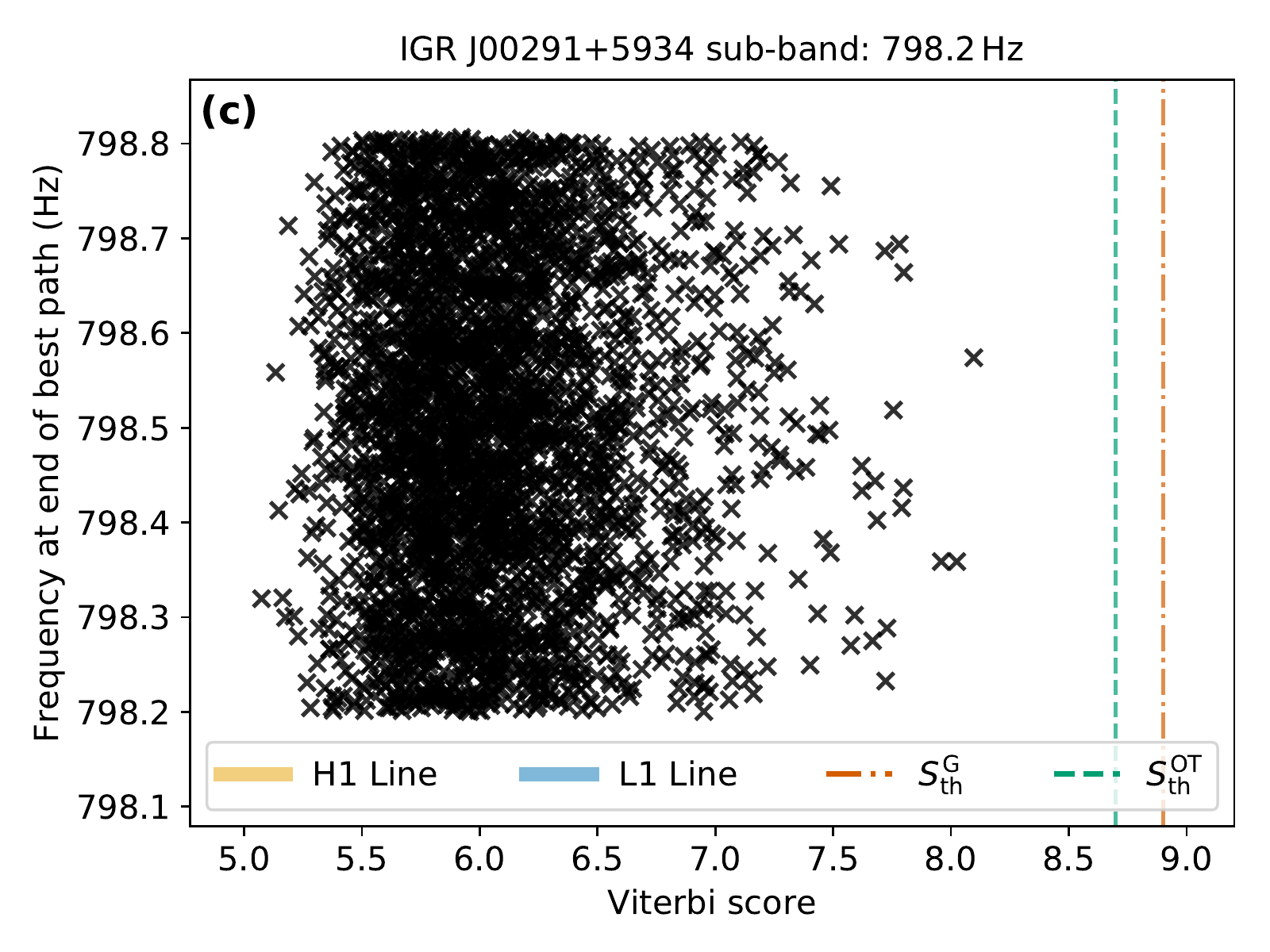}
\includegraphics[width=.49\textwidth]{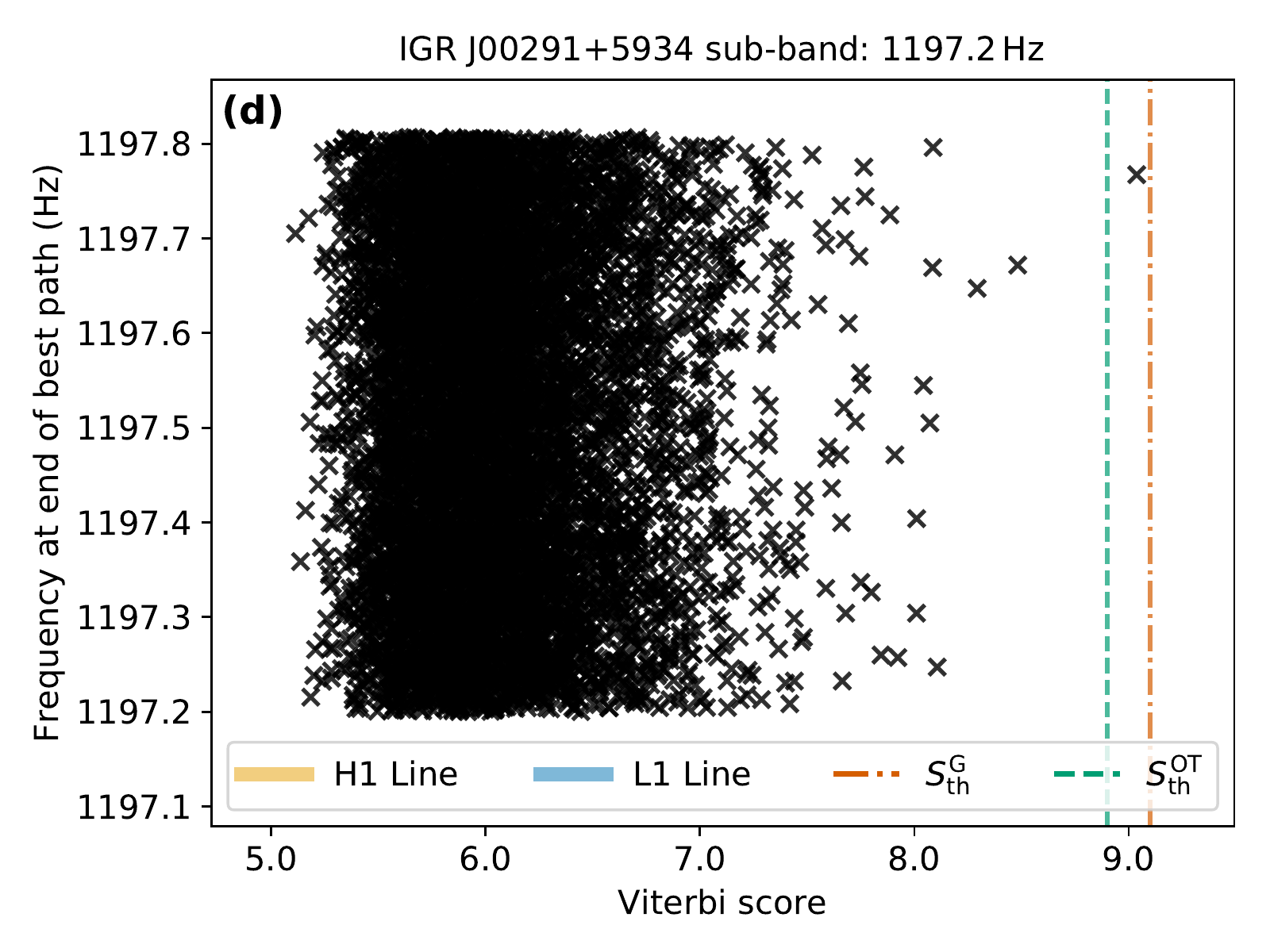}
\end{center}
\caption{\label{fig:IGRaSearch}
Search results for \IGRaName, laid out as in Fig.~\ref{fig:HETEaSearch}.
The reader is reminded that the shaded band in the panel (a) indicates the presence of an instrumental noise line and the sub-bands $\fstar/2$, $\fstar$, $4\fstar/3$, and $2\fstar$ are shown in panels (a), (b), (c), and (d) respectively. 
The instrumental noise line is due to a violin mode in the Hanford observatory which peaks at $299.60\,\Hz$ and has range $299.35$--$299.85\,\Hz$ (covering the entire plotted region). 
The veto procedure is applied to five templates (three in the $\fstar/2$ sub-band, one in the $2\fstar$ sub-band, and the highest scoring in the $\fstar$ sub-band). 
}
\end{figure*}

\begin{figure*}
\begin{center}
\includegraphics[width=0.48\textwidth]{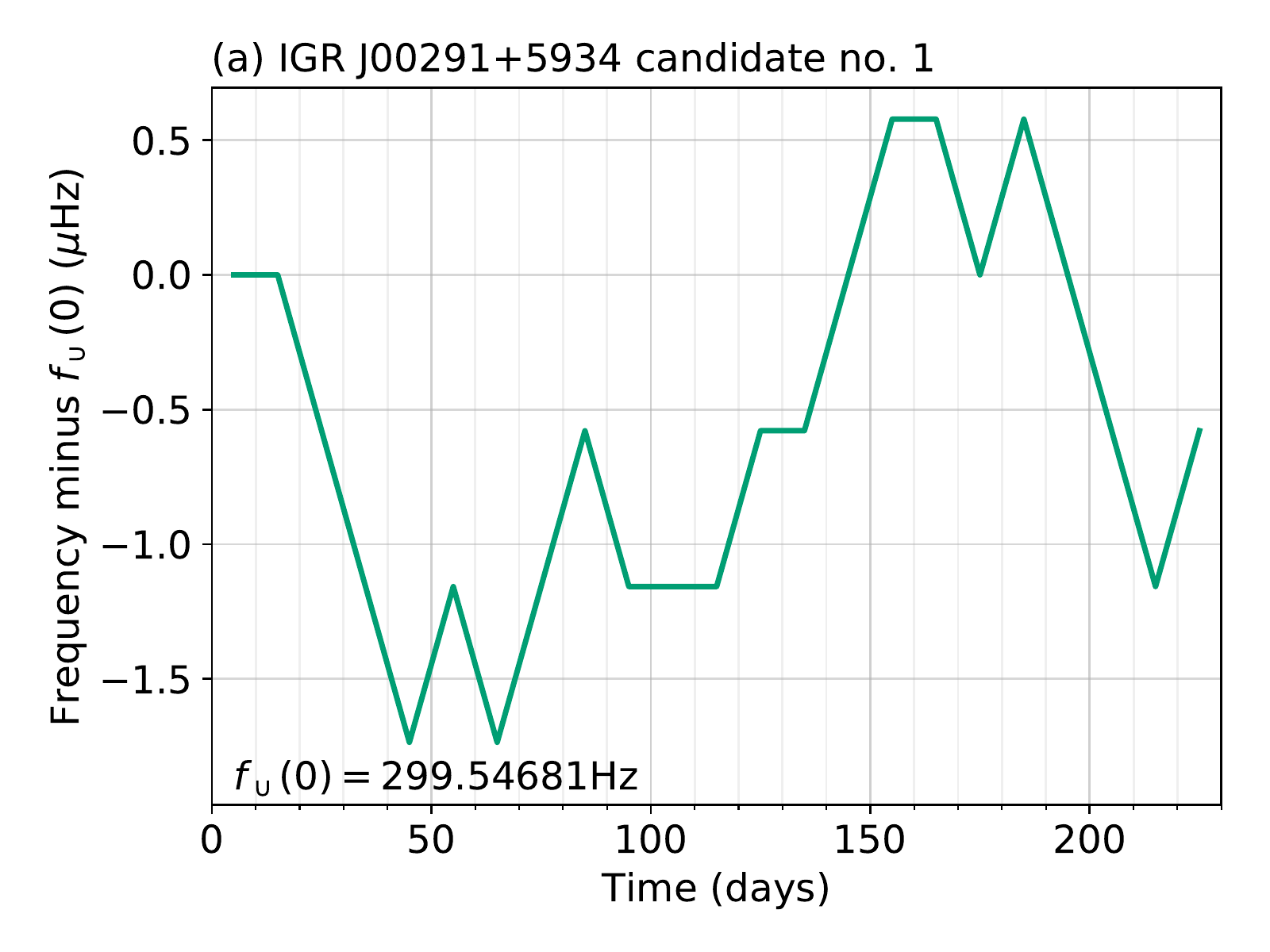}
\includegraphics[width=0.48\textwidth]{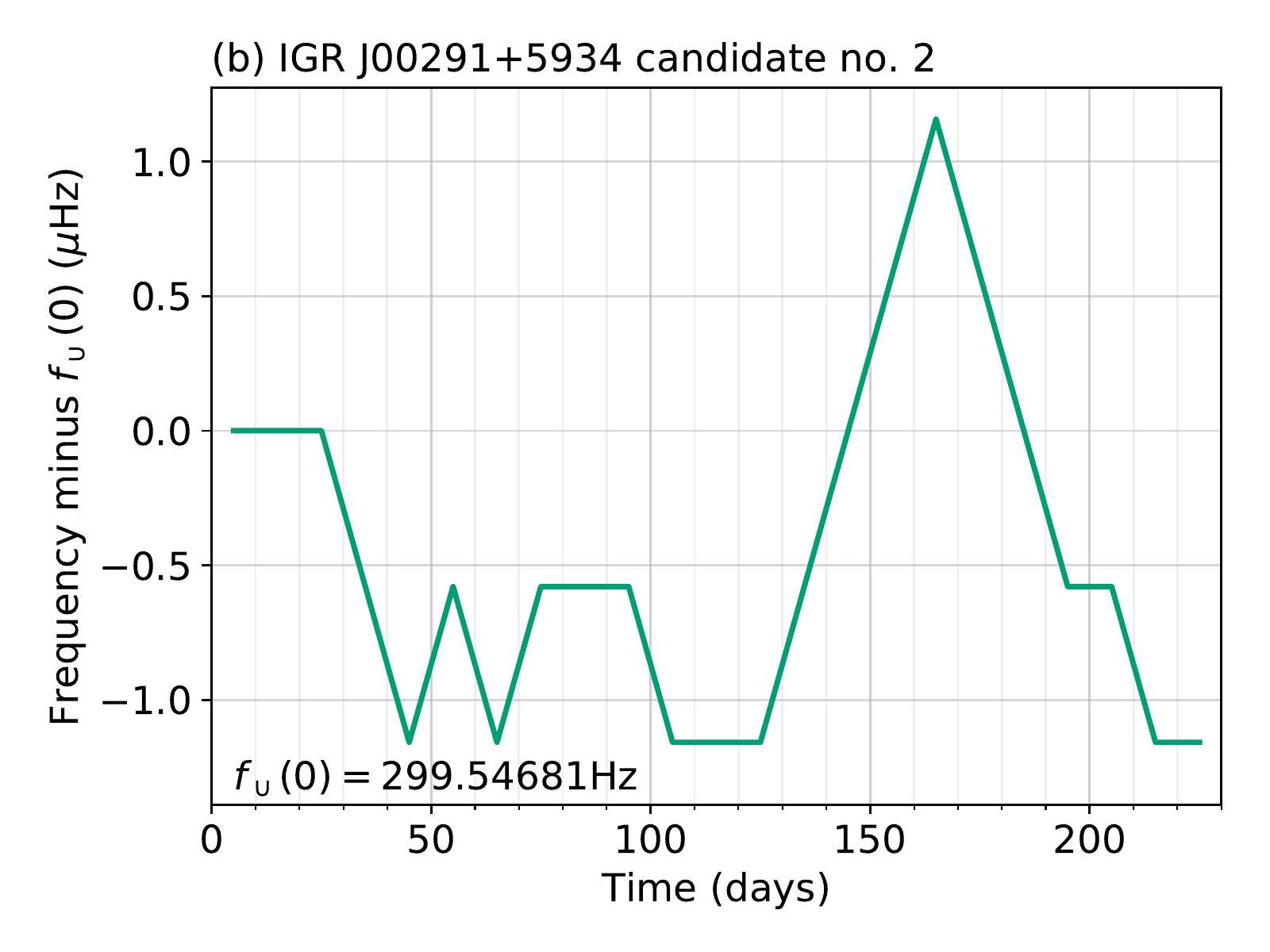}
\includegraphics[width=0.48\textwidth]{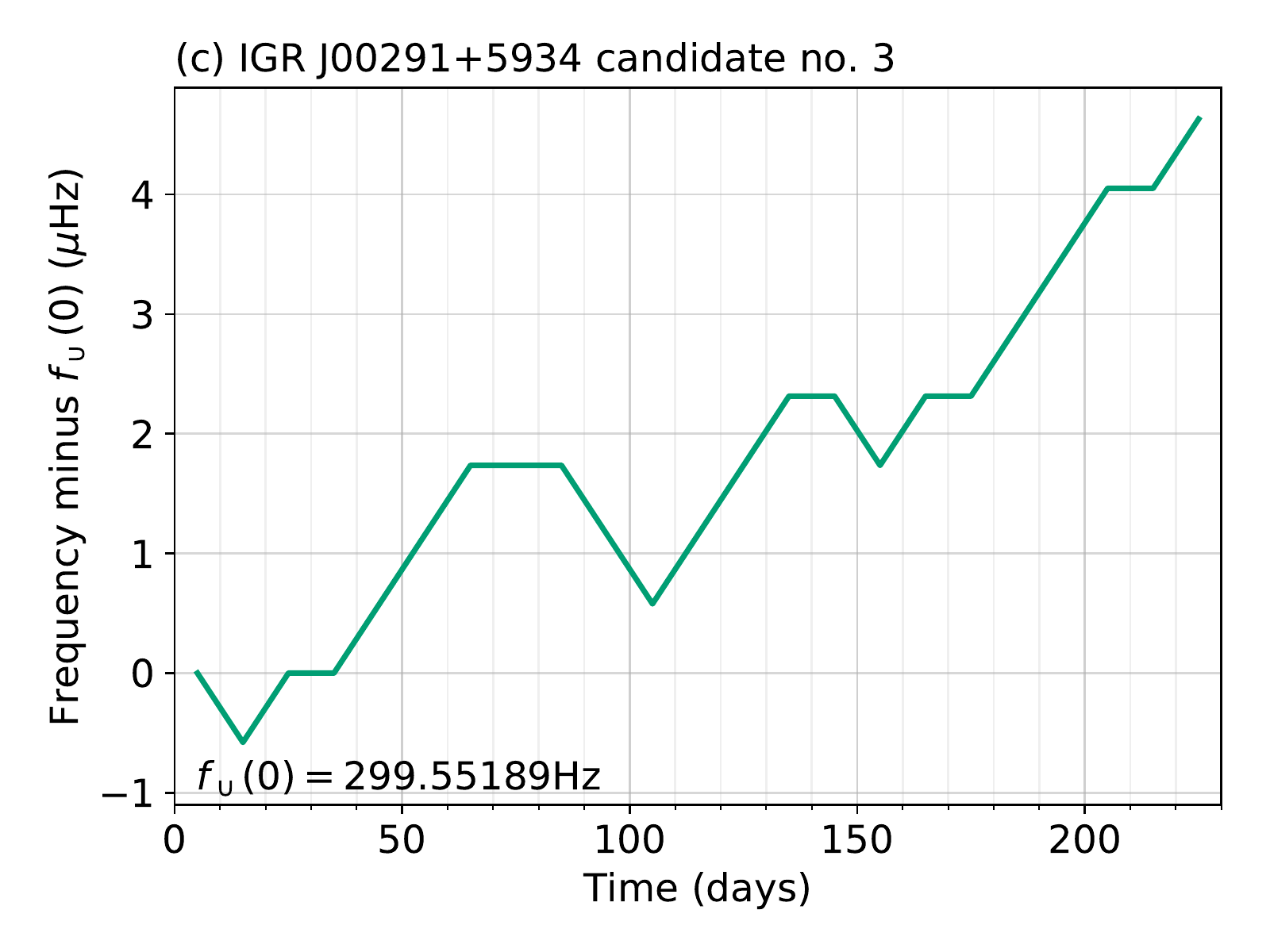}
\includegraphics[width=0.48\textwidth]{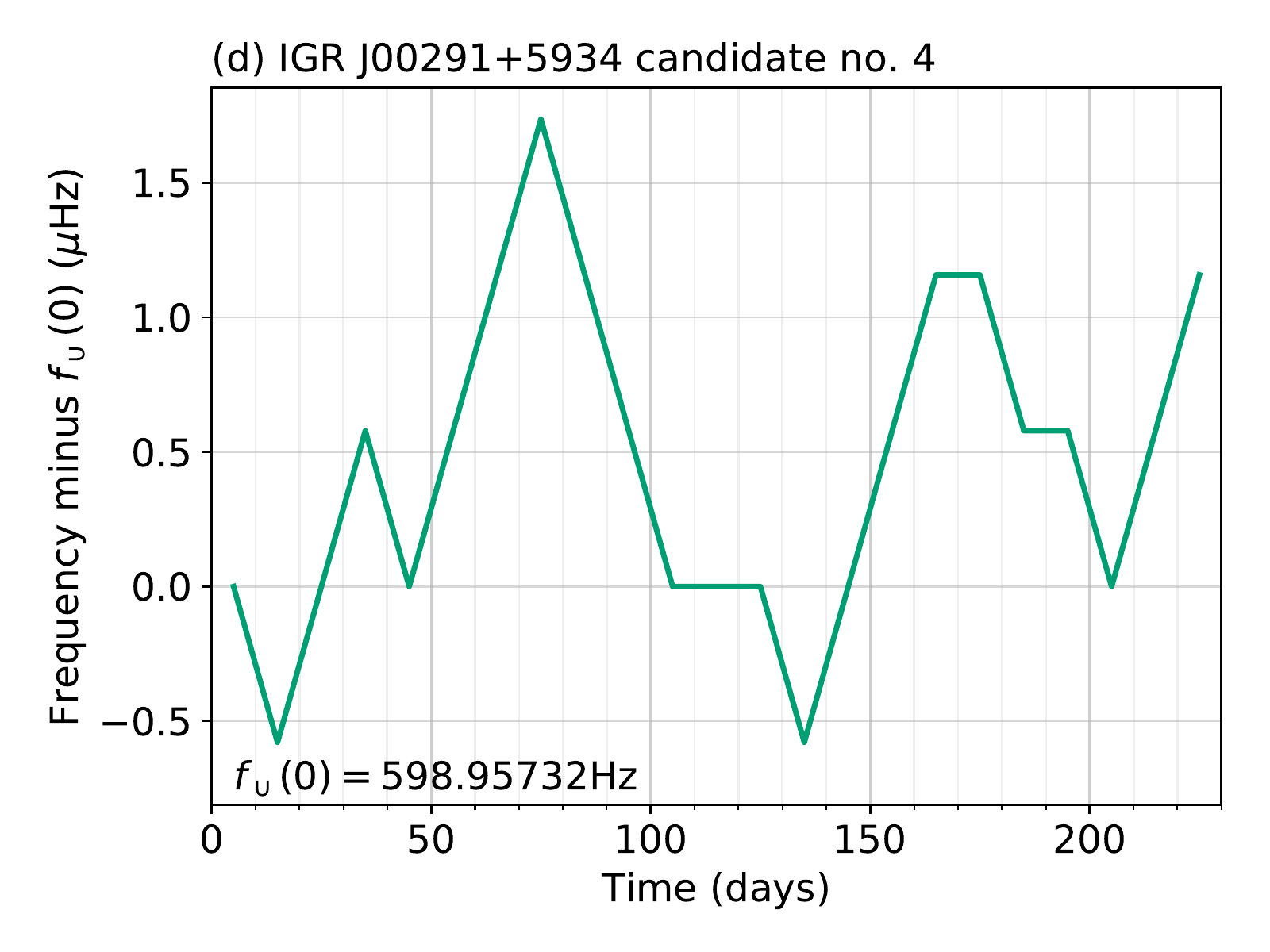}
\includegraphics[width=0.48\textwidth]{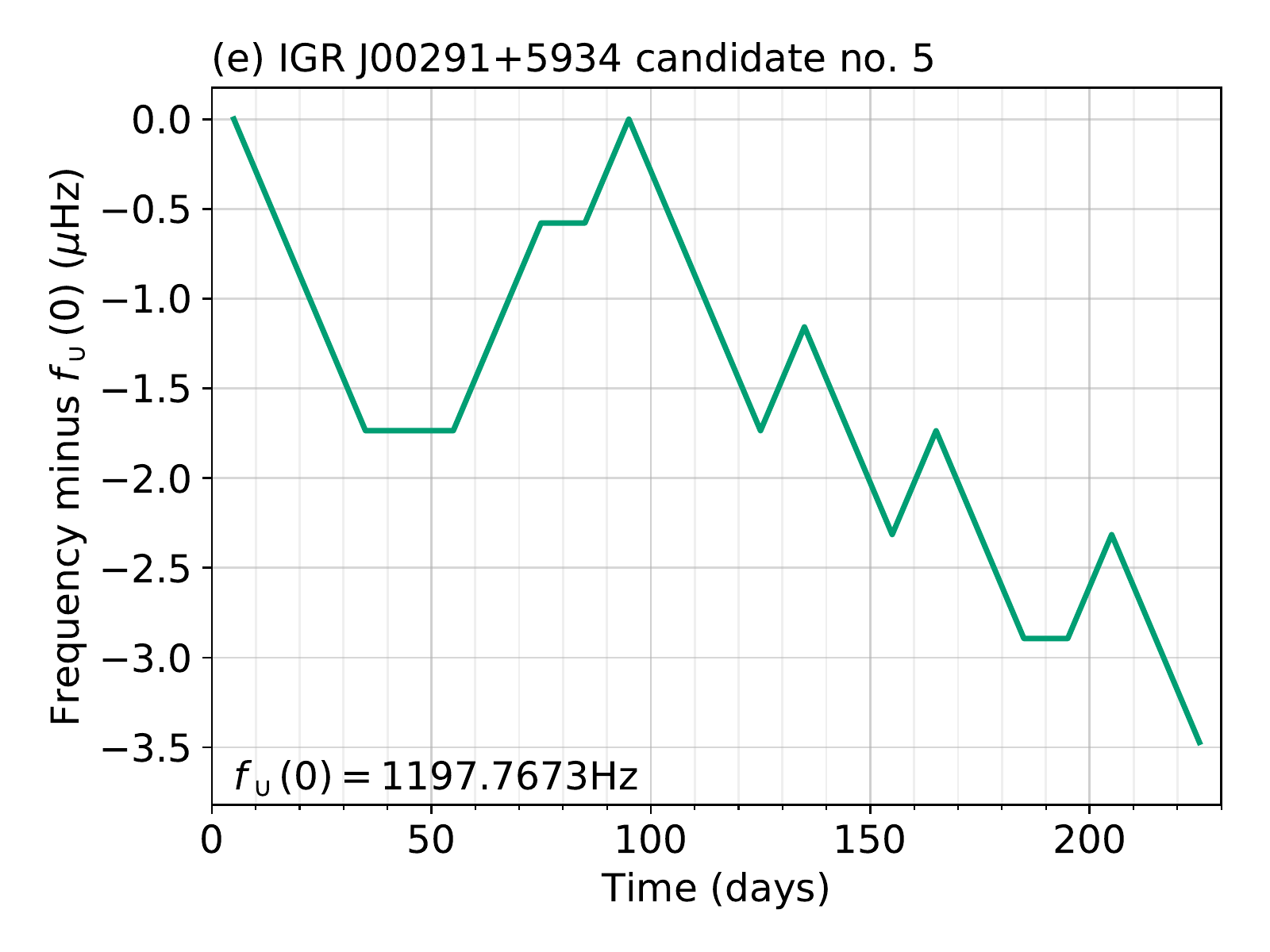}
\end{center}
\caption{\label{fig:IGRaPaths}
\IGRaName~candidate frequency paths laid out identically to Fig.~\ref{fig:HETEaPaths}.
}
\end{figure*}

\begin{figure*}
\begin{center}
\includegraphics[width=.49\textwidth]{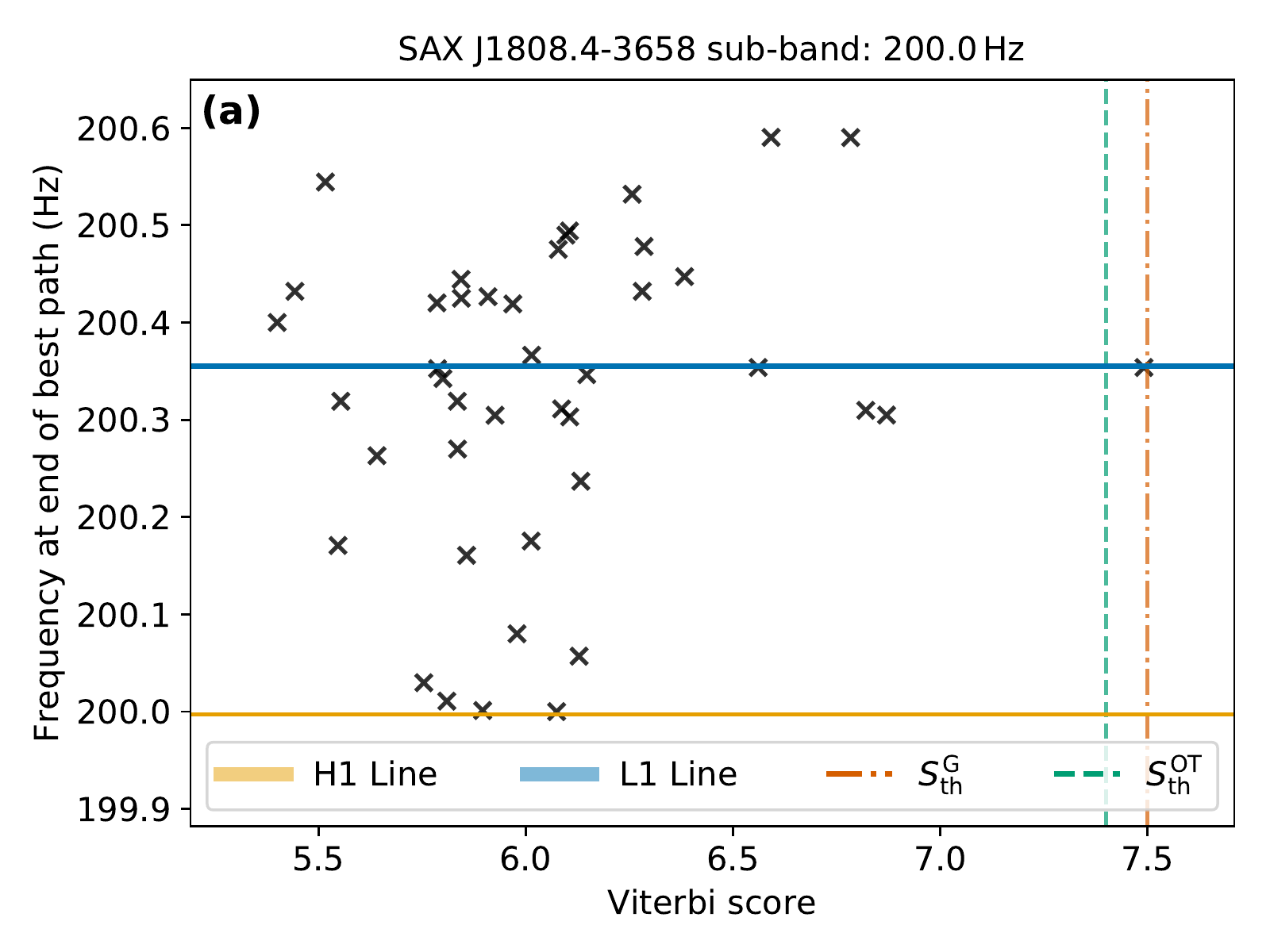}
\includegraphics[width=.49\textwidth]{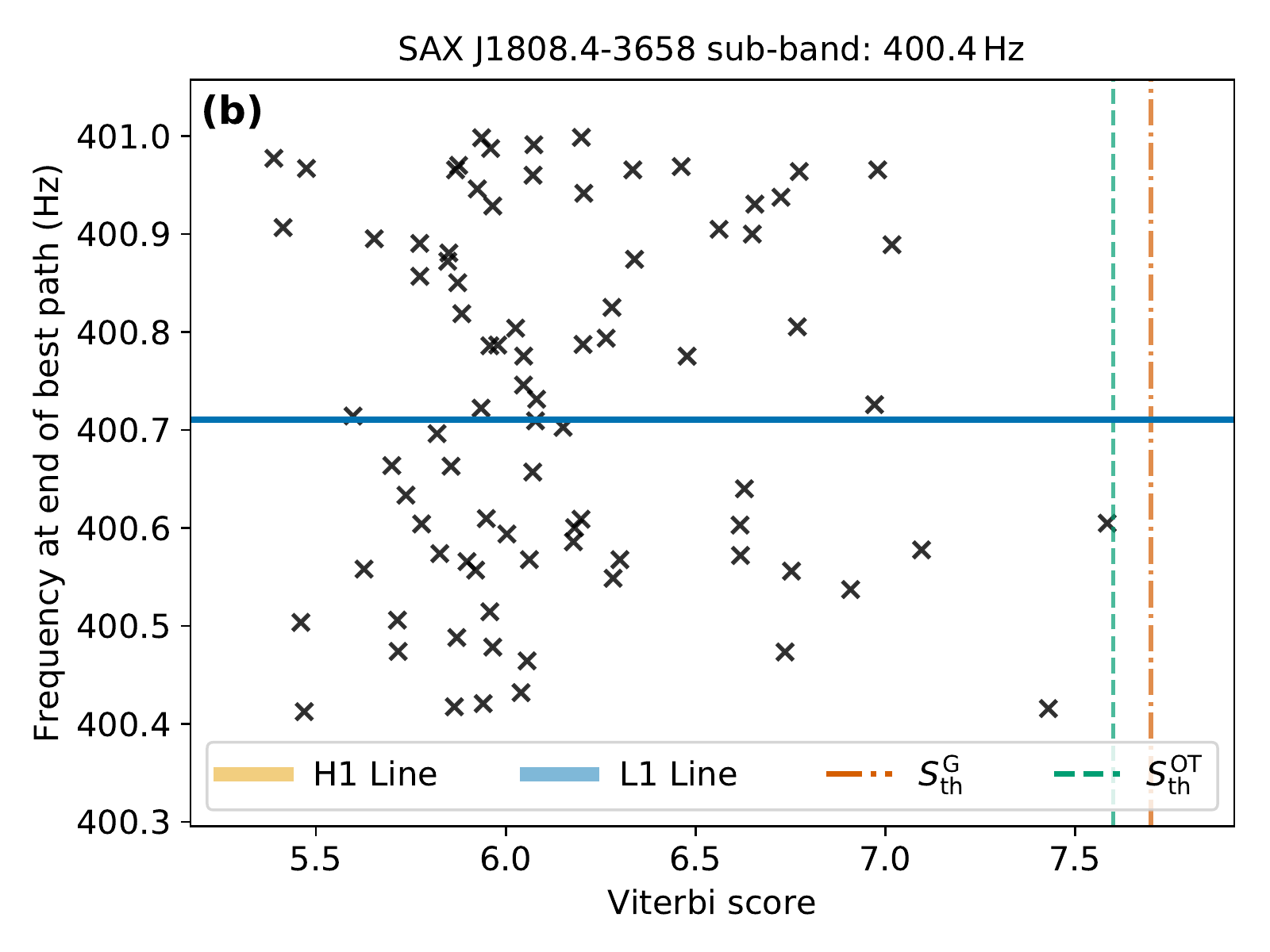}\\
\includegraphics[width=.49\textwidth]{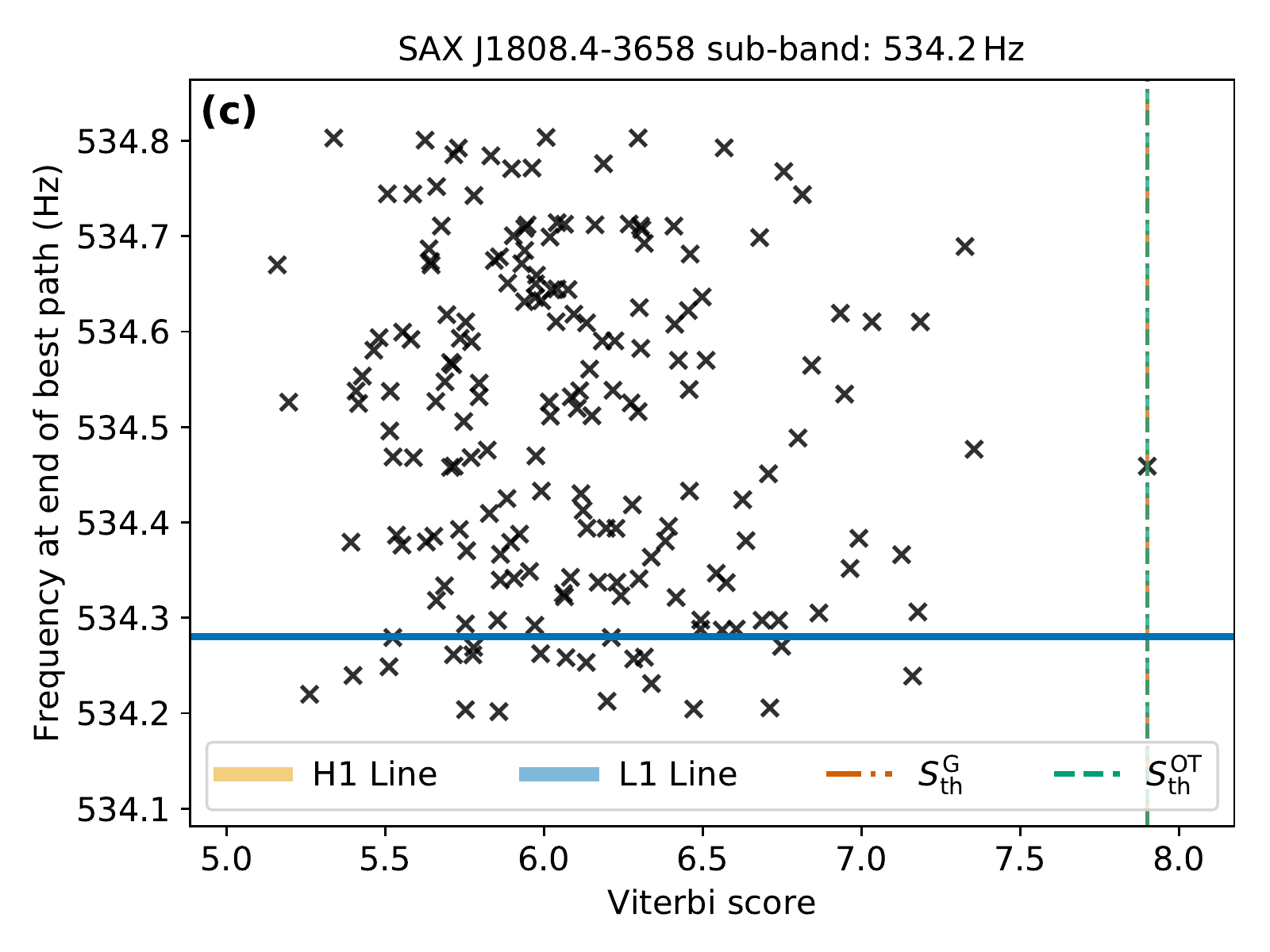}
\includegraphics[width=.49\textwidth]{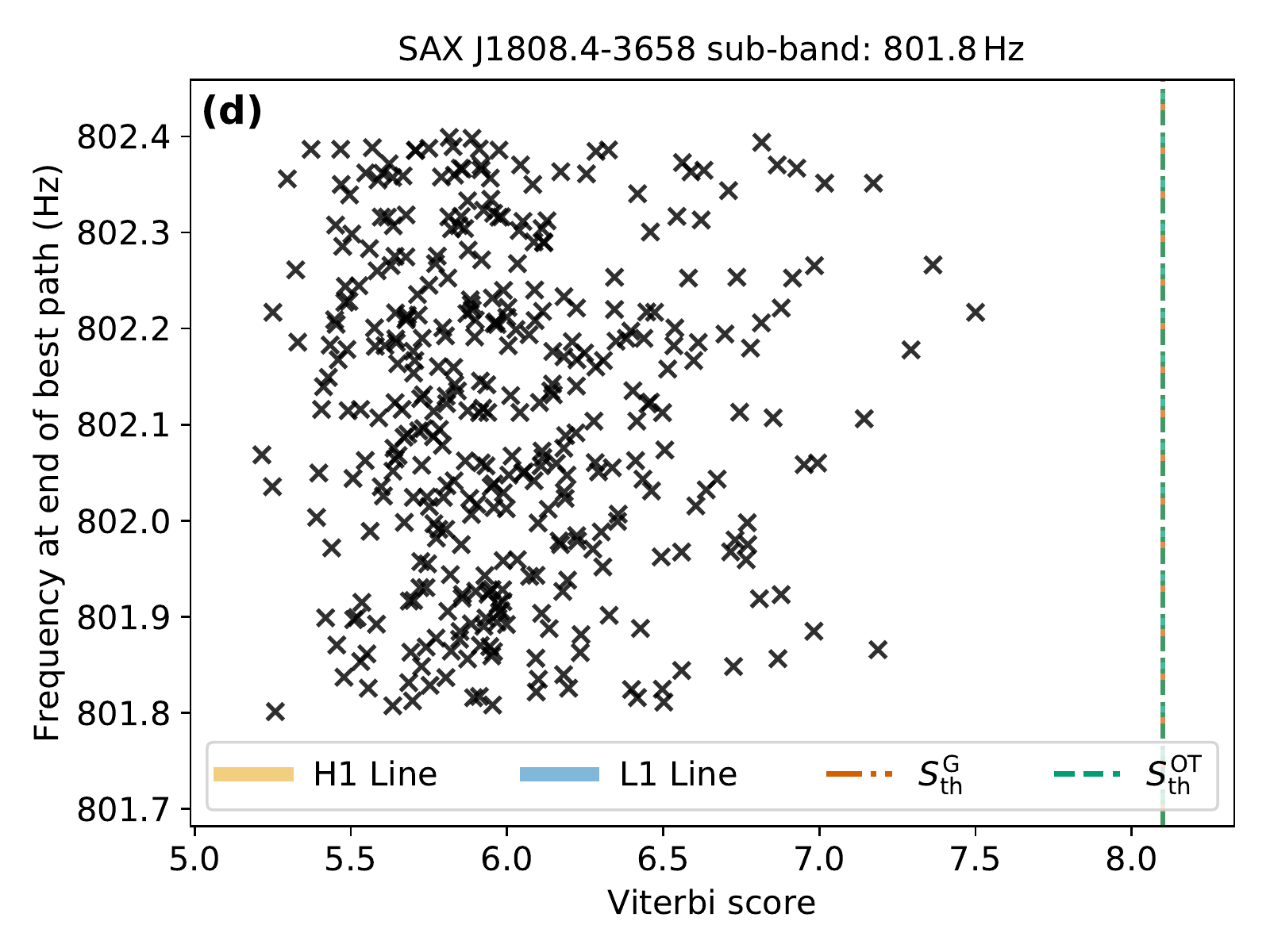}
\end{center}
\caption{\label{fig:SAXaSearch}
Search results for \SAXaName, laid out as in Fig.~\ref{fig:HETEaSearch}.
}
\end{figure*}

\begin{figure*}
\begin{center}
\includegraphics[width=0.48\textwidth]{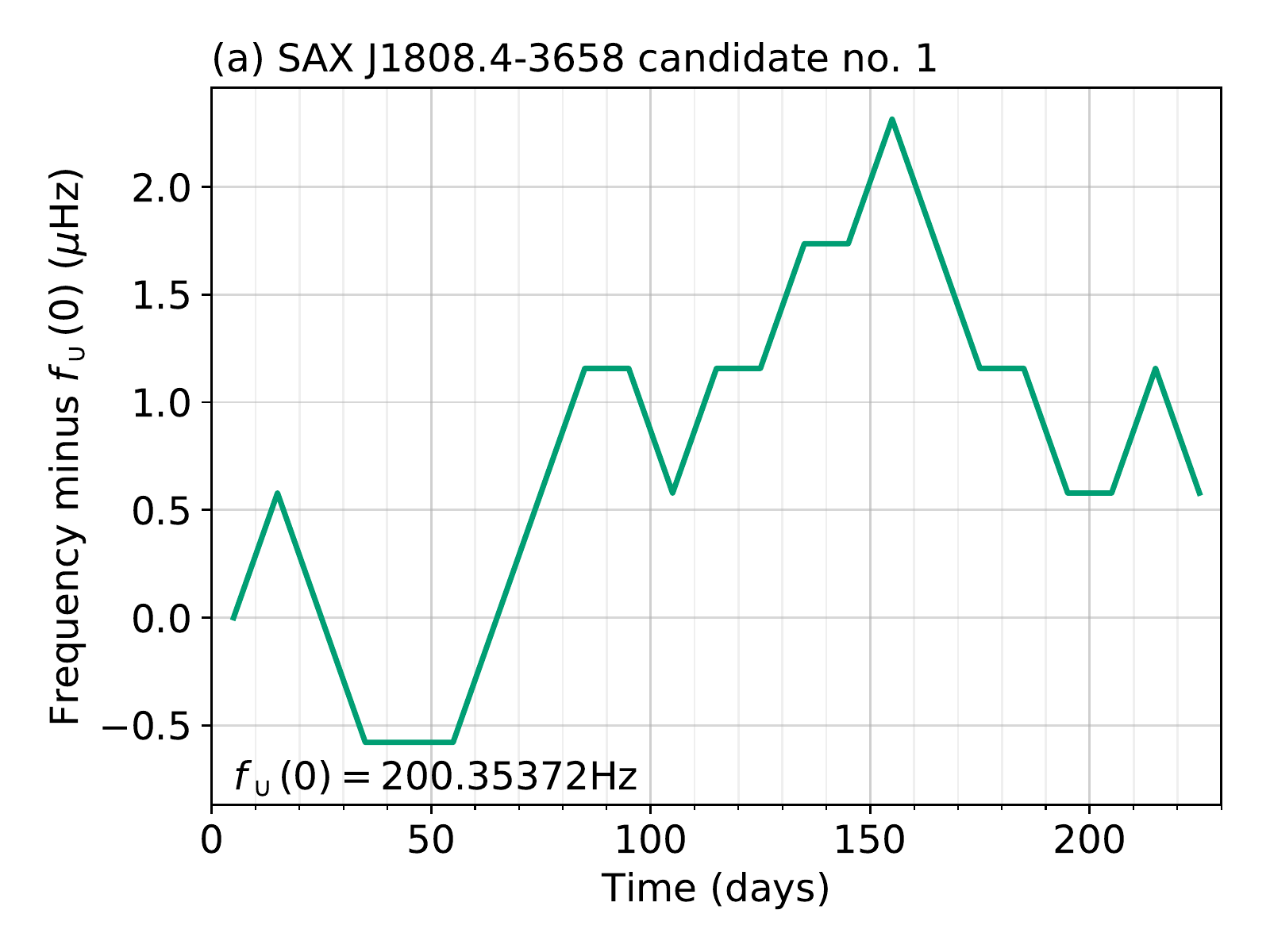}
\includegraphics[width=0.48\textwidth]{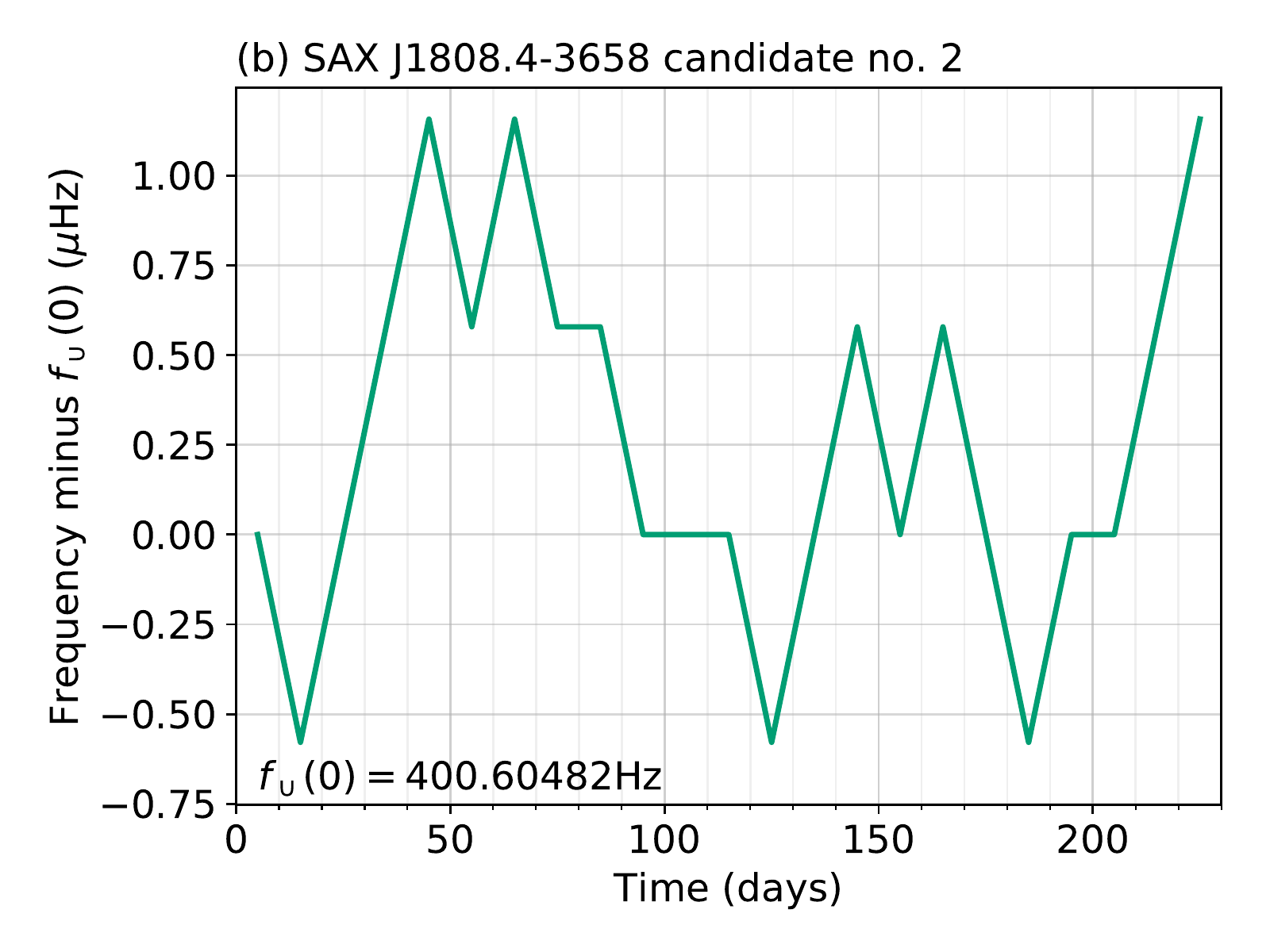}
\includegraphics[width=0.48\textwidth]{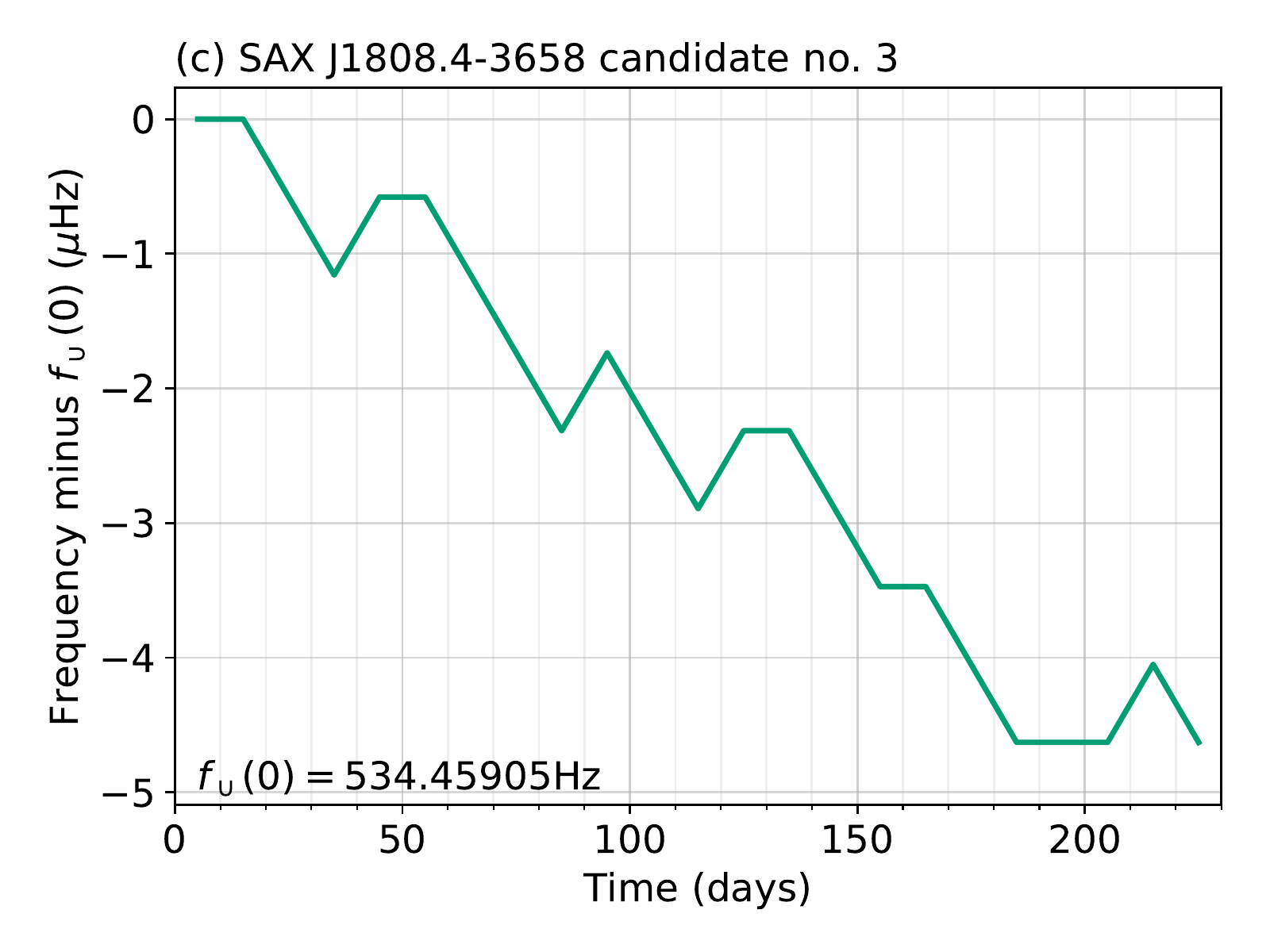}
\end{center}
\caption{\label{fig:SAXaPaths}
\SAXaName~candidate frequency paths laid out identically to Fig.~\ref{fig:HETEaPaths}.
}
\end{figure*}

\begin{figure*}
\begin{center}
\includegraphics[width=.49\textwidth]{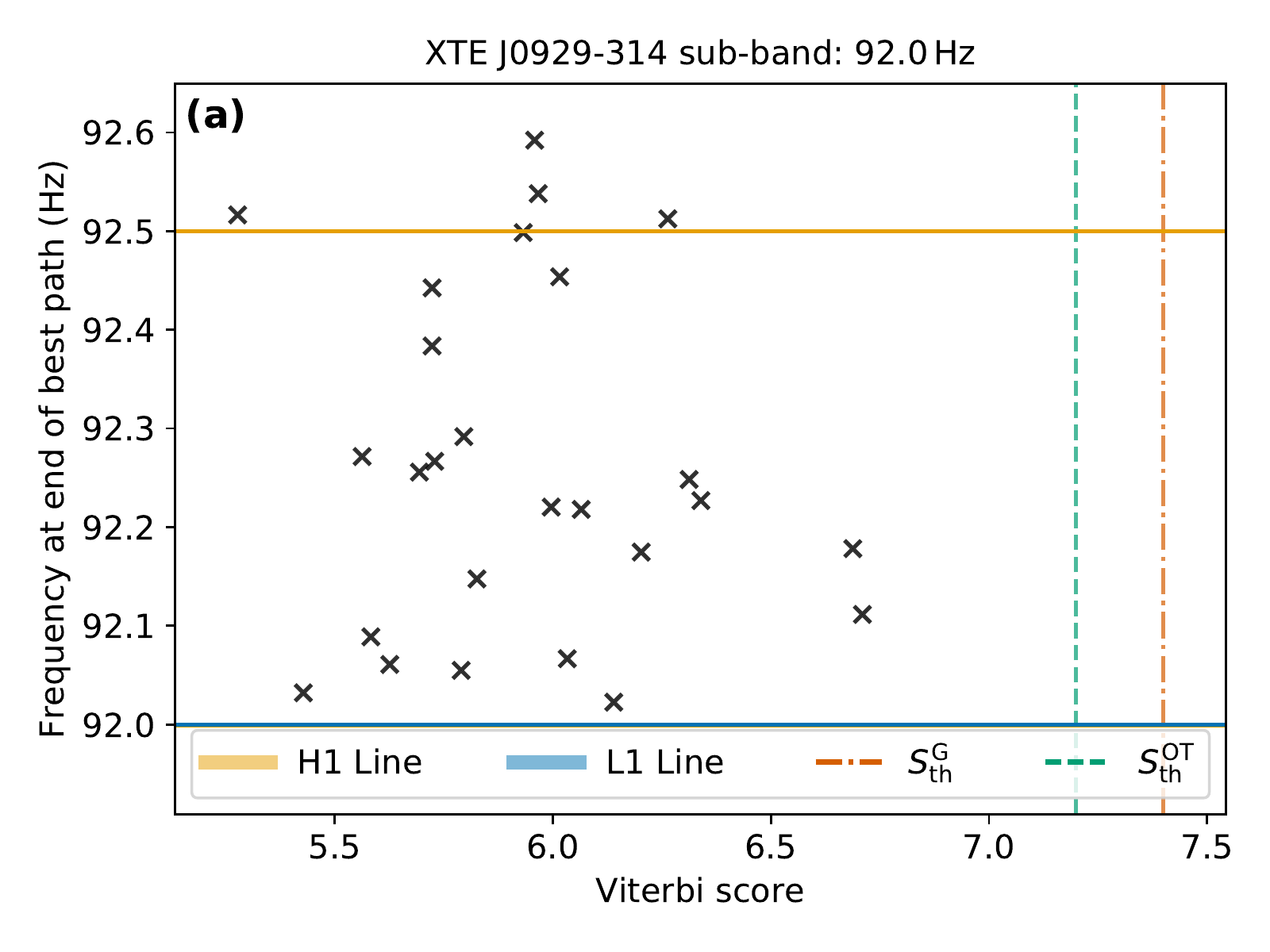}
\includegraphics[width=.49\textwidth]{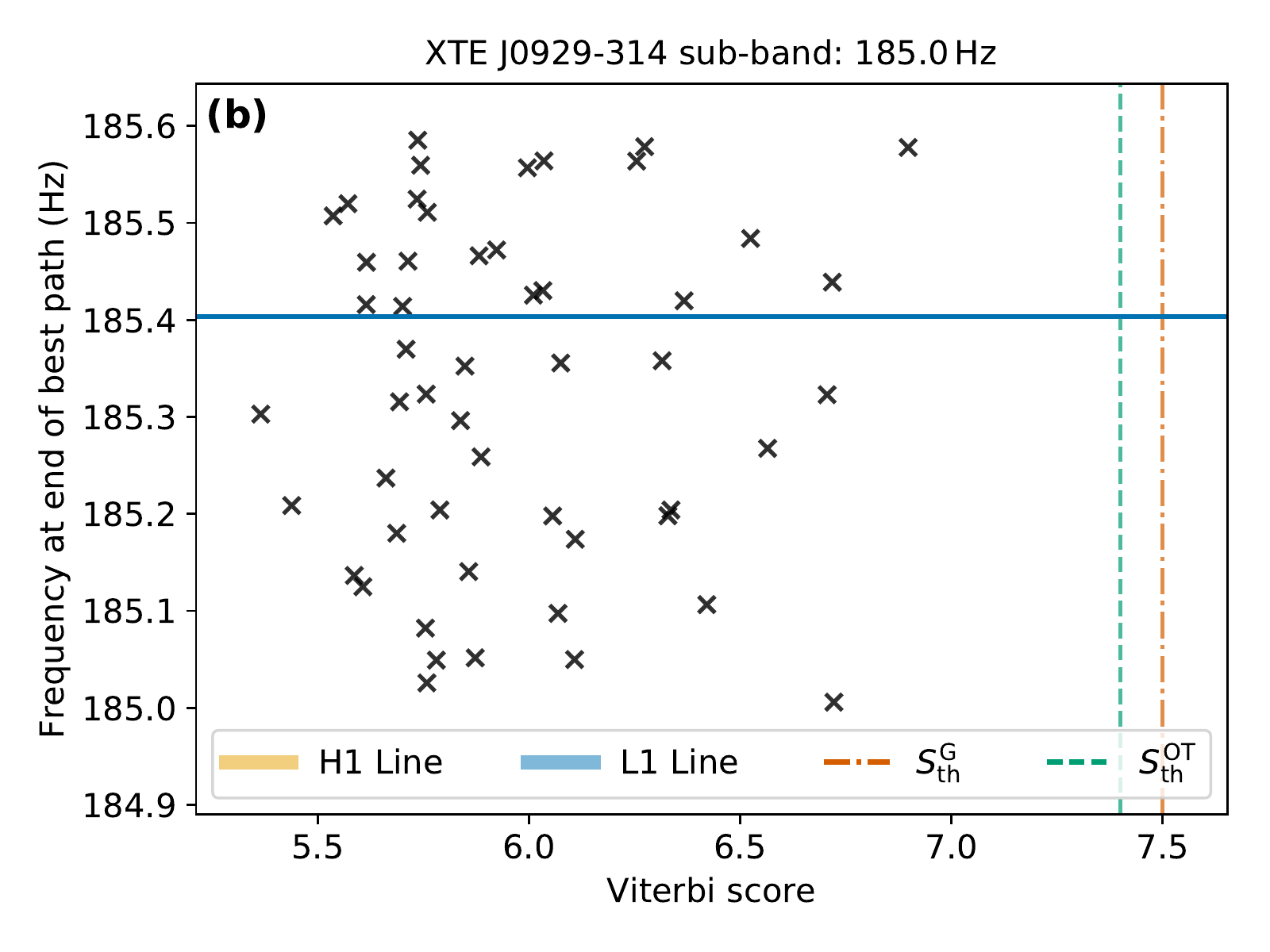}\\
\includegraphics[width=.49\textwidth]{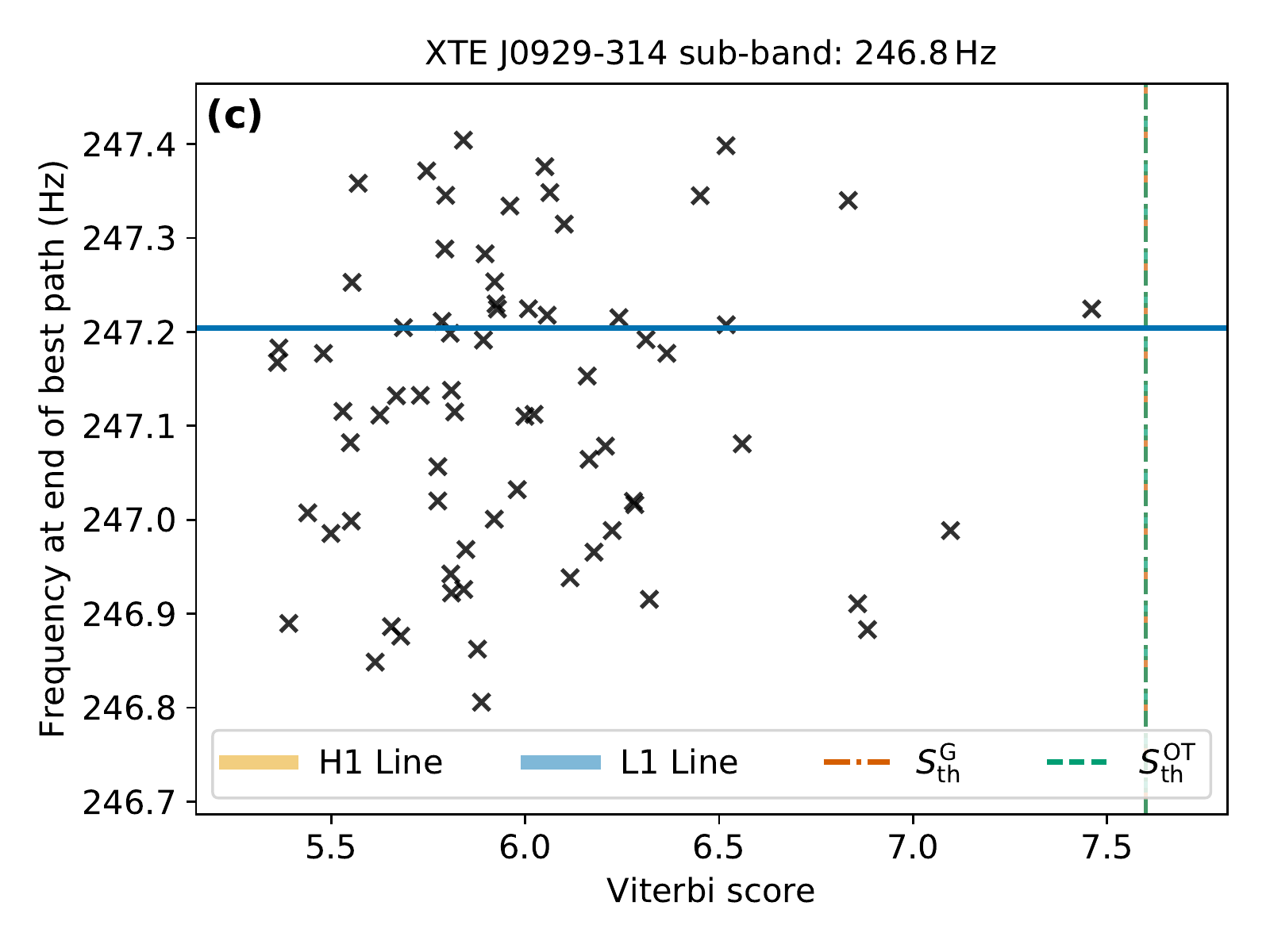}
\includegraphics[width=.49\textwidth]{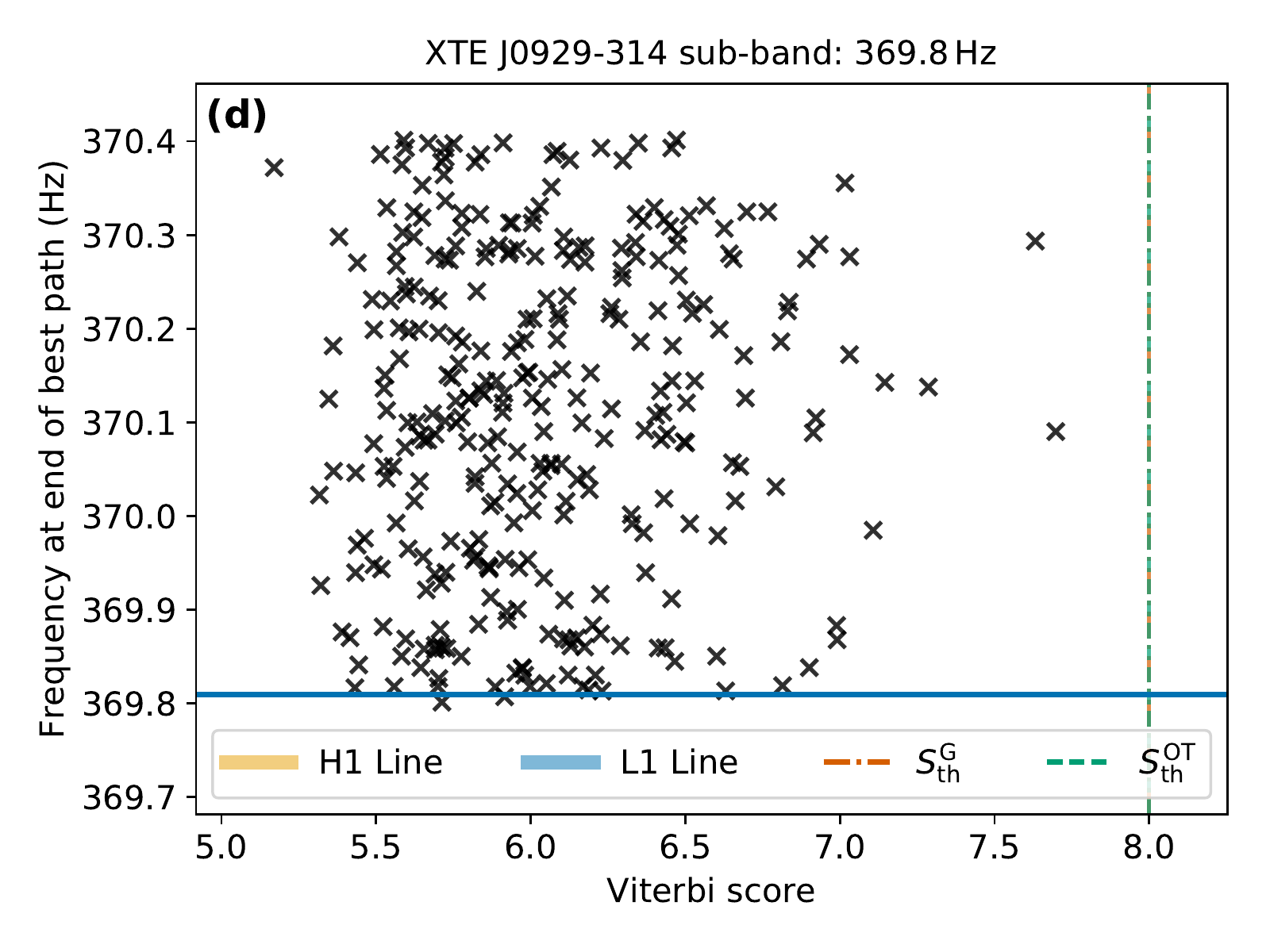}
\end{center}
\caption{\label{fig:XTEbSearch}
Search results for \XTEbName, laid out as in Fig.~\ref{fig:HETEaSearch}. 
}
\end{figure*}

\begin{figure*}
\begin{center}
\includegraphics[width=.49\textwidth]{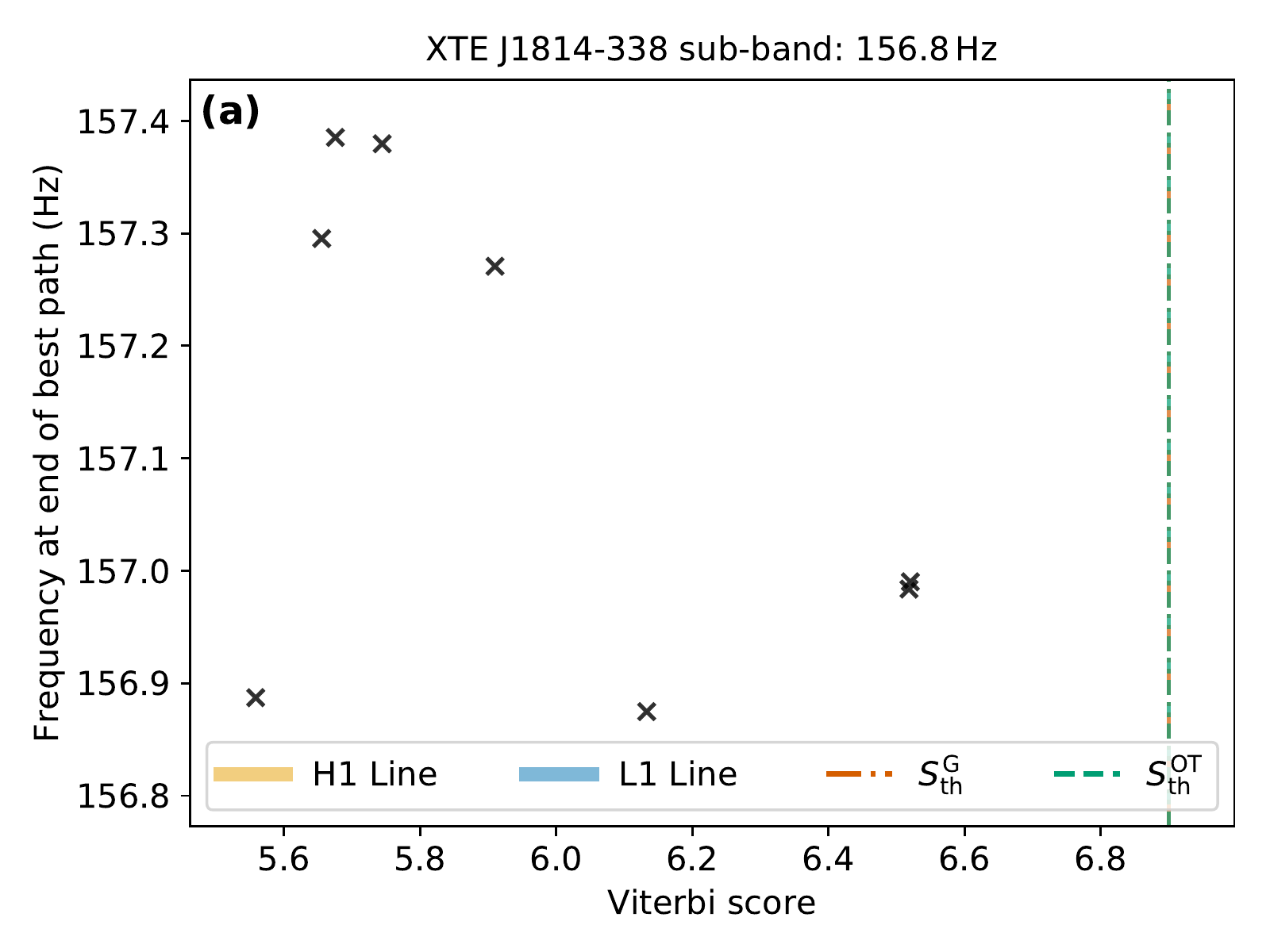}
\includegraphics[width=.49\textwidth]{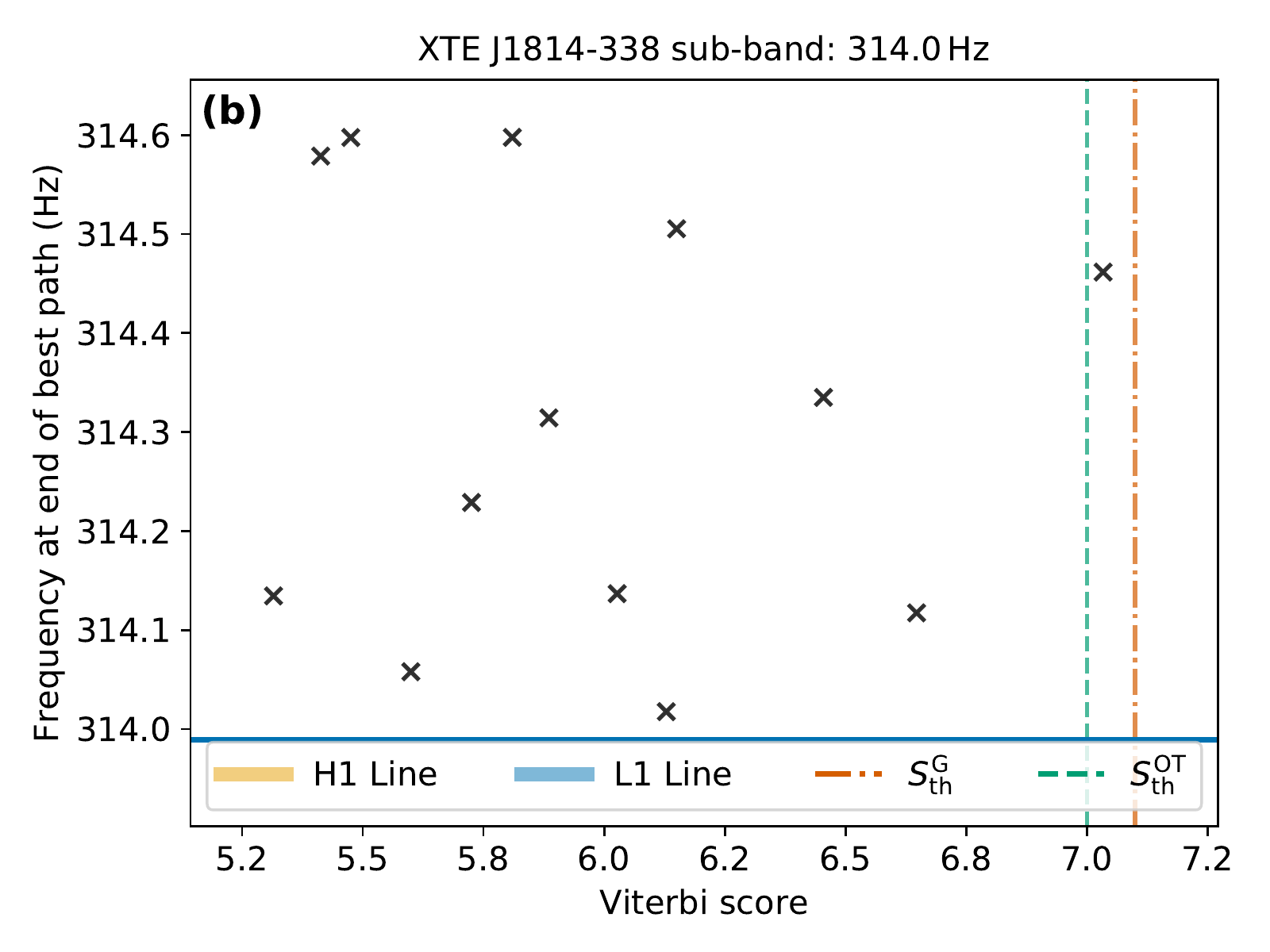}\\
\includegraphics[width=.49\textwidth]{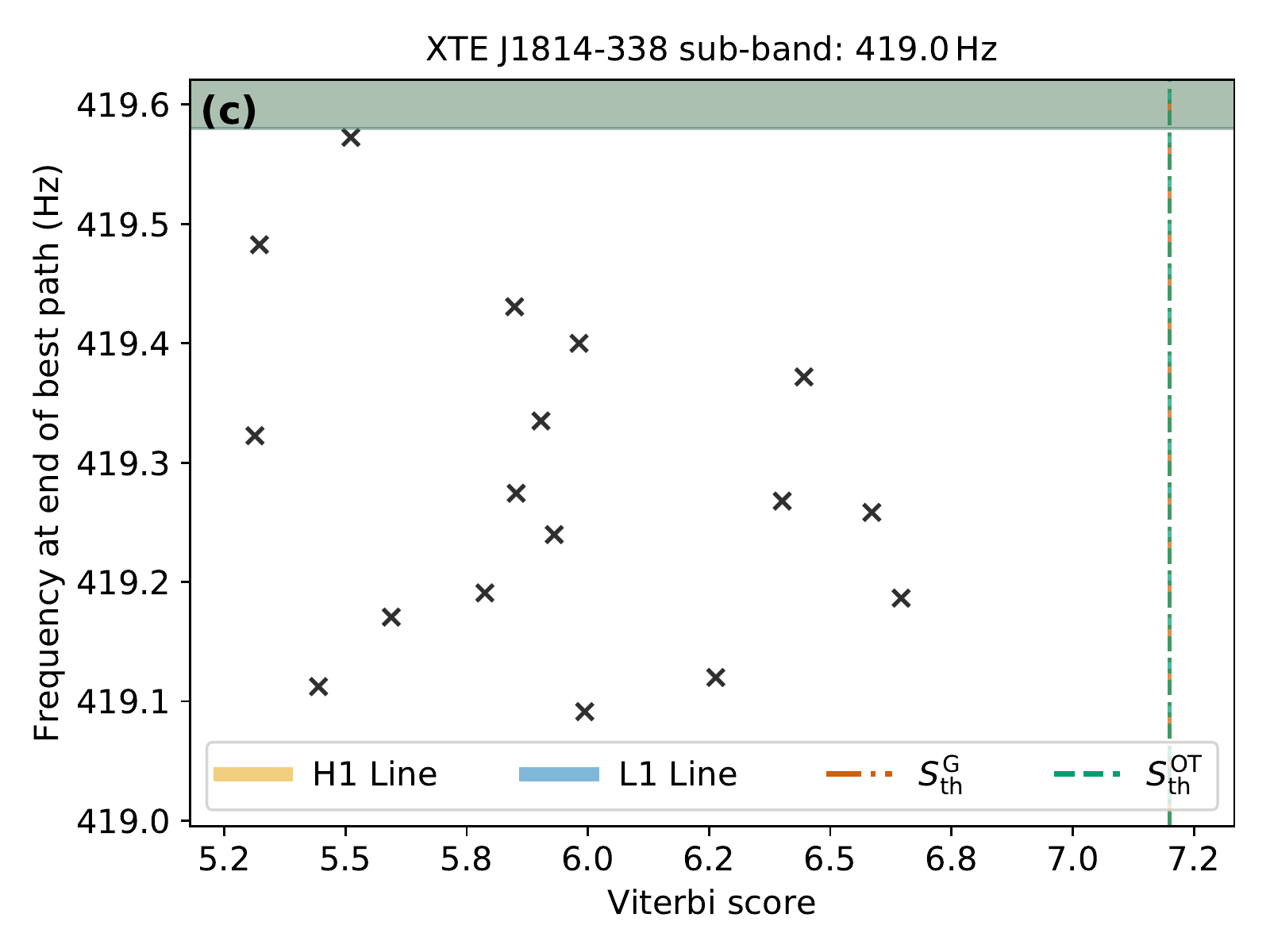}
\includegraphics[width=.49\textwidth]{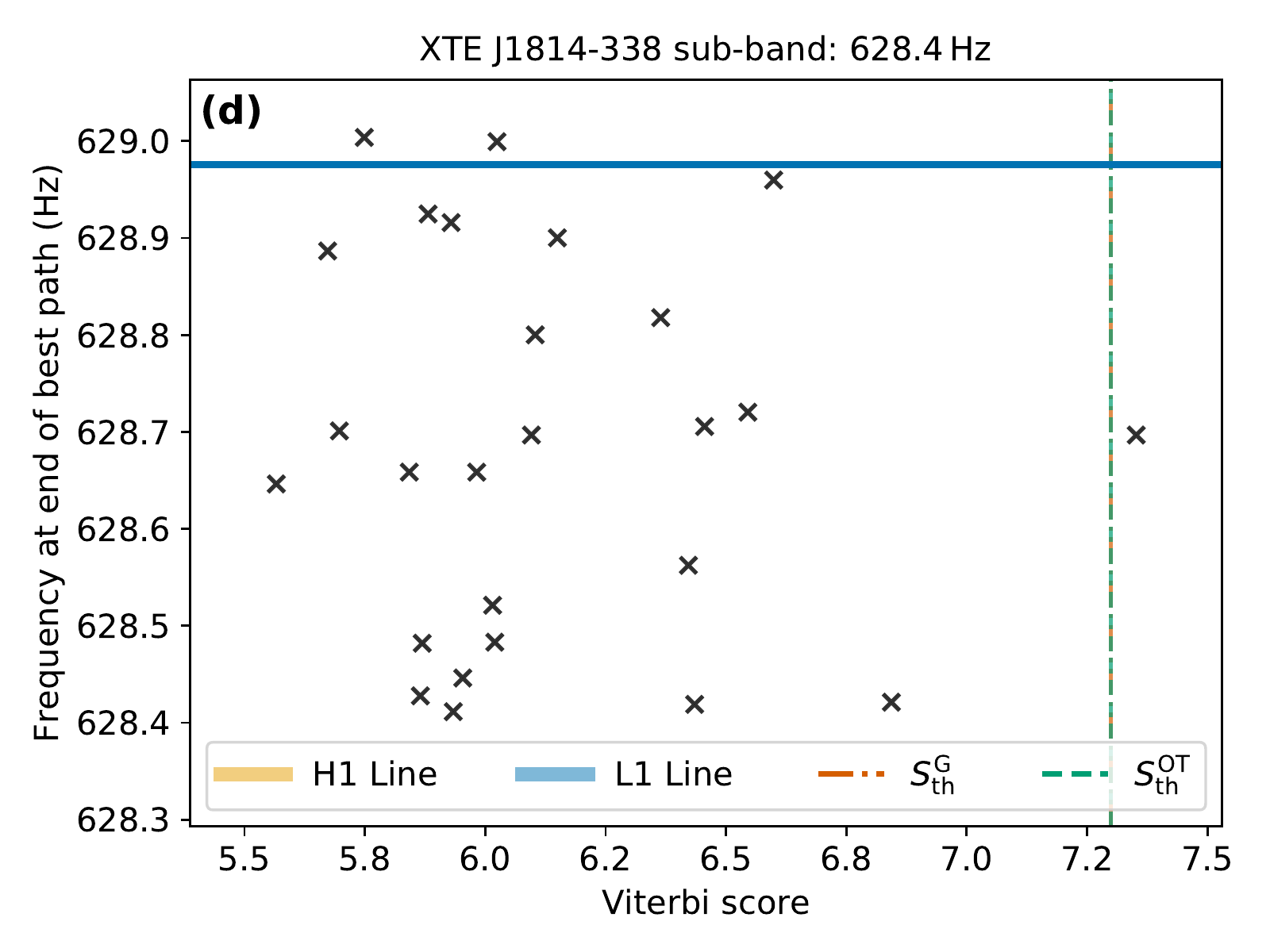}
\end{center}
\caption{\label{fig:XTEaSearch}
Search results for \XTEaName, laid out as in Fig.~\ref{fig:HETEaSearch}. 
The mixed-color shading at high frequencies in panel (c) is the overlap of instrumental lines from Hanford and Livingston. 
}
\end{figure*}

\begin{figure*}
\begin{center}
\includegraphics[width=0.48\textwidth]{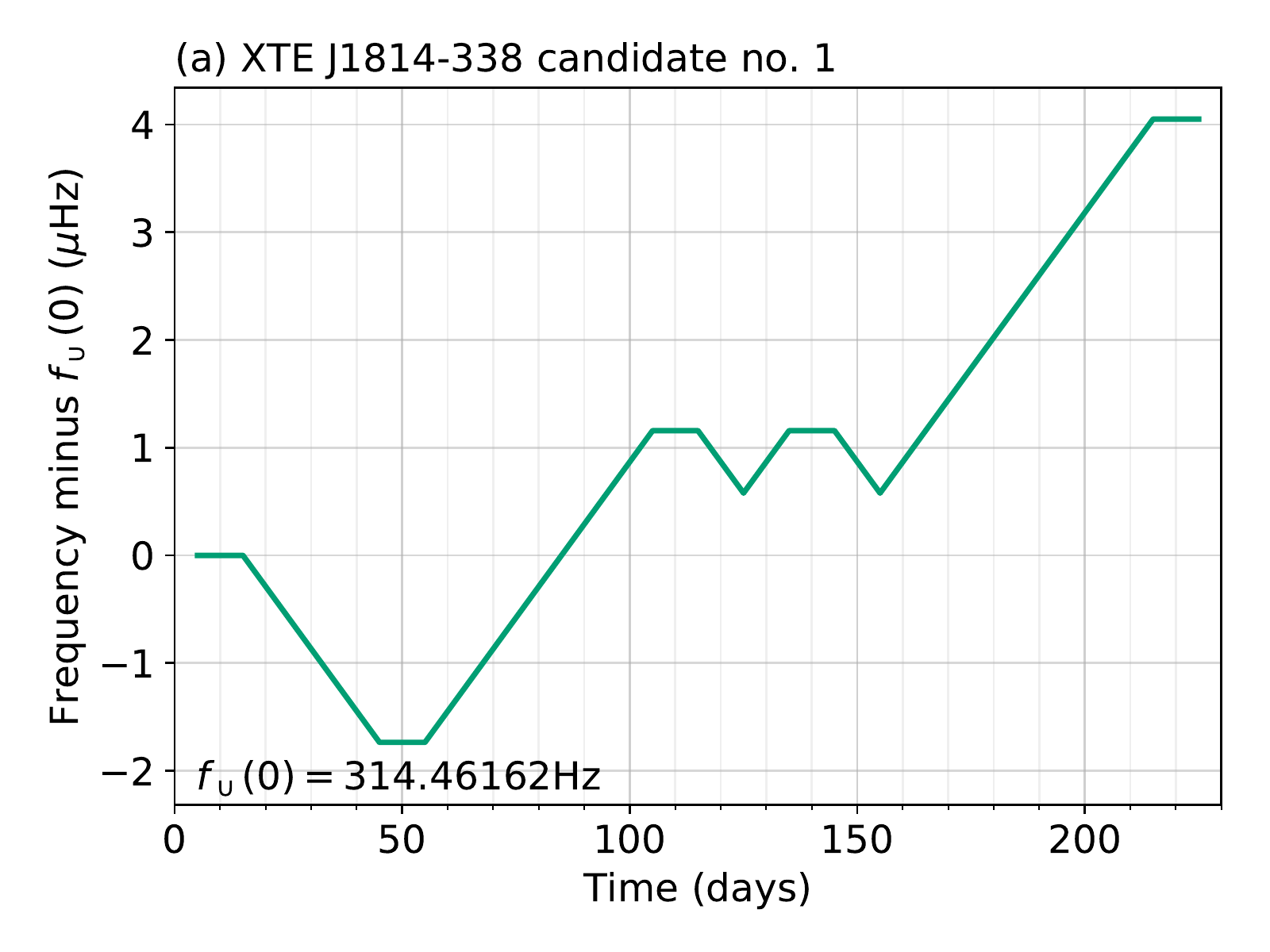}
\includegraphics[width=0.48\textwidth]{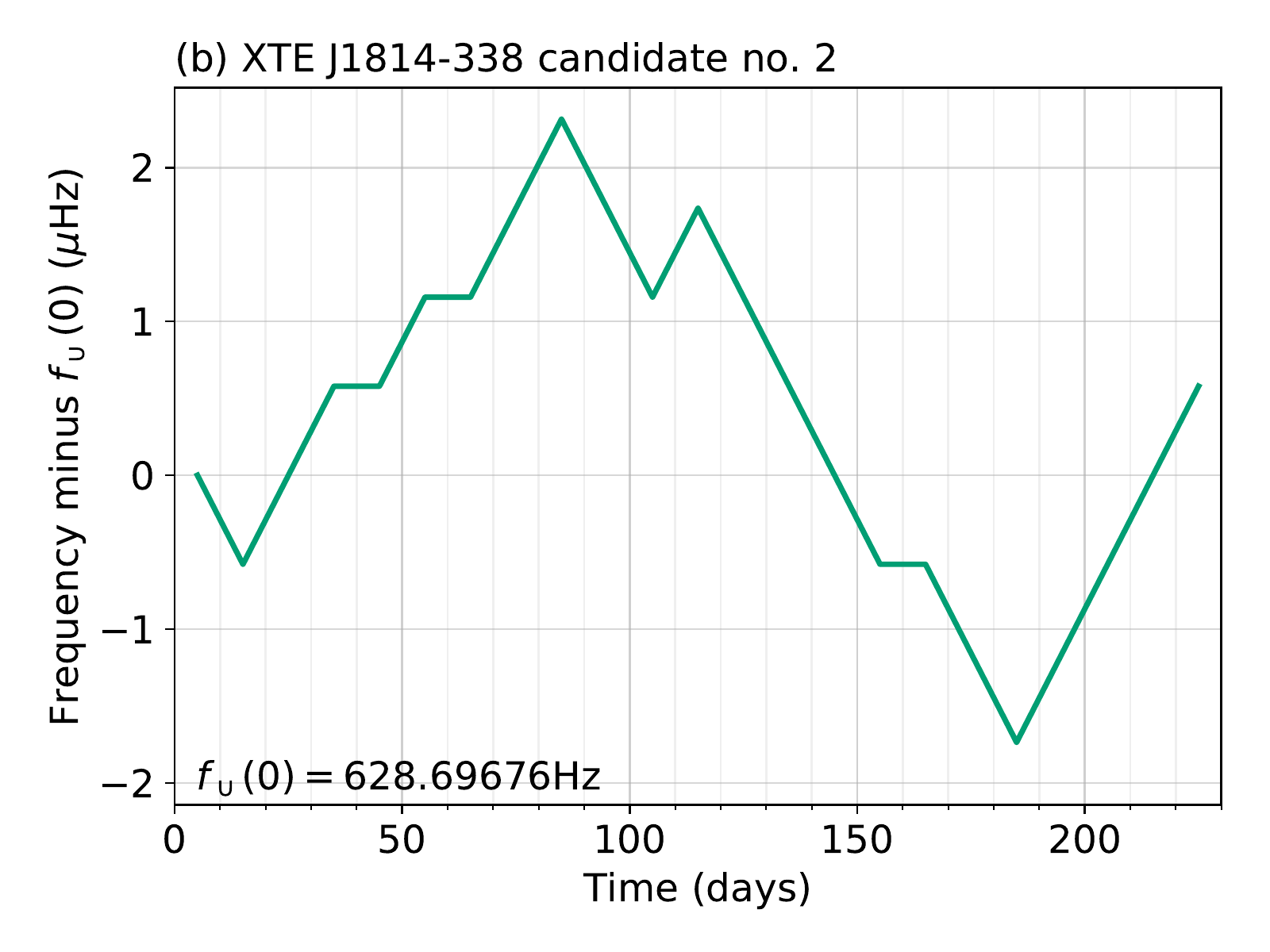}
\end{center}
\caption{\label{fig:XTEaPaths}
\XTEaName~candidate frequency paths laid out identically to Fig.~\ref{fig:HETEaPaths}.
}
\end{figure*}

Figure~\ref{fig:IGRaSearch} shows the search results for \IGRaName~identically laid out to the \HETEaName~results shown in Fig.~\ref{fig:HETEaSearch}.
In brief, the four panels of each figure show the $\{1/2, 1, 4/3, 2\}\fstar$ sub-bands, each marker indicates the terminating frequency of the best Viterbi path for each orbital template with the associated Viterbi score, the vertical lines show the $0.30$ FAP per-template thresholds, and the horizontal lines and shading indicate the presence of instrumental noise lines. 
Frequency paths for the five \IGRaName~templates listed in Table~\ref{tab:vetoOutcomes} are shown in Fig.~\ref{fig:IGRaPaths}. 
Search results and paths (for targets with candidates) are also shown for \SAXaName~(Figs.~\ref{fig:SAXaSearch} and~\ref{fig:SAXaPaths}), \XTEbName~(Fig.~\ref{fig:XTEbSearch}), and \XTEaName~(Figs.~\ref{fig:XTEaSearch} and~\ref{fig:XTEaPaths}).

%%%%%%%%%%%%%%%%%%%%%%%%%%%%%%%%%%%%%%%%%%%%%%%%%%%%%%%%%%%%%%%%%%%%%%%%%%%%%%%
%%%%%%%%%%%%%%%%%%%%%%%%%%%%%%%%%%%%%%%%%%%%%%%%%%%%%%%%%%%%%%%%%%%%%%%%%%%%%%%
%%%%%%%%%%%%%%%%%%%%%%%%%%%%%%%%%%%%%%%%%%%%%%%%%%%%%%%%%%%%%%%%%%%%%%%%%%%%%%%
%%%%%%%%%%%%%%%%%%%%%%%%%%%%%%%%%%%%%%%%%%%%%%%%%%%%%%%%%%%%%%%%%%%%%%%%%%%%%%%

%%%%%%%%%%%%%%%%%%%%%%%%%%%%%%%%%%%%%%%%%%%%%%%%%%%%%%%%%%%%%%%%%%%%%%%%%%%%%%%
%%%%%%%%%%%%%%%%%%%%%%%%%%%%%%%%%%%%%%%%%%%%%%%%%%%%%%%%%%%%%%%%%%%%%%%%%%%%%%%
%%%%%%%%%%%%%%%%%%%%%%%%%%%%%%%%%%%%%%%%%%%%%%%%%%%%%%%%%%%%%%%%%%%%%%%%%%%%%%%
%%%%%%%%%%%%%%%%%%%%%%%%%%%%%%%%%%%%%%%%%%%%%%%%%%%%%%%%%%%%%%%%%%%%%%%%%%%%%%%

%%%%%%%%%%%%%%%%%%%%%%%%%%%%%%%%%%%%%%%%%%%%%%%%%%%%%%%%%%%%%%%%%%%%%%%%%%%%%%%
%%%%%%%%%%%%%%%%%%%%%%%%%%%%%%%%%%%%%%%%%%%%%%%%%%%%%%%%%%%%%%%%%%%%%%%%%%%%%%%
%%%%%%%%%%%%%%%%%%%%%%%%%%%%%%%%%%%%%%%%%%%%%%%%%%%%%%%%%%%%%%%%%%%%%%%%%%%%%%%
%%%%%%%%%%%%%%%%%%%%%%%%%%%%%%%%%%%%%%%%%%%%%%%%%%%%%%%%%%%%%%%%%%%%%%%%%%%%%%%

%%%%%%%%%%%%%%%%%%%%%%%%%%%%%%%%%%%%%%%%%%%%%%%%%%%%%%%%%%%%%%%%%%%%%%%%%%%%%%%
%%%%%%%%%%%%%%%%%%%%%%%%%%%%%%%%%%%%%%%%%%%%%%%%%%%%%%%%%%%%%%%%%%%%%%%%%%%%%%%
%%%%%%%%%%%%%%%%%%%%%%%%%%%%%%%%%%%%%%%%%%%%%%%%%%%%%%%%%%%%%%%%%%%%%%%%%%%%%%%
%%%%%%%%%%%%%%%%%%%%%%%%%%%%%%%%%%%%%%%%%%%%%%%%%%%%%%%%%%%%%%%%%%%%%%%%%%%%%%%

\end{document}